\documentclass[fleqn,usenatbib]{mnras}

\usepackage{newtxtext,newtxmath}

\usepackage[T1]{fontenc}
\usepackage{anyfontsize}

\DeclareRobustCommand{\VAN}[3]{#2}
\let\VANthebibliography\thebibliography
\def\thebibliography{\DeclareRobustCommand{\VAN}[3]{##3}\VANthebibliography}

\usepackage{amsmath}	
\usepackage{float}
\usepackage{dsfont}
\usepackage{amsmath}
\usepackage{xcolor}
\usepackage{xparse}
\usepackage{listings}

\usepackage{graphicx} 
\usepackage{caption}
\usepackage{subcaption}
\usepackage{tabularray}
\usepackage[labelfont=bf]{caption}
\usepackage{gensymb}
\usepackage{setspace}
\usepackage{multirow}
\hypersetup{
  colorlinks=true,
    linkcolor=RoyalBlue,
    citecolor=RoyalBlue,
    filecolor=RoyalPurple,
    urlcolor=RoyalPurple
}
\usepackage[dvipsnames,svgnames,x11names]{xcolor}
\usepackage{comment}
\usepackage{textgreek}
\usepackage{xspace}
\usepackage{xargs}
\usepackage{multirow}

\newcommand{\Msun}{\ensuremath{\mathrm{M}_\odot}\xspace}
\newcommand{\jwst}{\textit{JWST}\xspace}

\newcommand{\Halpha}{\text{H\textalpha}\xspace}
\newcommand{\Hbeta}{\text{H\textbeta}\xspace}
\newcommand{\Hgamma}{\text{H\textgamma}\xspace}

\newcommandx{\permittedEL}[6][1=O,2=III,3=,4=,5=,6=]{\text{{#1}\,{\sc {#2}}{#3}{#4}{#5}{#6}}\xspace}
\newcommandx{\semiforbiddenEL}[6][1=O,2=III,3=,4=,5=,6=]{\text{{#1}\,{\sc {#2}}]{#3}{#4}{#5}{#6}}\xspace}
\newcommandx{\forbiddenEL}[6][1=O,2=III,3=,4=,5=,6=]{\text{[{#1}\,{\sc{#2}}]{#3}{#4}{#5}{#6}}\xspace}
\newcommand{\kms}{\ensuremath{\text{km s}^{-1}}\xspace}

\newcommand{\HeIIL}{\permittedEL[He][ii]}
\newcommand{\HeI}{\permittedEL[He][i]}

\newcommandx{\OIIL}[1][1=3728]{\forbiddenEL[O][ii][\textlambda][#1]}
\newcommand{\OIIall}{\forbiddenEL[O][ii][\textlambda][\textlambda][3727,][3729]}
\newcommand{\SIIall}{\forbiddenEL[S][ii][\textlambda][\textlambda][6716,][6731]}
\newcommand{\OIII}{\forbiddenEL[O][iii]}
\newcommand{\FeII}{\forbiddenEL[Fe][ii]}
\newcommandx{\OIIIL}[1][1=5007]{\forbiddenEL[O][iii][\textlambda][#1]}
\newcommand{\OIIIall}{\forbiddenEL[O][iii][\textlambda][\textlambda][4959,][5007]}
\newcommand{\NII}{\forbiddenEL[N][ii]}
\newcommandx{\NIIL}[1][1=6582]{\forbiddenEL[N][ii][\textlambda][#1]}
\newcommand{\NIIall}{\forbiddenEL[N][ii][\textlambda][\textlambda][6548,][6583]}

\newcommandx{\NeIIIL}[1][1=3869]{\forbiddenEL[Ne][iii][\textlambda][#1]}

\newcommand{\CIV}{\permittedEL[C][iv]}

\newcommand{\cliff}{\textit{The Cliff}\xspace}
\newcommand{\neigh}{ID~24647\xspace}

\newcommand{\orcidsymb}[2]{%
  \href{http://orcid.org/#2}{\textcolor{black}{#1}{\includegraphics[height=10pt]{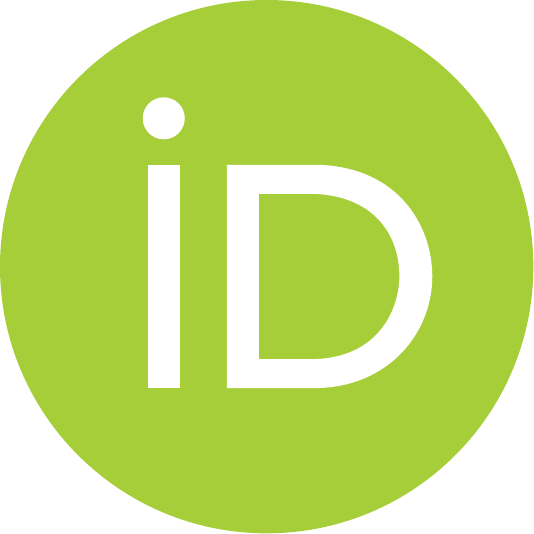}}}%
}

\title[The Cliff: A metal-poor LRD]{The Cliff: A Metal-Poor Little Red Dot Hosting an Overmassive Black Hole at $z = 3.55$}

\author[L. R. Ivey et al.]{\parbox[h]{\textwidth}{
\orcidsymb{L. R. Ivey,}{0009-0002-5105-1222}$^{1, 2}$\thanks{E-mail: li247@cam.ac.uk}
\orcidsymb{F. D'Eugenio,}{0000-0003-2388-8172}$^{1, 2}$
\orcidsymb{R. Maiolino,}{0000-0002-4985-3819}$^{1, 2, 3}$
\orcidsymb{Y. Isobe,}{0000-0001-7730-8634}$^{1, 2, 4}$,
\orcidsymb{I. Juodžbalis,}{0009-0003-7423-8660}$^{1, 2}$
\orcidsymb{S. Koudmani,}{0000-0002-1528-5091}$^{1, 5}$
\orcidsymb{M. Perna,}{0000-0002-0362-5941}$^{6}$
\orcidsymb{S. Zhang,}{0000-0003-1541-177X}$^{7, 8}$
\orcidsymb{V. Bromm,}{0000-0003-0212-2979}$^{7, 9}$
\orcidsymb{A. J. Bunker,}{0000-0002-8651-9879}$^{10}$
\orcidsymb{S. Carniani,}{0000-0002-6719-380X}$^{11}$
\orcidsymb{A. C. Fabian,}{0000-0002-9378-4072}$^{12}$
\orcidsymb{K. Inayoshi,}{0000-0001-9840-4959}$^{13}$
\orcidsymb{X. Ji,}{0000-0002-1660-9502}$^{1, 2}$ 
\orcidsymb{G. C. Jones,}{0000-0002-0267-9024}$^{1, 2}$
\orcidsymb{B. Liu,}{0000-0002-4966-7450}$^{14}$
\orcidsymb{R. Pascalau,}{0000-0001-9820-5773}$^{1, 2}$
\orcidsymb{P. Rinaldi,}{0000-0002-5104-8245}$^{15, 16}$
\orcidsymb{B. Robertson,}{0000-0002-4271-0364}$^{17}$
\orcidsymb{J. Scholtz,}{0000-0001-6010-6809}$^{1, 2}$
\orcidsymb{S. Tacchella}{0000-0002-8224-4505}$^{1, 2}$
}\vspace{0.4cm}
\\
$^{1}$Kavli Institute for Cosmology, University of Cambridge, Madingley Road, Cambridge, 
CB3 0HA, UK\\
$^{2}$Cavendish Laboratory, University of Cambridge, 19 JJ Thomson Avenue, Cambridge CB3 0HE, UK\\
$^{3}$Department of Physics and Astronomy, University College London, Gower Street, London WC1E 6BT, UK\\
$^{4}$Waseda Research Institute for Science and Engineering, Faculty of Science and Engineering, Waseda University, 3-4-1, Okubo, Shinjuku, Tokyo 169-8555, Japan \\
$^{5}$Centre for Astrophysics Research, Department of Physics, Astronomy and Mathematics, University of Hertfordshire, College Lane, Hatfield, AL10 9AB, UK \\
$^{6}$ Centro de Astrobiología (CAB), CSIC–INTA, Cra. de Ajalvir Km. 4, 28850 – Torrejón de Ardoz, Madrid, Spain
\\
$^{7}$Weinberg Institute for Theoretical Physics, Texas Center for Cosmology and Astroparticle Physics,
University of Texas at Austin, Austin, TX 78712, USA
\\
$^{8}$Department of Physics, University of Texas at Austin, Austin, TX 78712, USA
\\
$^{9}$Department of Astronomy, University of Texas at Austin, Austin, TX 78712, USA
\\
$^{10}$Department of Physics, University of Oxford, Denys Wilkinson
Building, Keble Road, Oxford OX1 3RH, UK
\\
$^{11}$Scuola Normale Superiore, Piazza dei Cavalieri 7, I-56126 Pisa, Italy
\\
$^{12}$Institute of Astronomy, University of Cambridge, Madingley Road, Cambridge CB3 0HA, UK
\\
$^{13}$Kavli Institute for Astronomy and Astrophysics, Peking University, Beijing 100871, China
\\
$^{14}$Institut f\"ur Theoretische Astrophysik, Zentrum f\"ur Astronomie, Universit\"at Heidelberg, Albert Ueberle Str. 2, D-69120 Heidelberg, Germany
\\
$^{15}$Space Telescope Science Institute, 3700 San Martin Drive, Baltimore, Maryland 21218, USA
\\
$^{16}$Steward Observatory, University of Arizona, 933 North Cherry Avenue, Tucson, AZ 85721, USA
\\
$^{17}$Department of Astronomy and Astrophysics, University of California, Santa Cruz, 1156 High Street, Santa Cruz, CA 95064, USA
}

\date{Accepted XXX. Received YYY; in original form ZZZ}

\pubyear{\the\year{}}

\begin{document}
\label{firstpage}
\pagerange{\pageref{firstpage}--\pageref{lastpage}}
\maketitle

\begin{abstract}
\jwst has revealed a large population of massive black holes (BHs) in the early Universe with unusual properties which mark them as distinct from low-redshift active galactic nuclei. Such findings have prompted the development of new models of BH formation and growth, and of their co-evolution with host galaxies. Linking the gas-phase metallicity of BH environments to seed masses is key to understanding which evolutionary pathways could explain the population of \jwst-discovered BHs. We present new high-resolution \jwst NIRSpec/IFU observations covering the rest-frame optical emission lines of a Little Red Dot (LRD) at $z=3.55$, known as \cliff, from the `Red Unknowns: Bright Infrared Extragalactic Survey' (RUBIES). We find evidence for low metallicity ($Z=0.017\pm0.004 \ Z_\odot$) based on the low narrow-line \OIIIL[5007]/\Hbeta ratio, supported by the non-detection of low-ionisation emission lines such as \OIIall and \NIIall. We find that the observed properties of \cliff, including its overmassive BH, can be reproduced by some simulations of black hole growth and evolution down to $z\sim3.5$. However, these simulation runs require high seed masses ($10^4 - 10^5\ M_\odot$) and appear as rarely in the simulation volume as in the RUBIES survey volume over redshifts $3<z<4$, highlighting the unusual nature of \cliff. Future simulations and numerical models will help to uncover how such a metal poor system managed to develop a massive black hole and persist to such low redshift.
\end{abstract}




\section{Introduction}

The James Webb Space Telescope (\jwst) has opened a new observational frontier for studying the formation and evolution of early black holes. 
The sensitivity and wavelength coverage of \jwst have enabled the detection of active galactic nuclei (AGN) at high redshift ($z>4$) with luminosities $L_{\mathrm{bol}}\sim 10^{42} - 10^{46} \ \mathrm{erg\,s^{-1}}$, rendering it possible to study low-luminosity AGN at higher redshifts than ever before \citep[e.g.,][]{Kocevski2023, Ubler2023, Harikane2023, Matthee2024, Maiolino2024_JADES, Maiolino2024_Nature, Kokorev2023, Greene2024, Taylor2025, Taylor2025_CAPERS, Juodzbalis2024_Nature, Juodzbalis2026_JADEScensus, Mazzolari2025, Chisholm2024, Scholtz2025, Adamo2025}.
The AGN most readily detected by \jwst are broad line AGN (BLAGN; i.e. Type 1 AGN), identified through the broad component of their permitted lines (usually \Halpha or \Hbeta). The absence of a corresponding broad component in the forbidden lines (e.g. \OIIIL) confirms an AGN Broad Line Region (BLR) as the origin for the broad emission, distinguishing it from potential outflows.
The low luminosities of \jwst-discovered AGN are likely due in part to lower black hole (BH) masses and/or lower accretion rates compared to the luminous high-redshift quasars previously identified by ground-based surveys \citep{Maiolino2024_JADES, Juodzbalis2024_Nature, Juodzbalis2026_JADEScensus}.

However, the \jwst-discovered low-luminosity AGN population has been found to exhibit a number of peculiarities. 
Crucially, these AGN are not simply the `high-redshift version' of local ($z<1$) AGN, nor are they `scaled-down versions' of high-redshift quasars, meaning \jwst AGN are likely a distinct population. 
Firstly, they show characteristic X-ray weakness \citep{Ananna2024, Yue2024_Stack, Maiolino2025_Chandra} and radio weakness \citep{Mazzolari2024, Mazzolari2025}, while both weak optical variability \citep[e.g.][]{Ji2025, Furtak2025, Naidu2025, Zhang2025_LRDvariability, Lin2026_variability} and non-variability \citep[e.g.][]{KokuboHarikane2025, Zhang2025} have been observed. 
The chemical enrichment associated with \jwst AGN is also poorly understood. Many \jwst AGN exhibit a lack of strong high-ionisation lines \citep[e.g.][]{Lambrides2024, Zucchi2026} or classical broad-line iron emission bumps \citep[e.g.][]{Trefoloni2025}. However, many \jwst AGN also exhibit significant nitrogen enhancement \citep{Ji2024, Isobe2025}, suggesting unexpectedly rapid chemical enrichment, potentially driven by extreme star formation or exotic stellar populations.

Many \jwst AGN also appear overmassive compared to the local black hole to stellar mass relation \citep[$M_\mathrm{BH}$--$M_\ast$; e.g.,][]{Maiolino2024_JADES, Furtak2024, Juodzbalis2024_Nature, Juodzbalis2026_JADEScensus, Chen2025, JonesKocevski2025, Ubler2023}. 
The presence of relatively massive BHs in early low-mass galaxies is remarkable \citep{Mezcua2024}, and raises fundamental questions about the co-evolution of black holes and their host galaxies \citep[e.g.][]{Smith_rev2019,Inayoshi_rev2020}. 
To a certain extent, the overmassive nature of \jwst AGN may be ascribed to selection effects \citep{Juodzbalis2026_JADEScensus, Li2025, Geris2026, Habouzit2025} or could point to $M_\mathrm{BH}-M_\mathrm{dyn}$ as a more fundamental relation between BHs and their host galaxies \citep{McClymont2026_BH, Danhaive2026_GEKO, Maiolino2024_JADES}. 
While selection effects alone cannot fully explain the observed shift in $M_\mathrm{BH}/M_\ast$ \citep{PacucciLoeb2024, JonesKocevski2025}, recent work by \cite{Ziparo2026} suggests that this evolution might not be a systematic shift in the mean $M_\mathrm{BH}-M_\ast$ relation (which remains consistent with local values at $z>4$), but rather a significant increase in its intrinsic scatter at higher redshifts.

Assuming certain colour selection criteria, approximately 10-30\% \citep{Hainline2025} of \jwst-discovered BLAGN at $z\sim5$ have been categorised as `Little Red Dots' \citep[LRDs;][]{Matthee2024, Hviding2025}.
LRDs exhibit a number of characteristic features, including rest-frame optical compactness, extreme redness and a V-shaped continuum \citep[e.g.][]{Matthee2024, Kocevski2025, Setton2025}. 
Other features commonly observed in LRDs \citep[though not exclusive to the LRD population, see e.g.][]{Brazzini2026} include exponential broad-line wings \citep[e.g.][]{Rusakov2026}, deep Balmer line absorption \citep[][]{Juodzbalis2024_Rosetta, Lin2026_COSMOS, DEugenio2025, DEugenio2026}, prominent Balmer breaks \citep[e.g.][]{Labbe2024, Baggen2024, Setton2025, deGraaff2025}, and X-ray weakness (e.g. \citealt{Yue2024_Stack,Maiolino2025_Chandra}, but see \citealt{Hviding2026} for an exception).

The Balmer absorption commonly seen in LRDs indicates high gas column densities along the line of sight \citep[][]{Juodzbalis2024_Rosetta, InayoshiMaiolino2025, Chang2026}, a feature which may also be related to their observed X-ray weakness \citep{Maiolino2025_Chandra}. 
The number density of LRDs has been found to decline sharply at $z<4$ \citep{Kocevski2025, Inayoshi2025_LRD}, though there are some luminous instances at lower redshifts \citep[e.g. $z\sim2-3$;][]{Juodzbalis2024_Rosetta, Loiacono2025, Wang2025, EuclidCollaboration2025, Ma2025_CountingLRD, Rinaldi2025} and a few local analogues \citep[e.g. $z\sim0.1$;][]{Lin2026, Ji2026}. This rapid evolution in LRD number density suggests that the obscuration and extreme gas column densities characteristic of LRDs may represent a transient phase of rapid black hole growth prevalent primarily in the early Universe.

A wide range of models have now been proposed for interpreting the peculiar observed properties of \jwst AGN, with fundamental differences in their initial seeding mechanisms. These can be broadly divided into two categories: (i) light seeds ($10^2-10^3 \ M_\odot$), possibly originating from Pop III stellar remnants \citep[e.g.][]{Nandal2025, Cammelli2025, Sanati2025, Prole2025}, and (ii) intermediate/heavy seeds ($10^4-10^6 \ M_\odot$) originating from the runaway merging of nuclear star clusters  \citep[e.g.][possibly resulting from the `feedback-free starbursts' proposed by \citealt{Dekel2025_LRD, Dekel2025}]{Partmann2025, Rantala2025}, or from the direct collapse of massive gas clouds \citep[e.g.][]{Natarajan2024, Jeon_DCBH2025, Pacucci2026}, or even primordial black holes \citep[e.g.][]{Zhang_PBH2025,DayalMaiolino2026}. Crucially, black hole seeding mechanisms are influenced by the gas-phase metallicity of the host galaxy. Pristine, low-metallicity environments are often cited as a prerequisite for heavy seed formation \citep[e.g.][]{Begelman2006, Ferrara2014, Latif2016, Maiolino2025_QSO1}, as such environments suppress gas fragmentation which would otherwise lead to normal star formation \citep{Inayoshi_rev2020}. By initiating growth from an already massive seed, a BH could outpace stellar assembly in its host galaxy even at early cosmic times \citep[$z\gtrsim4$; e.g.][]{Natarajan2024, Pacucci2023}. 
Conversely, higher metallicities could hinder heavy seed formation \citep{Latif2016} but may facilitate rapid growth via stellar mergers or super-Eddington accretion \citep{Inayoshi_rev2020}.

If LRD BH masses are verified as overmassive relative to their host galaxies \citep[see][for example]{Juodzbalis2025_QSO1}, they could be explained by invoking heavy seed models or even Primordial Black Holes \citep[PBHs; see e.g.][]{Dayal2024, DayalMaiolino2026, Zhang2025_hydrosim,Ziparo2025}. Alternatively, LRDs could represent an entirely new class of rapidly-accreting, low-mass BHs termed `black hole stars' \citep[][]{Naidu2025, Begelman2026} with cool gas envelopes \citep{Liu2025, Kido2025}. 
It has even been proposed that LRDs could simply be super-Eddington accreting black holes, which otherwise have the same structure as normal AGN \citep[][]{MadauMaiolino2026,Madau2026b}. Distinguishing between these pathways is essential to understanding the formation and early growth of BHs \citep[e.g.,][]{Jeon_LRDs2026}; in this regard, environmental metallicity may prove to be a valuable diagnostic for evaluating the plausibility of different seeding models.


A luminous, unlensed LRD at $z_\mathrm{spec} = 3.548$, also known as \cliff, was recently identified by the \textit{Red Unknowns: Bright Infrared Extragalactic Survey} \citep[RUBIES;][]{RUBIES2025, deGraaff2025}. 
The survey obtained \jwst Near Infrared Spectrograph (NIRSpec) microshutter assembly \citep[MSA;][]{Jakobsen2022,Ferruit2022} observations of \cliff with both the low-resolution PRISM/Clear ($0.6 - 5.3 \ \mu$m) and medium-resolution G395M/F290LP ($2.9 - 5.2 ~\mu$m) disperser/filter combinations. 
These observations revealed broad \Hbeta and \Halpha emission, typical of Type 1 AGN, along with the V-shaped continuum characteristic of LRDs. 
\cliff also shows an exceptionally strong Balmer break, for which \cite{deGraaff2025} exclude a stellar origin. 
\cite{deGraaff2025} also found \cliff hosts an AGN with mass of $\log(M_\mathrm{BH}/M_\odot) = 7.18^{+0.07}_{-0.06}$. 
Even with these medium-resolution observations they identify a significant absorption component in \Halpha, indicative of absorption by dense gas.

Another key and interesting feature of the original PRISM and G395M observations is the non-detection of the \OIIIL line. 
This could indicate a weak Narrow Line Region (NLR), which has previously been found for powerful high-redshift quasars \citep{Netzer2004}. 
In this paper, we revisit \cliff, now with recent high-resolution \jwst/NIRSpec Integral Field Unit (IFU) observations (G235H/F170LP; PID 9433), which finally reveal constraints on the narrow line kinematics and metallicity from \OIIIL. 
The high resolution spectrum shows a clear narrow component for both \Hbeta and \OIIIL, revealing the presence of either a NLR or ISM ionised by star-formation. 
In this work we analyse the high-resolution spectral data to assess the \OIIIL weakness more comprehensively and argue it is tracing low-metallicity gas within which the BH is embedded. 
We additionally discuss the implications of such a low-metallicity, high-$M_\mathrm{BH}$ system existing at $z\sim3.5$ in the context of chemical enrichment and black hole seeding.

The paper is organised as follows. 
In Section~\ref{sec:data} we describe the spectroscopic data and data reduction used in this work. 
We present our analysis of the data in Section~\ref{sec:Methods}, the results of which are then described in Section~\ref{sec:Results}.
In Section~\ref{sec:Discussion}, we discuss and interpret our findings in the wider context of other observations and simulations. 
Finally, our key observations and findings are summarised in Section~\ref{sec:Conclusions}. 
Throughout this work, unless stated otherwise, we assume a flat $\Lambda$CDM cosmology with $H_0 = 67.4 \mathrm{km} \ \mathrm{s}^{-1} \ \mathrm{Mpc}^{-1}$ and $\Omega_\mathrm{m} = 0.315$ \citep{Planck2020}. 
At $z = 3.548$, this means an on-sky separation of $1''$ corresponds to 7.453 kpc.
We additionally take solar metallicity to be defined by $[12+\log(\mathrm{O/H})]_\odot = 8.69$ \citep{Asplund2009}.

\section{Data}\label{sec:data}

In this work, we utilise NIRSpec-IFS observations from the Cycle 4 \jwst DDT programme 9433 (PIs R. Maiolino and F. D'Eugenio). 
The observations were taken with the high-resolution disperser/filter combination G235H/F170LP. We adopted a medium-cycling pattern with 10 dithers, 18 groups per integration and one integration per exposure, along with the NRSIRS2 readout mode \citep{Rauscher2017}. This yields a total on-source time of 3.7 h.

The data reduction uses the \jwst Science calibration pipeline v1.20.2 with CRDS jwst\_1464.pmap. To improve the quality of the datacube, we perform a number of further data processing steps, in addition to the default pipeline. The customised procedures used for flagging of residual cosmic ray snowballs, subtraction of pink noise (1/f noise), and removal of failed open MSA shutters signal are described in detail in \cite{Perna2023, Perna2025}. Stage~2 was executed using the bad-pixel self-calibration implemented in the standard pipeline, in which all exposures on a given detector are used to identify and flag bad pixels that may have been missed by the reference bad-pixel mask. For the final stage of the pipeline, the outlier rejection was performed using the standard \jwst pipeline implementation. Before combination, the cube was resampled to a scale of $0.05''$ per spaxel; the final cubes were combined using the `drizzle’ method.

\begin{figure}
\centering
   \includegraphics[width=\linewidth]{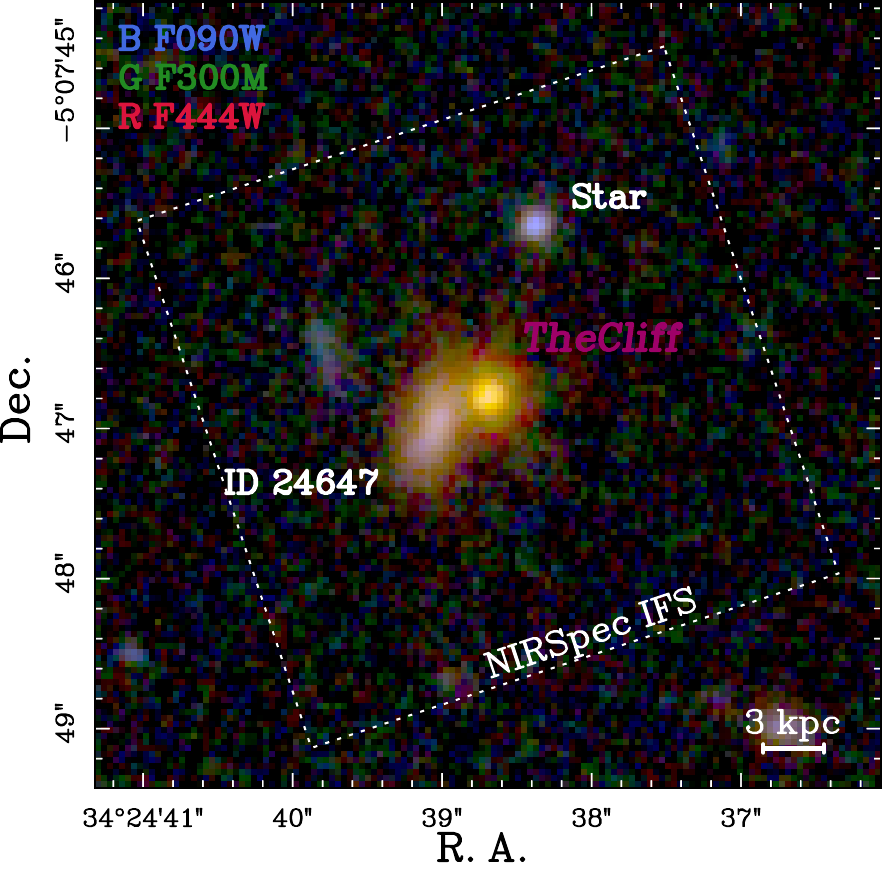}
\caption{False-colour RGB image of \cliff, illustrating the NIRSpec/IFS FoV. We label a foreground galaxy (\neigh, at $z_\mathrm{spec}=3.05$) to the east and a star to the north west. We use publicly available imaging from PRIMER (PID~1837) and MINERVA~\citep[PID~7814;][]{Muzzin2025}.}\label{fig:fov}
\end{figure}

We note that the NIRSpec field of view (FoV) of \cliff includes both a foreground galaxy at $z_\mathrm{spec}=3.05086\pm0.00002$ \citep[CANDELS~\neigh;][]{Grogin2011,Koekemoer2011},
and a star \citetext{see \citealp{deGraaff2025} and Fig.~\ref{fig:fov}}.
To perform background subtraction, we created a mask of source-free regions in the FoV, carefully excluding \cliff, the foreground galaxy and star. 
The background level is estimated independently for each spectral channel, performing $\sigma$-clipping with a threshold of $\sigma = 5$ from the median, with the local sky level then defined as the median of the resulting distribution. 
The final estimated background cube is smoothed with a median filter to avoid introducing additional noise, and is then subtracted from the flux data cube.

As noted by earlier studies \citep[e.g.][]{Ubler2023, RodriguezDelPino2024}, the \textsc{err} extension of the NIRSpec-IFS cubes is underestimated.
Hence, to obtain realistic uncertainties on our aperture-integrated spectra while retaining crucial information about outliers and the correlated noise between channels, we adopt the same procedure as previous works: rescaling \textsc{err} to match the $\sigma$-clipped r.m.s. noise of the integrated spectrum in regions free of line emission.

\section{Data Analysis}\label{sec:Methods}

\subsection{PSF and aperture determination}\label{sec:Size}

The optimal aperture from which an integrated spectrum of \cliff is extracted is not perfectly circular, as it is necessary to account for the shape of the IFU observations' point spread function (PSF). 
The PSF of IFU observations is non-trivial to determine, for two key reasons: (i) the observational PSF does not exactly follow the theoretical \jwst PSF, and
(ii) the PSF tends to be slightly elongated in the direction along the IFU slices. 
In the case of \cliff, we leverage the star present in the IFU FoV as a reference point-source to obtain a direct measurement of the PSF for our observations (see Appendix~\ref{sec:IFSPSF}).

To measure the PSF shape and full-width at half maximum (FWHM), we utilise the morphology modelling tool \textsc{pysersic} \citep{PashaMiller2023}. 
We generate a set of square narrow-band image cutouts centred on the star by median-stacking all wavelength slices within spectral windows spaced by 0.1~\textmu m in the NRS1 detector (where the star is brightest), and by 0.2~\textmu m in the NRS2 detector (which samples longer wavelengths). 
We model these synthetic images using a 2D Gaussian, implemented as a S\'ersic profile with fixed index $n=0.5$, along with a linear sky background. 
We adopt generous flat priors on all parameters, except for position angle (P.A.), which has a narrow flat prior within $\pm 5$\textdegree\ from the NIRSpec position angle.
We find that in NRS1 no constraints on the position angle are necessary, and we find a position angle that is fully consistent with the NIRSpec position angle, i.e. $\mathrm{P.A.} = 198\degree$. 
However, in NRS2 the position angle is loosely constrained, likely due to the combination of a mildly rounder shape and lower SNR.
For consistency, therefore, we adopt the same P.A. for all wavelengths.
As \textsc{pysersic} also requires an input PSF for the fit, we use a Gaussian with FWHM = 0.1~spaxels, a very small width representing a delta distribution; using a much smaller FWHM would incur artefacts in the fitting procedure. 
The PSF parameters are estimated with a simple maximum-likelihood estimator. 
We find that for our main spectral region of interest, the \Hbeta-\OIIIL complex, the corresponding PSF is well-described by an ellipse with axis ratio $q = 0.86$ and position angle $\mathrm{P.A.} = 198\degree$; the direction of elongation of the PSF is along the IFU slices.

\begin{figure}
\centering
   \includegraphics[width=\linewidth]{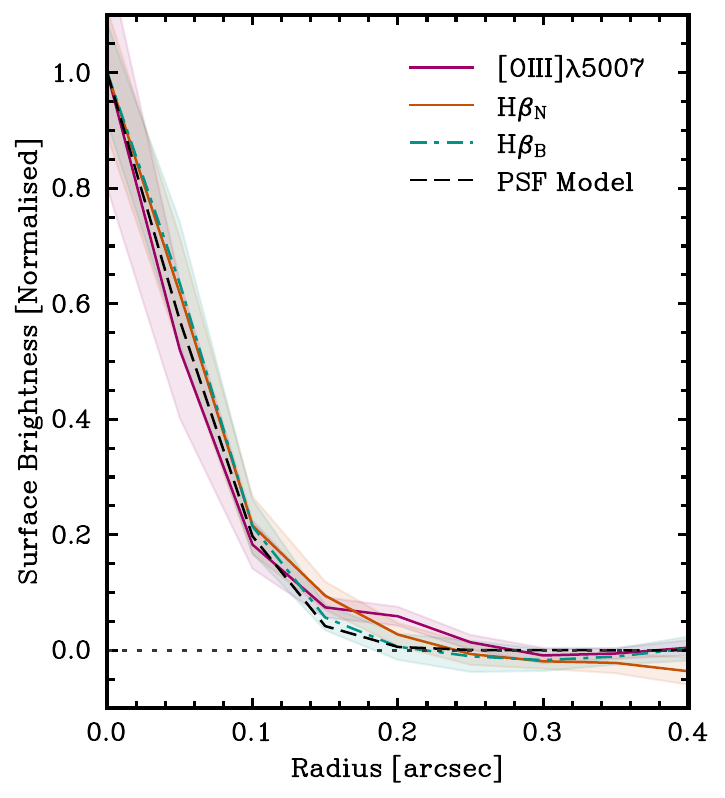}
\caption{Radial surface brightness profiles (normalised to the peak) of the narrow \Hbeta and \OIIIL (the solid orange and pink lines, respectively) and broad \Hbeta (dot-dashed green line) emission lines. 
The PSF model of \cliff (see Section~\ref{sec:Size} and Appendix \ref{sec:IFSPSF}) is shown by the black dashed line.
The shaded areas correspond to $1\sigma$ uncertainties on the mean for each profile. 
The emission lines have compact Gaussian cores, and do not show any significant extension beyond the scope of the PSF. While there are hints of extension out to a radius of $0.2''$ in \OIIIL[5007], beyond the scale of the PSF, it should be noted that the wings of the PSF are not well-constrained and therefore this cannot be interpreted as conclusive evidence for spatial extension of \cliff. 
Overall, these profiles suggest \cliff is spatially unresolved in NIRSpec/IFS.}
\label{fig:surfacebrightnessprofiles}
\end{figure}

\begin{figure}
\centering
   \includegraphics[width=\linewidth]{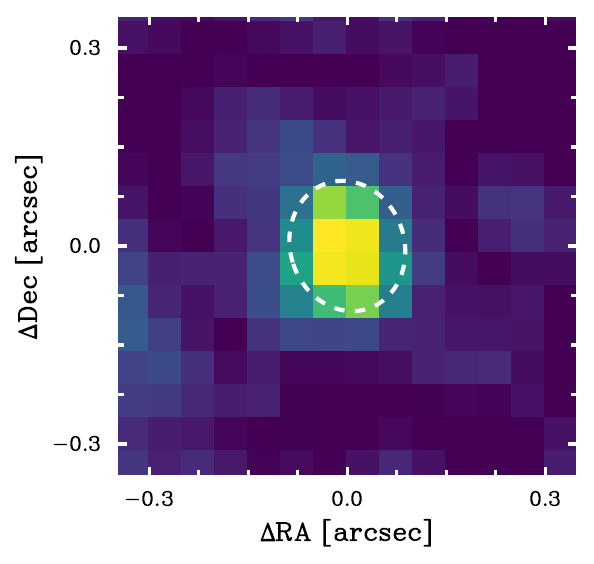}
\caption{Map of narrow \Hbeta emission, obtained by collapsing the three central channels of the continuum- and background-subtracted line (see Section~\ref{sec:fitting_Hb_OIII} for the line shapes). 
This is overlaid with a white dashed ellipse illustrating the central aperture used for extracting the spectra shown in Figs.~\ref{fig:full_spectra}, \ref{fig:spectra} and \ref{fig:Halphafit}. 
While this map may include some contribution from the broad \Hbeta component, this has no impact on our results as it is not used for any quantitative measurements.}
\label{fig:aperture}
\end{figure}

\begin{figure*}
        \centering
	\includegraphics[width=\linewidth]{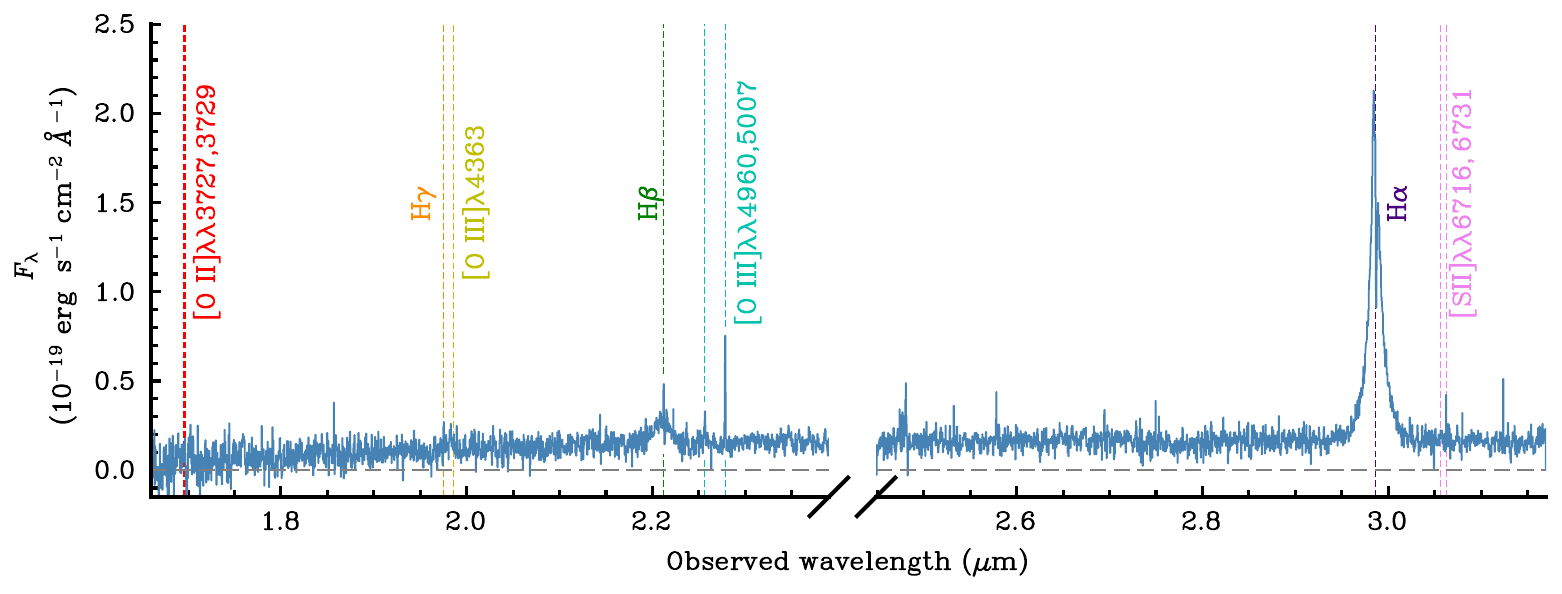}
    \caption{Full, background-subtracted NIRSpec-IFU G235H/F170LP integrated spectrum of \cliff, extracted from the aperture shown in Fig.~\ref{fig:aperture}. 
    The locations of key emission lines are labelled with coloured vertical lines (except \NIIall, to avoid overcrowding near \Halpha). 
    This spectrum has not been corrected for aperture losses.}
    \label{fig:full_spectra}
\end{figure*}

To determine the size of the optimal integrated spectrum extraction aperture, it is important to first investigate whether \cliff shows any evidence for spatially-extended emission. 
In Fig.~\ref{fig:surfacebrightnessprofiles}, we present the radial surface brightness profiles of the \Hbeta and \OIIIL emission lines, compared to the profile of our best-fit PSF model (dashed black line; also refer to Appendix~\ref{sec:IFSPSF}).  
The profile of narrow \Hbeta (henceforth denoted $\Hbeta_\mathrm{N}$, solid orange line) is obtained by simply collapsing the three spectral channels around the peak of $\Hbeta_\mathrm{N}$ (a range of $\pm 75~\kms$); similarly, the profile of \OIIIL (solid pink line) is also obtained by collapsing the narrow line over the three spectral channels around its peak. 
The profile of broad \Hbeta (henceforth denoted $\Hbeta_\mathrm{B}$; dot-dashed teal line) is extracted from the spectral regions $2.204-2.210\ \mu$m and $2.214-2.221 \ \mu$m (i.e. including emission from both the blue and red wings). 
The comparison of the profiles demonstrates that the bulk of \Hbeta and \OIIIL emission comes from a very compact, unresolved region. 
Unlike in Abell2744-QSO1 \citep[hereafter QSO1;][]{Maiolino2025_QSO1}, there is no conclusive evidence for an extended component of \Hbeta or \OIIIL emission in \cliff, indicating that \cliff is ultimately compact and spatially unresolved at our resolution. Further evidence for this conclusion arises from morphological fitting of the star in the IFU FoV: \cliff appears to show the same spatial profile as the star, i.e. any apparent extension arises due to the PSF.

We determine the optimal spectral extraction aperture for \cliff, using a curve of growth analysis, which allows us to maximise the  signal-to-noise ratio (SNR) of the resulting integrated spectrum while accurately accounting for aperture losses. 
As \Halpha is detected with the greatest SNR, we utilise the curve of growth of the broad \Halpha emission to derive overall aperture corrections. 
To account for wavelength-dependent aperture losses, the \Halpha correction is then scaled to determine the relevant aperture loss correction (ALC) factors for shorter-wavelength lines such as \Hbeta and \OIIIL (see Appendix~\ref{sec:ALC}). 
The final extraction aperture is an ellipse defined by the same axis ratio and P.A. as the PSF, with major radius of $0.1''$, and centred on the continuum centroid of \cliff.
The final integrated spectrum extraction aperture used for \cliff in this work is shown by the white dashed ellipse in Fig.~\ref{fig:aperture}, overlaid onto a map of $\Hbeta_\mathrm{N}$ emission.
The full integrated spectrum of \cliff extracted from this $r_\mathrm{maj}=0.1''$ aperture is presented in Fig.~\ref{fig:full_spectra}. For this aperture, the ALC factor for \Halpha is found to be $1.88\pm0.05$.

\subsection{Spectral analysis}\label{sec:line_measure_methods}
To analyse the emission lines, we adopt a Bayesian approach, using different models for different regions of the spectrum as discussed across the following subsections. 
To integrate the posterior distribution of each fit, we use the Markov Chain Monte Carlo (MCMC) method with the software \textsc{emcee} \citep{emcee2013}. 
We run 15000 steps for each chain, with 50 percent burn-in steps.
To initialise the chains, we first identify the minimum-$\chi^2$ solution using the relevant model and ordinary least-squares minimisation. 
For the MCMC fit, we adopt Gaussian priors informed by the minimum-$\chi^2$  fit, with strong boundaries only set when physically motivated (e.g. to ensure non-negative emission line fluxes). 
All chains are visually inspected for convergence. 
The final best-fit parameters and their uncertainties are calculated as the median value and 68 percent confidence interval of the posterior distribution. 
To account for any uncorrelated uncertainties, all quantities derived from the spectral fitting are calculated from the posterior distribution. 
The key spectral properties determined from the model fits described in this section, together with their uncertainties, are summarised later in Table~\ref{table:fit_properties}.

\subsubsection{The \texorpdfstring{\Hbeta-\OIIIL}{HbO3} complex}\label{sec:fitting_Hb_OIII}

\begin{figure*}
        \centering
	\includegraphics[width=\linewidth]{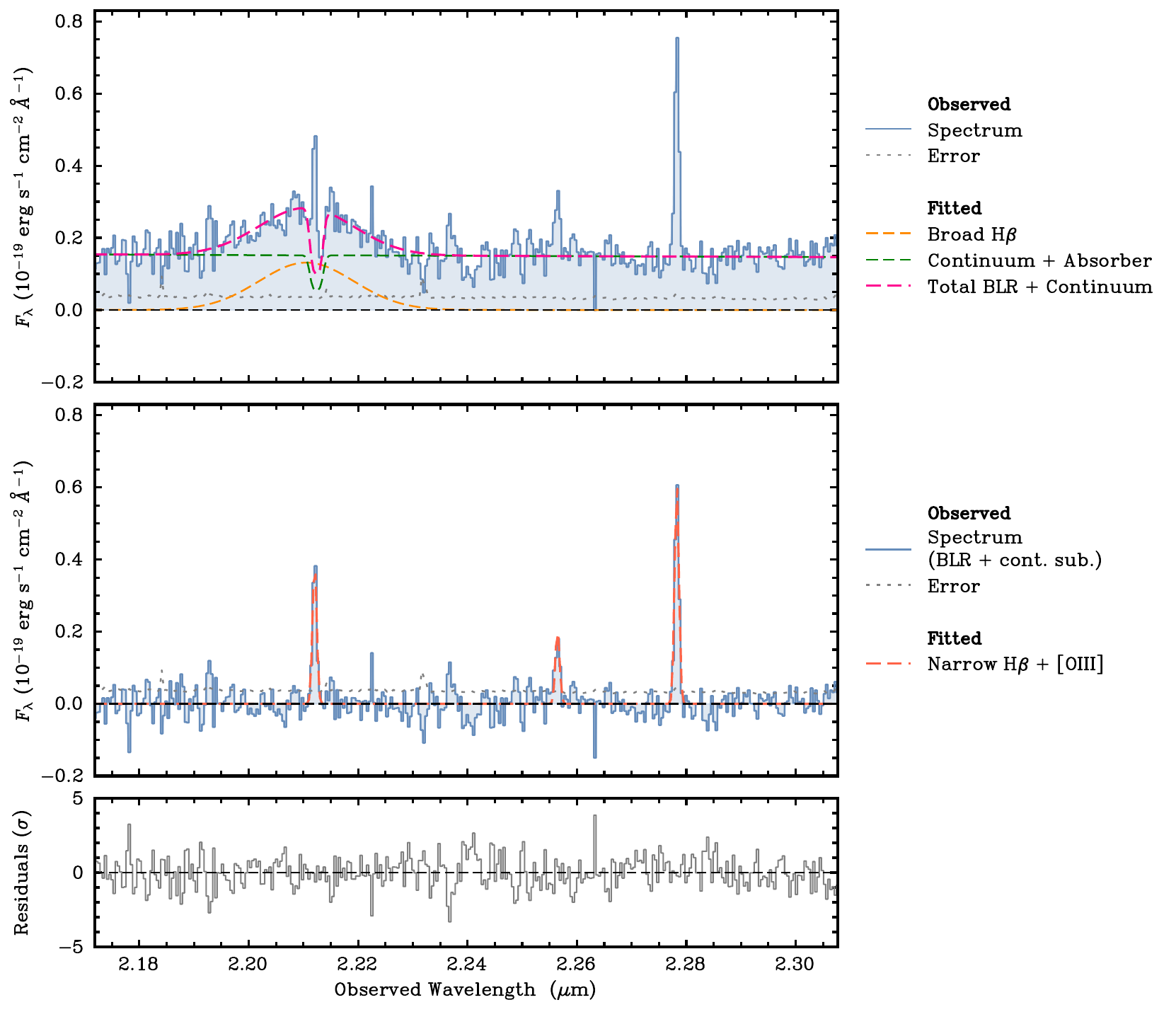}
    \caption{The integrated spectrum of \cliff presented in Fig.~\ref{fig:full_spectra}, highlighting the \Hbeta-\OIIIL spectral region. 
    \textbf{Top panel}: The integrated spectrum (blue solid line), with the fitted broad component of \Hbeta (orange dashed line), and the continuum and \Hbeta absorption (green dashed line). 
    The pink dashed line shows the total fit to the BLR and continuum emission. 
    \textbf{Middle panel}: The same spectrum, where the broad \Hbeta, continuum and absorption components have now been subtracted to highlight the narrow components of \Hbeta and \OIIIall. Note that the feature observed between \Hbeta and \OIIIL[4959] was not found to correspond to any known emission line (e.g. \HeI or \FeII) and is therefore deemed to be a noise feature.
    \textbf{Bottom panel}: Residuals from the fit, in units of $\sigma$.}
    \label{fig:spectra}
\end{figure*}

Fig.~\ref{fig:spectra} shows the integrated spectrum of \cliff, now highlighting the \Hbeta and \OIIIall emission lines. 
This integrated spectrum reveals the narrow and broad components of \Hbeta emission, together with a broad and slightly-redshifted absorption component; the absorption in \Hbeta was not visible in the earlier PRISM observations presented by \cite{deGraaff2025}. 
We also note the detection of the weak and narrow \OIIIall lines in our observations, which were not significantly detected by \cite{deGraaff2025} in the low resolution spectrum. 
In this work, the narrow \OIIIL emission provides valuable constraints for the kinematics of the narrow \Hbeta line in our spectral fits of the G235H spectrum, meaning it is possible to spectrally disentangle the $\Hbeta_\mathrm{B}$ and absorption components from the $\Hbeta_\mathrm{N}$ component. 

To fit the \Hbeta-\OIIIL complex, we adopt an approach similar to that outlined by \cite{Ji2025} or \cite{Maiolino2025_QSO1}.
We fit a single Gaussian profile to describe the narrow components of each emission line, tying their redshift centroids and intrinsic FWHMs ($z_N$ and $\mathrm{FWHM}_\mathrm{N}$, respectively) to common values to reduce the number of free parameters. 
We allow the amplitude of each Gaussian profile to vary freely, but the \OIIIL/\OIIIL[4959] flux ratio is fixed to 2.99 \citep{Dimitrijevic2007}. 
The broad line component is modelled as a Gaussian profile with its own independent redshift and intrinsic FWHM ($z_B$ and $\mathrm{FWHM}_\mathrm{B}$, respectively).
In order to retrieve intrinsic line FWHMs from the spectral data, we must account for the wavelength-dependent line spread function (LSF) of \jwst/NIRSpec.
The nominal, pre-launch resolution LSF \citep{Jakobsen2022} from the JDOCS\footnote{Available at the \hyperlink{https://jwst-docs.stsci.edu/jwst-near-infrared-spectrograph/nirspec-instrumentation/nirspec-dispersers-and-filters}{jwst-docs website}.} is too conservative because \jwst provides a lower than expected wavefront error.
Hence, for our deconvolution, we assume the nominal LSF improved by a factor of 0.7 \citep[roughly in agreement with e.g.][]{Shajib2025}. 
The intrinsic FWHM of any individual emission line is therefore given by:
\begin{equation}
\mathrm{FWHM_{int}} = \sqrt{\mathrm{FWHM_{obs}^2}- (0.7\times\mathrm{FWHM_{pre-launch \ LSF}})^2}.
\label{eq:fwhm}
\end{equation}
The continuum is modelled with a power-law, with the amplitude and exponent of the power law as additional free parameters.

Finally, we add a \Hbeta absorption component, associated with the broad-line region.
The absorption component is described by 4 parameters: velocity offset $v_\mathrm{abs}$ relative to the narrow line, central optical depth $\tau_0$, covering fraction $C_f$ and velocity dispersion $\sigma_\mathrm{abs}$. 
The residual intensity at wavelength $\lambda$ is given by
\begin{equation}\label{eq:abs}
\begin{aligned}
    I(\lambda)/I_0(\lambda) &= 1 - C_f + C_f\exp(-\tau(k; \lambda)), {\rm with} \\
    \tau(k; \lambda) &= \tau_0(k)f[v(\lambda)], 
\end{aligned}
\end{equation}
where $I_0(\lambda)$ is the spectral flux density before absorption (consisting of both the continuum and $\Hbeta_\mathrm{B}$ emission), $\tau_0(k)$ is the optical depth at the centre of the line ($k = \Hbeta$ here), and $f[v(\lambda)]$ is the velocity distribution of the absorbing atoms, assumed to be a Gaussian probability distribution \cite[see e.g.][]{DEugenio2025} and parameterised by $v_\mathrm{abs}$ and $\sigma_\mathrm{abs}$.

The middle panel of Fig.~\ref{fig:spectra} shows the integrated spectrum, now with the broad \Hbeta, continuum and absorption components subtracted.  
The BLR-subtracted spectrum highlights the prominence of $\Hbeta_\mathrm{N}$ and the relative weakness of \OIIIL. 
The inferred ratio between \OIIIL and $\Hbeta_\mathrm{N}$ is low, $F(\OIIIL)/F(\Hbeta_\mathrm{N}) = 1.59^{+0.26}_{-0.24}$, as reported in Table~\ref{table:flux_ratios}. 
While this ratio is not as low as in the case of QSO1 \citep[$F(\OIIIL)/F(\Hbeta_\mathrm{N}) \sim 0.55$;][]{Maiolino2025_QSO1} for example, we will discuss the \OIIIL weakness and its implications for metallicity in Section~\ref{sec:Evidence_for_low_metal}. 

Due to the relatively low signal to noise of the \Hbeta-\OIIIL complex, we note as a caveat that we cannot meaningfully perform statistical tests of different models for the broad \Hbeta line profile (e.g. exponential wings, double Gaussian BLR, Lorentzian).

\subsubsection{The \texorpdfstring{\Halpha}{Ha}profile}\label{sec:Fitting_Ha}

\begin{figure*}
        \centering
	\includegraphics[width=\linewidth]{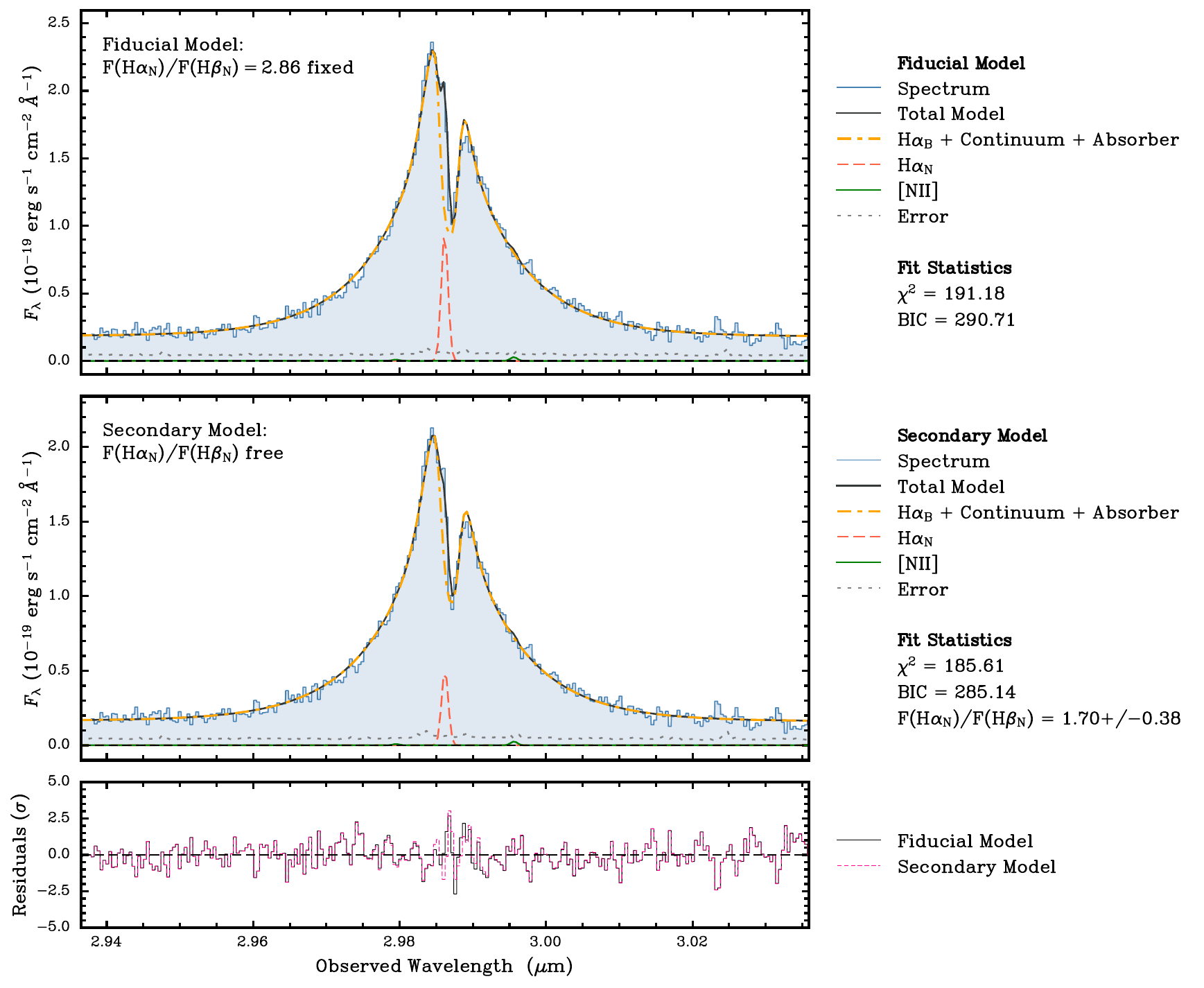}
    \caption{Integrated spectrum of \cliff around \Halpha, fitted with two variations on the scattering model as described in Section~\ref{sec:Fitting_Ha}. 
    \textbf{Top panel}: Our \textit{fiducial \Halpha model}, with the $\Halpha_\mathrm{N}/\Hbeta_\mathrm{N}$ flux ratio restricted to Case B as part of the fit, i.e. $F(\Halpha_\mathrm{N})/F(\Hbeta_\mathrm{N}) = 2.86$. 
    Components shown are the integrated spectrum (blue solid line), the overall \Halpha model (solid black line), the total fit to the BLR and continuum emission (dot-dashed orange line), the narrow component of \Halpha (dashed red line) and the \NII doublet (solid green line). 
    \textbf{Middle panel}: The same overall model, but now with the $\Halpha_\mathrm{N}/\Hbeta_\mathrm{N}$ flux ratio allowed to deviate from Case B (our \textit{secondary \Halpha model}).
    Individual components are shown with the same line styles as in the top panel. 
    \textbf{Bottom panel}: A comparison of the fit residuals for the \textit{fiducial \Halpha model} (black solid line) and \textit{secondary \Halpha model} (dotted pink line) in units of $\sigma$.}
    \label{fig:Halphafit}
\end{figure*}

Fig.~\ref{fig:Halphafit} shows the integrated spectrum of \cliff around the \Halpha emission line. 
The high-resolution IFU spectrum reveals the complexity of \cliff's \Halpha profile: deep absorption, no visibly distinct narrow component, and potentially asymmetric broad wings. 
While a single- or double-Gaussian BLR model is often used to approximate the \Halpha line breadth (as was the case for our \Hbeta fit), such models result in a poor fit to the wings of the profile.
In contrast, a `scattering model' \citep[see e.g.][]{Rusakov2026} can account for exponential wings in the line profile, which are often attributed to Doppler broadening due to electron or resonant scattering \citep[e.g.][]{Laor2006}.
Even with the higher resolution of the G235H spectrum (compared to earlier PRISM or G235M observations), we still lack the spectral resolution required to decisively disentangle the individual \Halpha spectral components.
As a result, we deem the exponential model to be a more appropriate parameterisation of the \Halpha profile for this work (compared to a double Gaussian model, for example). For simplicity, and for comparison with some previous studies, we model the exponential profile in the context of the scattering scenario. However, we note that while the exponential profile is often ascribed to electron scattering by a ionised medium, \citet{Scholtz2026_LRD} and \citet{Madau2026b} have recently shown that the same profile can be ascribed to normal BLR stratification, or to the overlap of multiple non-exponential components.
A full exploration of the \Halpha emission line profile, and the nature of its absorber, is beyond the scope of this work but has been presented in \cite{Scholtz2026_LRD}.

To model the \Halpha profile, we first define the narrow component of \Halpha (henceforth $\Halpha_\mathrm{N}$) and the \NIIall doublet as Gaussians with their kinematic parameters (i.e. $z_N$ and FWHM\textsubscript{N}) constrained to those from the \Hbeta-\OIIIL fit, with the \NIIL[6548]/\NIIL[6583] flux ratio fixed to 0.327 \citep[see][]{Dojcinovic2023}. In line with the `scattering scenario' parameterisation, we adopt a BLR profile comprised of an intrinsic Gaussian profile $G_\mathrm{BLR}(\lambda)$ described by its own independent redshift ($z_\mathrm{B}$) and FWHM ($\mathrm{FWHM}_\mathrm{B}$), convolved with an exponential kernel defined by
\begin{equation}
E(\lambda_0, W;\lambda) \propto \exp\left({-\frac{|\lambda-\lambda_0|}{W}}\right),
\label{eq:E}
\end{equation}
where $\lambda_0$ is the central wavelength (assumed to be the same as that of the Gaussian BLR) and $W$ is the exponential width \citep{Laor2006}. The overall BLR emission is therefore given by:
\begin{equation}
    \mathrm{BLR} = f_\mathrm{scatt}E(\lambda)\ast G_\mathrm{BLR}(\lambda)+(1-f_\mathrm{scatt})G_\mathrm{BLR}(\lambda),
    \label{eq:BLR}
\end{equation}
where $f_\mathrm{scatt}$ is the fraction of scattered light. $f_\mathrm{scatt}$ can be related to the optical depth of the scattering medium ($\tau_\mathrm{thom}$) by $1-e^{-\tau_\mathrm{thom}}$. The exponential width is related to both $\tau_\mathrm{thom}$ and the temperature of the scattering gas ($T$) by
\begin{equation}
    W = (428\times\tau_\mathrm{thom} + 370)\times(T/10^4 \ \mathrm{K})^{0.5};
    \label{eq:W}
\end{equation}
see \cite{Laor2006} and \cite{Rusakov2026}. Note that within our analysis $T$ is treated as a nuisance parameter, and is not interpreted as a physical measurement of the system's temperature.
Once again, we model the continuum with a power law.
Finally, we include an absorption component with the same $C_f$ as that fitted to \Hbeta, but its other parameters ($v_\mathrm{abs, \Halpha}, \tau_0(\Halpha)$ and $\sigma_\mathrm{abs, \Halpha}$) allowed to vary independently.

Supplying constraints and Gaussian priors for this model based on results from the \Hbeta-\OIIIL fit ensures the resulting \Halpha model is physically motivated.
This is particularly important as the absorption component is only redshifted by approximately 2 spectral pixels, and there hence exists significant degeneracy between the narrow line and absorption components.

We investigate this degeneracy and its impact further by fitting two variations on our overall \Halpha model.
Firstly, in our \textit{fiducial \Halpha model} (top panel of Fig.~\ref{fig:Halphafit}), the $\Halpha_\mathrm{N}/\Hbeta_\mathrm{N}$ flux ratio is fixed to the typical Case B value of 2.86, appropriate for $T=10,000$~K and $n_e\sim10^2 - 10^3 \ \mathrm{cm}^{-3}$, and assuming no reddening \citep[][here accounting for relative ALCs between \Halpha and \Hbeta]{Osterbrock2006}\footnote{We note that the Balmer decrement is insensitive to density variation of $10^2-10^6 \ \mathrm{cm}^{-3}$; see e.g. \cite{StoreyHummer1995}.}.
The second variation is our \textit{secondary \Halpha model} (middle panel of Fig.~\ref{fig:Halphafit}), in which the flux of $\Halpha_\mathrm{N}$ is allowed to freely vary. 
We make a statistical comparison of the two models through the Bayesian Information Criterion \citep[BIC; see e.g.][]{Schwarz1978}, defined as
\begin{equation}
    \text{BIC} = \chi^2 + k \ln n,
    \label{eq:BIC}
\end{equation}
where $\chi^2$ is the chi-squared fit statistic, $k$ is the number of free parameters, and $n$ is the number of data points. 
One model is generally considered to be statistically preferred over another if the relative change in BIC ($\Delta\mathrm{BIC}$) is $<-10$.
With the \textit{secondary \Halpha model}, we obtain a slightly better fit in terms of $\mathrm{BIC}$, and measure $F(\Halpha_\mathrm{N})/F(\Hbeta_\mathrm{N}) = 1.70\pm0.38$, much lower than the Case B ratio (though marginally consistent within $3\sigma$). 
However, $\Delta\mathrm{BIC}$ between the \textit{fiducial} and \textit{secondary} \Halpha models is only $\sim -5$, which is not statistically significant enough to argue the \textit{fiducial} model should be ruled out. 
Hence, the observed \Halpha profile could be regarded as consistent with a Case B Balmer decrement.

According to our best fit, \NIIall is undetected ($\mathrm{SNR}<0.3$), and we therefore determine the $3\sigma$ upper limit on the \NIIall flux to be $F(\NII) < 0.18 \times 10^{-18} \ \mathrm{erg} \ \mathrm{s}^{-1} \ \mathrm{cm}^{-2}$ (see Table~\ref{table:flux_SNR}).
We will utilise this constraint later in Section~\ref{sec:Evidence_for_low_metal}.

In principle, \Halpha could have been included in a simultaneous fitting with the \Hbeta-\OIIIL complex. 
However, the high flux and severe blending of \Halpha could have driven increased uncertainty in the \Hbeta-\OIIIL fit parameters and effectively reduced the SNR of these lines, introducing major caveats for our analysis. 
Instead, with our approach we prioritise securing an understanding of the narrow-line kinematics, applying these measurements as a prior on $\Halpha_\mathrm{N}$ to better inform the deblending of the \Halpha profile while ensuring confidence in our analysis of $\Hbeta_\mathrm{N}$ and \OIIIL[5007].

\subsubsection{Constraints on other emission lines}\label{sec:Fitting_other_lines}

\begin{figure}
        \centering
	\includegraphics[width=\linewidth]{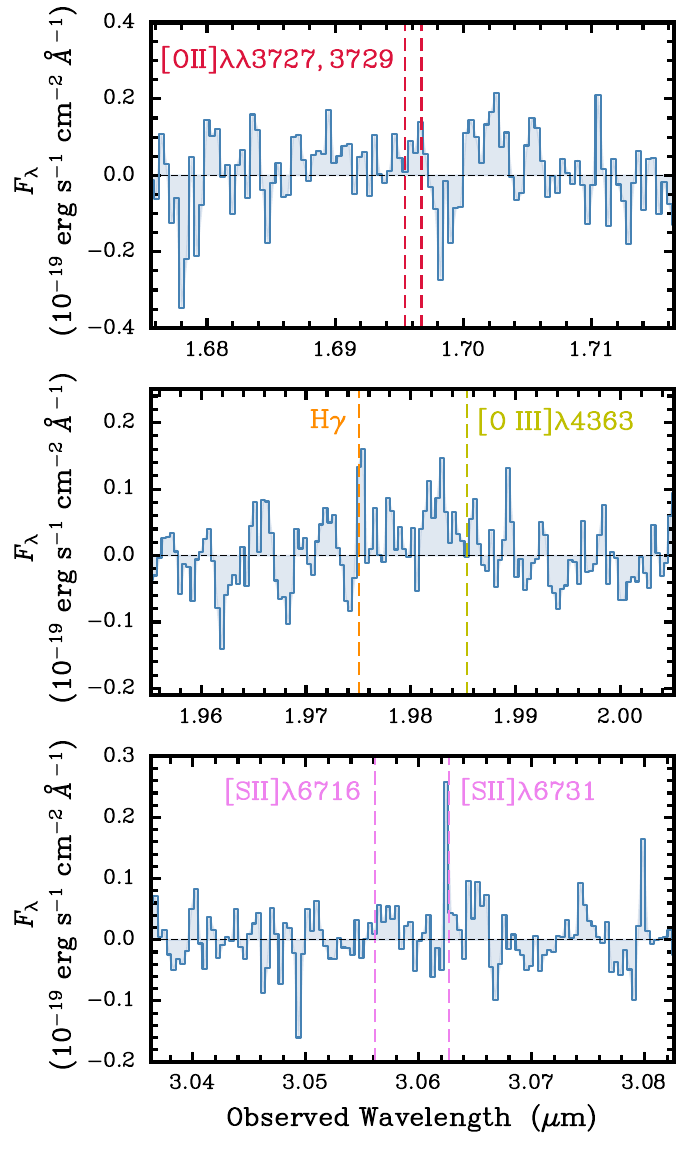}
    \caption{Portions of the integrated spectrum of \cliff, illustrating, from top to bottom, the non-detections of the following emission lines: \OIIall, \OIIIL[4363], and \SIIall, respectively.
    The spectra presented here are continuum-subtracted to further emphasize the non detection of these lines.
    We do however note a marginal detection of \Hgamma in the second panel ($\mathrm{SNR} = 2.9$).}
    \label{fig:undetectedlines}
\end{figure}

Some additional key lines for our study are the low-ionisation forbidden line doublets \OIIall and \SIIall (see Section~\ref{sec:Evidence_for_low_metal}), along with the temperature-sensitive auroral \OIIIL[4363] line. 
Fig.~\ref{fig:undetectedlines} shows integrated spectrum zoom-ins for each of these emission lines, illustrating a lack of any significant detection (all have $\mathrm{SNR} \lesssim 1.5$; see Table~\ref{table:flux_SNR}). 
The upper limits on the fluxes of these emission lines provides valuable constraints on the inter-stellar medium (ISM) properties of \cliff (see Section~\ref{sec:Evidence_for_low_metal} for further discussion). 

Hence, we fit each of these emission lines with \textsc{lmfit} \citep{Newville2016_LMFIT} and simple single-Gaussian models to obtain upper limits on their integrated fluxes, again fixing the intrinsic FWHM\textsubscript{N} of each line to the value from the \Hbeta fit. 
For the \OIIall and \SIIall doublets, the flux ratios are dependent on ISM conditions (e.g. the \OIIall flux ratio depends on density), and their flux ratios are therefore allowed to vary over the permitted range identified with \textsc{PyNeb} \citep{Luridiana2015_PyNeb}. 
As we note a tentative detection of \Hgamma (see middle panel of Fig.~\ref{fig:undetectedlines}), we fit \Hgamma and \OIIIL[4363] together, obtaining an upper limit on \OIIIL[4363] and a tentative detection of \Hgamma ($\mathrm{SNR}\sim2.9$; see Section~\ref{sec:Dust_extinction} for further discussion of \Hgamma).

{\renewcommand{\arraystretch}{1.6}
\begin{table}
\normalsize
\centering
 \begin{tabular}{c |c | c| c} 
 \hline
 \hline
& Parameter &  \multicolumn{2}{c}{Value} \\ 
 \hline
\multirow{8}{*}{\Hbeta-\OIIIL} & $z_N$ &  \multicolumn{2}{c}{$3.54903\pm0.00004$} \\
 & FWHM\textsubscript{N}  & $90\pm9$ & km s\textsuperscript{-1}\\
 & $z_B$ &  \multicolumn{2}{c}{$3.546\pm0.001$} \\
 & FWHM\textsubscript{B} 
 & $2770\pm210$ & km s\textsuperscript{-1}\\
 & $v_\mathrm{abs}$   & $46^{+15}_{-14}$ & km s\textsuperscript{-1}\\
 & $\tau_0(\Hbeta)$ &  \multicolumn{2}{c}{$3.22^{+2.26}_{-1.36}$} \\
 & $C_f$ &   \multicolumn{2}{c}{$0.67^{+0.17}_{-0.13}$}\\
 & $\sigma_\mathrm{abs}$ & $47\pm21$  & km s\textsuperscript{-1} \\
\hline 
 \multirow{7}{*}{\Halpha} 
 & $z_\mathrm{B, \Halpha}$ &  \multicolumn{2}{c}{$3.5486\pm0.0001$} \\
 & FWHM\textsubscript{B, \Halpha}  & $480\pm50$ & km s\textsuperscript{-1} \\
 & $T$  & \multicolumn{2}{c}{$0.35^{+0.07}_{-0.06} \times 10^4 \ \mathrm{K}$} \\
 & $\tau_\mathrm{thom}$  &  \multicolumn{2}{c}{$2.22^{+0.19}_{-0.15}$} \\
 & $v_\mathrm{abs, \Halpha}$   & $61\pm4$ & km s\textsuperscript{-1}\\
 & $\tau_0(\Halpha)$ &   \multicolumn{2}{c}{$2.83^{+0.26}_{-0.24}$} \\
 & $\sigma_\mathrm{abs, \Halpha}$  & $52\pm6$& km s\textsuperscript{-1}  \\
 \hline
 \hline
 \end{tabular}
 \caption{Summary of best-fit \Hbeta-\OIIIL and \Halpha model parameters of \cliff, reporting the median and 16\textsuperscript{th}-84\textsuperscript{th} percentile range of the posterior probability distributions.
 The `N' and `B' subscripts denote values associated with narrow and broad components, respectively. 
 The free parameters for the emission line amplitudes and power-law continuum are not listed here. 
 \textbf{Top segment:} Results from the \Hbeta-\OIIIL fit described in Section~\ref{sec:fitting_Hb_OIII}. 
 \textbf{Bottom segment:} Results from the \textit{Fiducial} \Halpha fit described in Section~\ref{sec:Fitting_Ha}. 
 Parameters that were fixed to values from the \Hbeta-\OIIIL model as part of the \Halpha fit are not repeated in this segment.}
 \label{table:fit_properties}
\end{table}
}

The integrated fluxes of all emission lines measured in this work, including their SNRs, are presented in Table~\ref{table:flux_SNR}.
A summary of all flux ratios used in this work, together with their abbreviations and measured values or limits, is presented in Table~\ref{table:flux_ratios}.

{\renewcommand{\arraystretch}{1.4}
\begin{table}
\normalsize
\centering
 \begin{tabular}{c || c c} 
 \hline
 \hline
 Emission Line & Integrated Flux & SNR \\ 
  & $10^{-18} \ \mathrm{erg} \ \mathrm{s}^{-1} \ \mathrm{cm}^{-2}$ &  \\ 
 \hline
 $\Hbeta_\mathrm{N}$ & $0.65^{+0.09}_{-0.08}$ & 12.9\\
 $\Hbeta_\mathrm{B}$ & $4.81^{+0.56}_{-0.45}$ & 18.8 \\
 \OIIIL & $1.02\pm0.10$ & 16.3\\
 $\Halpha_\mathrm{N}$ & $1.85\pm0.04$ & 17.8\\
 $\Halpha_\mathrm{B}$ & $60.4^{+10.6}_{-9.3}$ & 133 \\
 $\Hgamma^{*}$ & $0.21\pm0.06$ & 2.9 \\
 \hline
 \OIIall & $<0.45$  & $< 0.7$\\
 \OIIIL[4363] & $<0.15$ & $<1.5$ \\
 \NIIall & $<0.18$ & $<0.3$\\
 \SIIall & $<0.36$ & $<0.3$ \\
 \hline
 \hline
 \end{tabular}
 \caption{Integrated fluxes and SNRs for the emission lines studied in this work. 
 All fluxes have been corrected for aperture losses. \Halpha fluxes refer to those determined from our fiducial Case B model (see Section~\ref{sec:Fitting_Ha}); the low uncertainty arises because this flux is fixed based on the \Hbeta flux. 
 \Hgamma is marked with an asterisk as this represents a tentative detection; see Section~\ref{sec:Dust_extinction} for further discussion. 
 The lower portion of the table refers to undetected lines and therefore corresponds to upper limits on their fluxes, which are quoted at the $3\sigma$ upper limit.}
 \label{table:flux_SNR}
\end{table}
}

{\renewcommand{\arraystretch}{1.6}
\begin{table}
\normalsize
\centering
 \begin{tabular}{c c|| c} 
 \hline
 \hline
 Flux Ratio & Abbreviation & Value \\ 
  & [for $\log(\mathrm{Ratio})$]&  \\
 \hline
 $\frac{\OIIIL}{\Hbeta_\mathrm{N}}$ & R3 & $1.59^{+0.26}_{-0.24}$ \\
 $\frac{\Halpha_\mathrm{N}}{\Hbeta_\mathrm{N}}$ & - &  $2.86^{+0.48}_{-0.36}$\\
 $\frac{\Hgamma_\mathrm{N}}{\Hbeta_\mathrm{N}}$ & - &  $0.32^{+0.11}_{-0.10}$\\
 $\frac{\OIIIL}{\OIIall}$ & O32 & $>2.3$ \\
 $\frac{\OIIIL[4363]}{\OIIIL}$ & - & $<0.15$ \\
 $\frac{\NIIall}{\Halpha_\mathrm{N}}$ & N2 & $<0.09$ \\
 $\frac{\OIIIL/\Hbeta_\mathrm{N}}{\NIIall/\Halpha_\mathrm{N}}$ & O3N2 & $>18$ \\
 $\frac{\SIIall}{\Halpha_\mathrm{N}}$ & S2 & $<0.19$ \\
 $\frac{\OIIIL/\Hbeta_\mathrm{N}}{\SIIall/\Halpha_\mathrm{N}}$ & O3S2 & $>8.3$\\
 \hline
 \hline
 \end{tabular}
 \caption{Flux ratios and their abbreviations for the narrow emission lines. 
 Flux ratios are calculated based on aperture loss-corrected flux values (see Table~\ref{table:flux_SNR}). 
 Uncertainties are bootstrapped from uncertainties on the fluxes and are quoted at the 68 percent confidence level. 
 Upper or lower limits on flux ratios are quoted at the $3\sigma$ level.
 For further discussion of these calibrations, see e.g. \citealt{Curti2020}.}
 \label{table:flux_ratios}
\end{table}
}

\section{Results}\label{sec:Results}

\subsection{Dust attenuation}\label{sec:Dust_extinction}
\cliff exhibits a complex \Halpha profile, with strong degeneracy between the narrow component and absorber. 
Consequently, it was not possible to determine a well-constrained $\Halpha_\mathrm{N}/\Hbeta_\mathrm{N}$ Balmer decrement; indeed, detailed exploration of the \Halpha and \Hbeta profiles is beyond the scope of this paper. 
Consequently, within this work we adopt the Case B ratio of  $F(\Halpha)/F(\Hbeta) = 2.86^{+0.48}_{-0.21}$ derived from our fiducial \Halpha model, which assumes no dust.

Furthermore, we investigate the $\Hgamma_\mathrm{N}/\Hbeta_\mathrm{N}$ Balmer decrement as a second check on the validity of Case B. 
In the middle panel of Fig.~\ref{fig:undetectedlines}, we show a tentative detection ($\mathrm{SNR} \approx 2.9$) of \Hgamma emission. Like the other Balmer lines, \Hgamma emission should have its own broad component, which would show absorption comparable in depth to that seen in \Hbeta and deeper than that of \Halpha \citep[e.g.,][]{Naidu2025, DEugenio2026}. 
However, due to the low SNR in this region of the spectrum, it is only possible to meaningfully fit the narrow component of \Hgamma (henceforth $\Hgamma_\mathrm{N}$). 
As a consequence of the degeneracy between absorption and narrow components seen in the other Balmer lines, this suggests that our measurement of $\Hgamma_\mathrm{N}$ flux is likely an underestimate (unlike the other emission lines discussed in Section~\ref{sec:Fitting_other_lines}). 
However, our measurement yields $F(\Hgamma)/F(\Hbeta) = 0.32^{+0.11}_{-0.10}$, consistent with the intrinsic Case B value of 0.466 \citep[for a typical electron temperature of $T_e \sim 10^4 \ \mathrm{K}$;][]{Osterbrock2006} within 1.3$\sigma$. 

Consequently, we cannot make a compelling claim for Case B violation, and henceforth adopt the assumption of no significant dust attenuation in \cliff. 
This conclusion is supported by the  previous studies which find no dust attenuation in the narrow lines of other LRDs \citep[e.g.][]{Lin2026, Ji2026, DEugenio2025_Irony, Nikopoulos2025}.

We additionally note the $\Halpha_\mathrm{B}/\Hbeta_\mathrm{B}$ Balmer decrement is $\sim 13$ in this source. This high value may be associated with strong dust attenuation in the BLR \citep[$A_V=4$~mag assuming Case-B recombination and the SMC attenuation curve;][]{Gordon2003}, which can be achieved with relatively low amounts of dust concentrated on small spatial scales \citep{Pacucci2026,MadauMaiolino2026}. Alternatively, the high $\Halpha_\mathrm{B}/\Hbeta_\mathrm{B}$ flux ratio could arise from collisional excitation in high-density regions \citep[see e.g.][]{Zu2025}. Regardless, the peculiar conditions of the BLR cannot apply to the narrow-line emitting gas, on account of its narrow spectral profile ($\mathrm{FWHM}_\mathrm{N} < 100 \ \kms$) and significantly lower Balmer decrement, implying the narrow-line emitting gas is physically distinct from the BLR.

\subsection{Metallicity}\label{sec:metallicity}

\subsubsection{Evidence for low metallicity in \cliff}\label{sec:Evidence_for_low_metal}

Following the standard approach for high-redshift sources within the literature, we infer the metallicity of \cliff using strong-line metallicity diagnostics calibrated based on high-$z$ star-forming galaxies (SFGs). 
In this work, we adopt the calibrations of Isobe et al. (in prep), which are based on stacks of SFGs over $1<z<10$ ($z_\mathrm{median} \sim 3-5$), with linear extrapolations for metallicities $12 + \log(\mathrm{O/H}) < 7$. 

\begin{figure}
\centering
   \includegraphics[width=\linewidth]{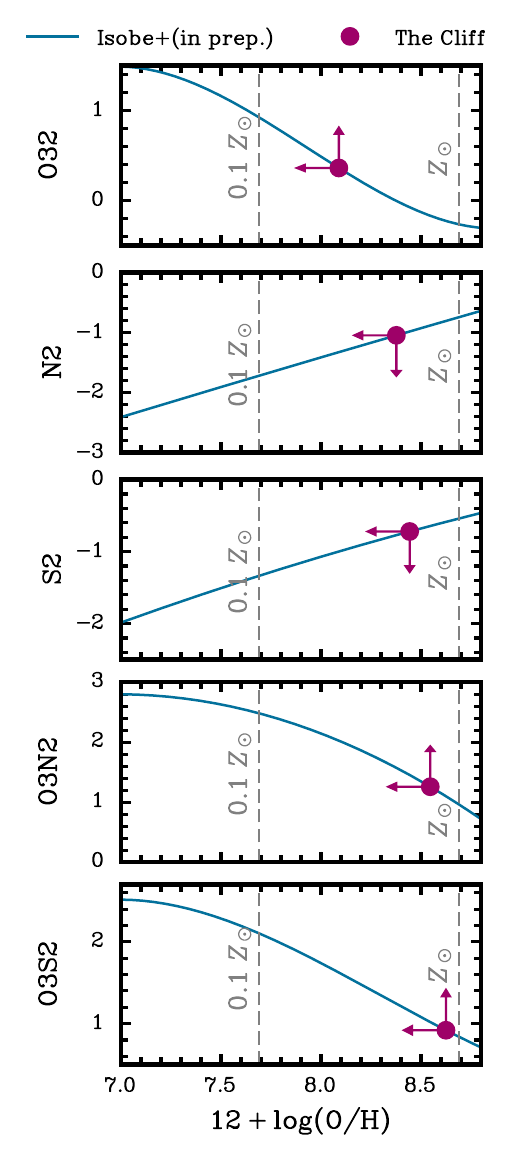}
\caption{Metallicity constraints on \cliff inferred from upper limits on the non-detected emission lines (\OIIall, \SIIall and \NIIall). The solid blue line in each panel illustrates the relevant diagnostic calibration from Isobe et al. (in prep.).
The magenta symbols indicate the upper or lower limits obtained for flux ratios in \cliff (see Table~\ref{table:flux_ratios}), which altogether exclude super-solar metallicity as an explanation for the measured $\OIIIL[5007]/\Hbeta_\mathrm{N}$ flux ratio (see Fig.~\ref{fig:R3metalconstraint}). 
The dashed vertical gray lines correspond to metallicities of $0.1\ Z_\odot$ and $Z_\odot$, further emphasizing that a super-solar metallicity for \cliff is inconsistent with our observations.}
\label{fig:undetected_lines_constraint}
\end{figure}

The $\OIIIL/\Hbeta_\mathrm{N}$ flux ratio, denoted by R3, is often used to infer metallicity at high-$z$ where other emission lines are not available \citep[e.g.][]{Curti2020, Sanders2025}. 
Crucially, the R3 metallicity diagnostic has two branches, and a given R3 value could therefore correspond to both low-metallicity and high-metallicity solutions. 
Hence, to constrain the metallicity inferred from our measured R3 value, we must utilise the upper limits on the low-ionisation line fluxes determined earlier (see Section~\ref{sec:Fitting_other_lines}). 
The O32, N2, S2, O3N2 and O3S2 flux ratios (defined in Table~\ref{table:flux_ratios}) all have monotonic dependences on metallicity, and our observational limits on these flux ratios therefore provide upper limits on the metallicity of \cliff. 
Furthermore, Isobe et al. (in prep.) have provided revised calibrations for these diagnostics, updating them for higher redshifts. 
We present these constraints from the low-ionisation line flux ratios in Fig.~\ref{fig:undetected_lines_constraint}. 
Altogether, these diagnostics correspond to a metallicity upper limit of $Z\lesssim 0.25 \ Z_\odot$. 
While this upper limit is not particularly constraining by itself, it is important for removing the potential degeneracy associated with the double-branched R3 diagnostic.

\begin{figure}
\centering
   \includegraphics[width=\linewidth]{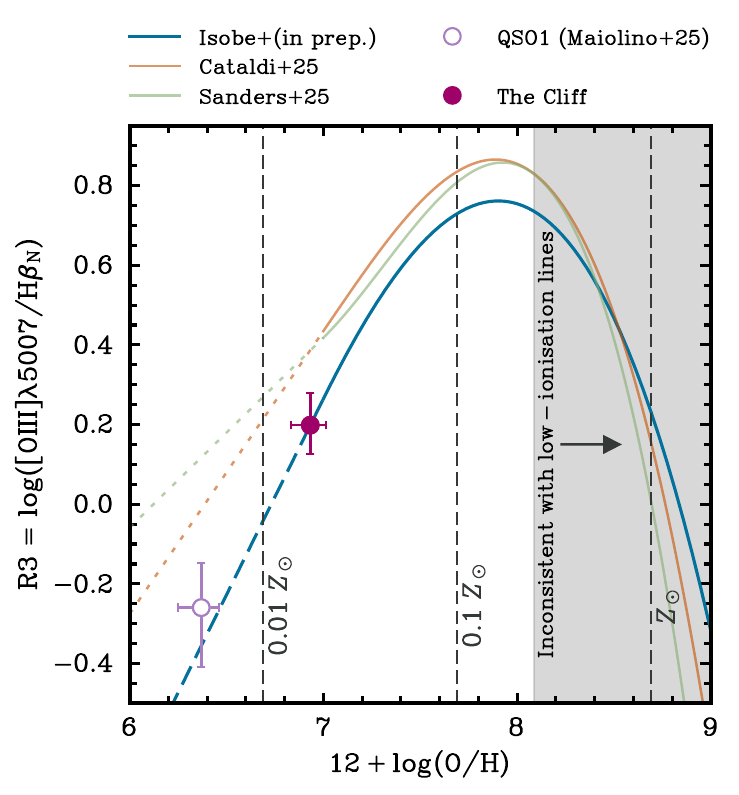}
\caption{Metallicity of \cliff as inferred from the measured $\OIIIL/\Hbeta_\mathrm{N}$ flux ratio.
The blue line illustrates the R3 calibration of Isobe et al. (in prep.), with the dashed segment of the line indicating the region for which the calibration is linearly extrapolated. 
The filled magenta point shows our favoured low-metallicity solution for \cliff as derived from this calibration. 
The gray shaded area illustrates the high-metallicity branch ruled out by the low ionisation lines (Fig.~\ref{fig:undetected_lines_constraint}). 
The turquoise point shows central region values for QSO1 from \citet{Maiolino2025_QSO1}. 
The orange and green lines show the high-redshift strong-line metallicity calibrations of \citet{Cataldi2025} and \citet{Sanders2025} respectively, with dashed segments indicating where these calibrations are extrapolated. 
All error bars show uncertainties at the $1\sigma$ level.}
\label{fig:R3metalconstraint}
\end{figure}

Fig.~\ref{fig:R3metalconstraint} shows the location of \cliff on the R3 versus metallicity diagram (the filled magenta point), with the R3 calibration of Isobe et al. in prep shown by the solid blue line. 
The Isobe et al. (in prep.) R3 calibration provides a metallicity measurement of $0.017\pm0.004 \ Z_\odot$ for \cliff (see Table~\ref{table:galaxyproperties}). 
The upper branch solution would imply a super-solar metallicity, which would be implausible at such high redshift and for such a low mass system, but which is anyway ruled out by the weakness of the low-ionisation lines (see Fig.~\ref{fig:undetected_lines_constraint}), as discussed above. The metallicity range excluded by the non-detection of the low ionisation lines is marked with a gray shaded region in Fig.~\ref{fig:R3metalconstraint}.


We also note that the calibrations of Isobe et al. (in prep.) yield a higher metallicity for a given low-metal branch R3 value than other high-redshift strong-line metallicity calibrations \citep[e.g.][]{Cataldi2025, Sanders2024} by $\sim0.2-0.3$ dex, so our choice of calibration provides the most conservative low-metallicity measurement.

The metallicity inferred from the R3 diagnostic is presented in Table~\ref{table:galaxyproperties}, along with other properties of \cliff measured later in this section. 
We note as a caveat that in making this metallicity measurement we are adopting an extrapolation of the calibration to metallicities $12+\log(\mathrm{O/H}) < 7$, similar to other studies of high-redshift, low-luminosity systems \citep[e.g.][]{Maiolino2025_QSO1}. As an additional caveat, global measurements of metallicity at high redshift may be subject to systematic uncertainties driven by inhomogeneous ISM conditions \citep{Belfiore2017, Katz2022, Moreschini2026}. The measured emission line ratios can be skewed by localised high-excitation regions or dense clumps, potentially complicating the interpretation of galaxy-integrated scaling relations, and we cannot exclude the possibility this scenario applies to \cliff.

\subsubsection{Accounting for potential AGN excitation}\label{sec:Accounting_for_AGN}

The extremely narrow line widths of $\Hbeta_\mathrm{N}$ and \OIIIL measured in \cliff ($\mathrm{FWHM_N} \sim 90$ \kms) hint that these narrow components are powered by weak star formation in the host galaxy \citep[see e.g.][]{DEugenio2025}.
For the metallicity analysis above, we have therefore assumed that the narrow emission lines are entirely excited by star formation. 
In this section, we will now discuss the possible scenario in which the narrow lines are actually excited by the AGN and are therefore part of the NLR.

The calibrations obtained by Isobe et al. (in prep.) for high-$z$ SFGs are not expected to change significantly in the case of ionisation by AGN rather than star formation.
On average, high-$z$ SFGs are characterised by high ionisation parameters ($\log U > -2.5$; see Section~\ref{sec:peculiar_ionisation} for further discussion of ionisation parameter), and tend to have harder ionising spectra than local galaxies \citep{Sanders2025, Cameron2023}, primarily because of their reduced metal content \citep{Cameron2024, Strom2017}.  
High-$z$ AGN are characterised by softer ionising spectra than local AGN, as indicated by the weakness of the high-ionisation lines \citep{Lambrides2024,Zucchi2026}.
Overall, we expect the ionisation properties of the ISM in high-$z$ SFGs and AGN NLRs to be similar \citep{Kocevski2023,Ubler2023,Juodzbalis2026_JADEScensus}, which is further evidenced by the significant overlap between high-$z$ AGN and SFG populations in traditional diagnostic diagrams \citep[e.g. the BPT diagrams;][]{Ubler2023, Maiolino2024_JADES, Juodzbalis2026_JADEScensus}.
Hence, the low metallicity inferred from the Isobe et al. (in prep.) calibrations may hold even in the case of AGN photoionisation of the narrow lines.

If anything, the harder-ionisation spectrum from an AGN would amplify the $\OIIIL/\Hbeta_\mathrm{N}$ flux ratio for a given metallicity and ionisation parameter relative to SFGs (an illustration of this effect may be found in Appendix B of \cite{Maiolino2025_QSO1}, for example). 
For the low-metallicity R3 branch, in the case of AGN photoionisation, a given $\OIIIL/\Hbeta_\mathrm{N}$ flux ratio would yield an even lower metallicity than that inferred by assuming SF photoionisation (by $\sim 0.2-0.3$ dex). 
Once again, the assumptions we have adopted in our analysis yield the most conservative metallicity estimate.

\subsubsection{Accounting for density}\label{sec:Accounting_for_density}

\begin{figure}
\centering
   \includegraphics[width=\linewidth]{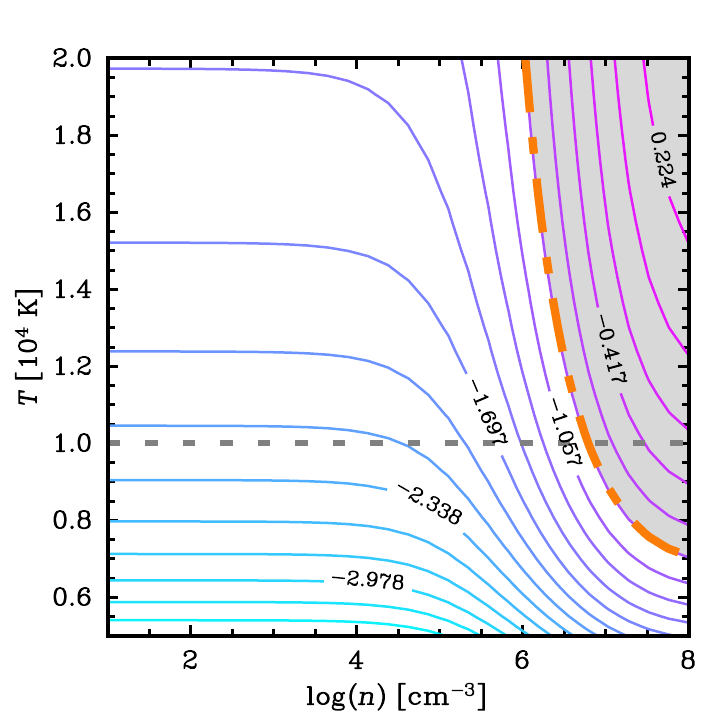}
\caption{The \OIIIL[4363]/\OIIIL flux ratio as a function of temperature ($T$) and density ($n$). 
The $3\sigma$ upper limit on the flux ratio, $F(\OIIIL[4363])/F(\OIIIL) < 0.15$ (orange dot-dashed line), rules out densities $n> 6\times10^6 \ \mathrm{cm}^{-3}$ for reasonable temperatures $T > 10^4 \ \mathrm{K}$, as indicated by the gray shaded region. 
Note the values of $F(\OIIIL[4363])/F(\OIIIL)$ quoted on each curve are logarithmic.}
\label{fig:density_constraint}
\end{figure}

\begin{figure}
\centering
   \includegraphics[width=\linewidth]{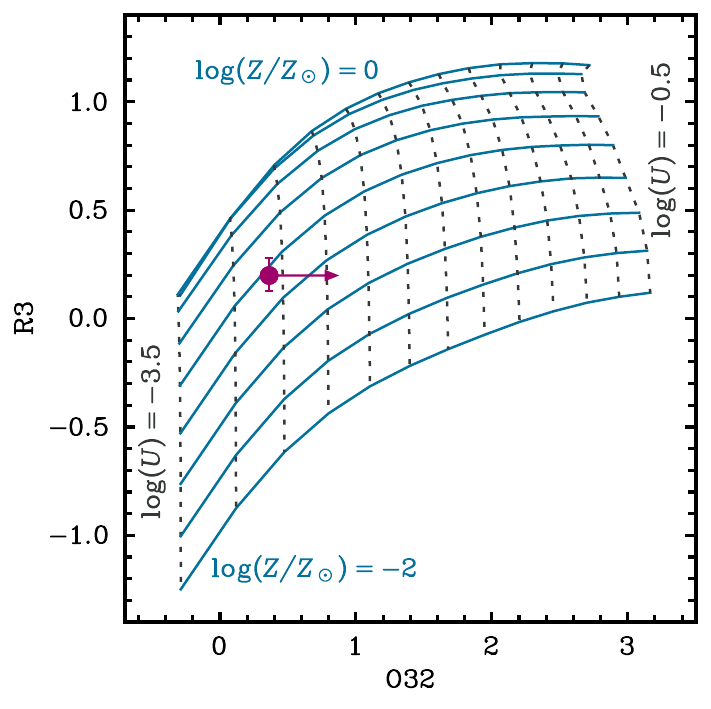}
\caption{Joint constraints on $Z$ and the ionisation parameter $U$ from the measured R3 ratio and lower limit on O32. The solid and dotted lines correspond to the results of AGN photoionisation models from \citealt{Isobe2025}, calculated for an assumed density of $n=10^4 \ \mathrm{cm}^{-3}$. The blue lines illustrate curves of constant $Z$, decreasing from $\log(Z/Z_\odot) = 0$ (at the top of the plot) to $\log(Z/Z_\odot) = -2$ (at the bottom). Similarly, the dotted gray lines illustrate curves of constant $U$, increasing from $\log(U) = -3.5$ to $\log(U) = -0.5$ from left to right across the plot. The position of \cliff is illustrated by the magenta circle, with the lower limit on O32 (see Table~\ref{table:flux_ratios}) providing a constraint on the minimum ionisation parameter.}
\label{fig:photoion}
\end{figure}

A potential alternative interpretation for the weakness of \OIIIL relative to $\Hbeta_\mathrm{N}$ is that the gas density is so high that collisional de-excitation starts to affect the population of the \OIIIL levels. 

The critical density of the \OIIIL line is about $5\times 10^5~\mathrm{cm}^{-3}$. As discussed in \citet{Martinez2025} and \citet{Maiolino2025_QSO1}, the emissivity of the \OIIIL line is only marginally affected at densities of $10^6~\mathrm{cm}^{-3}$ (a decrease by a factor of only $\sim$1.5), while it becomes prominent at densities of $10^7~\mathrm{cm}^{-3}$ (suppressed by a factor of $\sim 10$).
Although the densities of the ionised ISM are found to increase on average in high-$z$ galaxies, they are still typically found to be $<10^5 \ \mathrm{cm}^{-3}$ \citep[][]{Isobe2023, Topping2025}.


The non-detection of the \OIIIL[4363] line (middle panel of Fig.~\ref{fig:undetectedlines}), which has a significantly higher critical density than \OIIIL, provides a key constraint on the density. 
Fig.~\ref{fig:density_constraint} illustrates the variation of the \OIIIL[4363]/\OIIIL flux ratio as a function of temperature and density, as estimated with \textsc{PyNeb} \citep{Luridiana2015_PyNeb}. 
For reasonable temperatures typical of photoionised gas emitting \Hbeta \citep[$T > 10^4 \ \mathrm{K}$, e.g.][]{Osterbrock2006}, the inferred upper limit of $F(\OIIIL[4363])/F(\OIIIL)<0.15$ yields a density upper limit of $n<7\times10^6 \ \mathrm{cm}^{-3}$. 
This is actually a stricter upper limit on the density than that found for QSO1 \citep[$n<10^7 \ \mathrm{cm}^{-3}$;][]{Maiolino2025_QSO1}, and excludes extreme suppression of \OIIIL as a consequence of collisional excitation. The $10^4~\mathrm{K}$ limit is quite conservative, as most high-$z$ galaxies are characterised by higher temperatures \citep{Sanders2025,Cataldi2025}, hence the upper limit on the density is likely close to $10^6~\mathrm{cm}^{-3}$.

Additional arguments against the high density scenario, following the methodology presented in \cite{Maiolino2025_QSO1}, are discussed in Appendix~\ref{sec:density2}.


\subsubsection{Excluding peculiar ionisation scenarios}\label{sec:peculiar_ionisation}

Another potential scenario to explain the weakness of \OIIIL relative to \Hbeta, without invoking low density and metallicity, is to ascribe the effect to a gas ionisation state so low that $O^{++}$ emission is suppressed. 
Quantitatively, photoionisation models indicate that to suppress the \OIII/\Hbeta flux ratio to $\sim1$, the corresponding ionisation parameter would need to be $\log(U)\lesssim-2.5$ \citep[e.g.][]{Nagao2006, BaronNetzer2019, Cameron2023}.
This is in significant tension with established findings that high-$z$ galaxies are typically characterised by high ionisation parameters \citep[$\log(U)\gtrsim -2.0$; e.g.][]{Cameron2023,Reddy2023, Tang2023}. 
Of course, a single source could have lower than average $\log(U)$, but in this case the rarity of \cliff would suggest a link between the spectral properties of the AGN and host galaxy.
Assuming densities below the critical density of \OIIall (i.e. $n = 10^4 \ \mathrm{cm}^{-3}$), we utilise our measurements of R3 and O32 and the AGN photoionisation models of \cite{Isobe2025} to lay a joint constraint on $U$ and $Z$, as illustrated in Fig.~\ref{fig:photoion}.
The lower limit we have obtained for O32 provides a lower limit on the ionisation parameter $U$ as $\log(U) \gtrsim -3$. 
Furthermore, \cliff sits just below the model of $\mathrm{\log(Z/Z_\odot)} = -1$ (i.e. $12+\log(\mathrm{O/H}) \lesssim 7.69$), providing a further conservative upper limit on the metallicity of \cliff, which notably still sits on the low-metallicity branch of the R3 diagnostic regardless of high-$z$ strong line calibration adopted (see Fig.~\ref{fig:R3metalconstraint}).

A similar scenario, without invoking low metallicity or low ionisation parameter, is a lack of photons energetic enough to even ionise $O^+$ \citep[$\sim35$ eV; see e.g.][]{Wenaker1990, Martin1993, Draine2011}. 
However, this scenario would also not explain the absence of \OIIall from the spectrum, as oxygen and hydrogen have similar ionisation potentials \citep{Osterbrock2006}. 
Similarly, \NIIall and \SIIall are typically expected to be strongly detected in such low-ionisation environments \citep{Rhea2025}.
A potential caveat of this interpretation is that the abundance of nitrogen, and thus the strength of its associated emission lines, drops significantly faster than oxygen in low-metallicity regimes due to their different enrichment channels \citep[][]{Nicholls2017}.
Nevertheless, as previously established (Sections~\ref{sec:Fitting_Ha} and \ref{sec:Fitting_other_lines}), neither of these doublets are significantly detected in the spectrum of \cliff, let alone comparable to \Hbeta or \OIIIL, further disfavouring a simple low-ionisation interpretation.

Finally, we consider a scenario in which the ionisation parameter is so high that oxygen primarily exists as $O^{+++}$. Such a scenario has never been observed, not even in the most extreme AGN \citep{Dors2020}. 
Such a high level of ionisation would also result in other high ionisation lines, such as \CIV and \HeIIL \citep{Wang2025_HeII}, appearing very strong; however, there is no evidence for these lines in the \cliff \citep{deGraaff2025}.

By ruling out both extreme high- and low- ionisation parameter scenarios, we conclude that the low $\OIIIL/\Hbeta_\mathrm{N}$ flux ratio in \cliff most likely originates from a metal-poor ISM.

{\renewcommand{\arraystretch}{1.4}
\begin{table}
\normalsize
\centering
 \begin{tabular}{c || c c} 
 \hline
 \hline
 Galaxy Property & & Value \\ 
 \hline
 \multirow{2}{*}{Metallicity} & $12 + \log(\mathrm{O/H})$  & $6.93^{+0.07}_{-0.09}$ \\
  & $\log(Z/Z_\odot)$ & $-1.75^{+0.07}_{-0.09}$ \\
  \hline
 $\log(M_\mathrm{dyn}/M_\odot)$ & & $8.41\pm0.12$  \\
 $\log(M_\mathrm{*}/M_\odot)$ & & $<8.41$ \\
 \hline
 \multirow{2}{*}{$\log(M_\mathrm{BH}/M_\odot)$} & Fiducial (from \Hbeta) & $7.35\pm0.24$\\
  & Scattering Scenario & $6.32\pm0.22$ \\
  \hline
 \multirow{2}{*}{$\log(M_\mathrm{BH}/M_\ast)$} & Fiducial (from \Hbeta) & $>-1.06$\\
 & Scattering Scenario & $>-2.07$\\
 \hline
 \hline
 \end{tabular}
 \caption{Summary of the properties of \cliff as derived in this work. 
 The metallicity was calculated using the strong-line R3 calibration of Isobe et al. in prep. (see Section~\ref{sec:Evidence_for_low_metal}). 
 The fiducial BH mass is determined from the average of three different single-epoch \Hbeta-based virial calibration BH mass estimates.
 $M_\mathrm{dyn}$ is taken as a conservative upper limit on $M_\ast$.
 Uncertainties are quoted at the 68 percent confidence level.}
 \label{table:galaxyproperties}
\end{table}
}

\subsection{Potential low metallicity satellite}\label{sec:satellites}

\begin{figure*}
\centering
   \includegraphics[width=\linewidth]{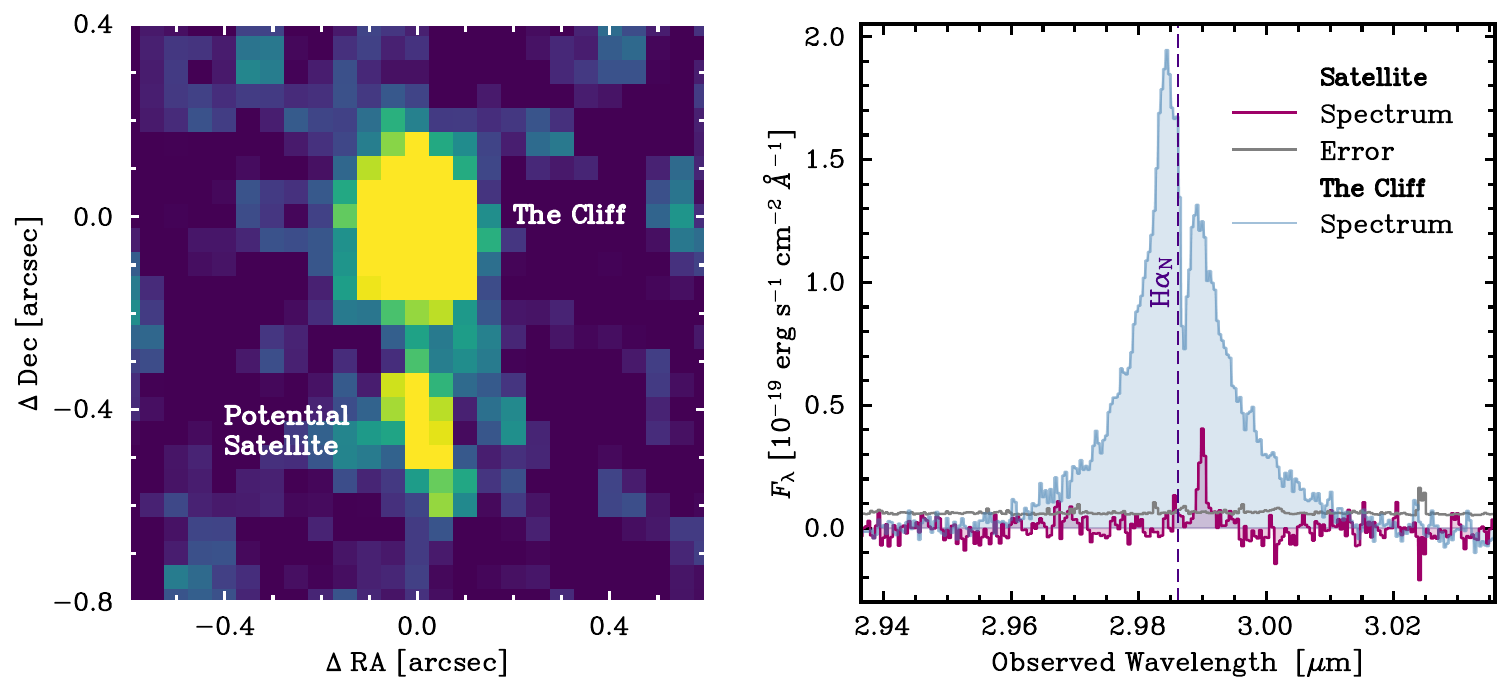}
\caption{Image and integrated spectrum of the potential southern satellite of \cliff. \textbf{Left panel:} The continuum-subtracted data cube collapsed over the $2.9888-2.9907\ \mu$m spectral channels, highlighting both \cliff and the potential southern \Halpha satellite (both labelled). \textbf{Right panel}: Comparison of continuum-subtracted integrated spectra ofThe  \cliff and the satellite. The dashed vertical line indicates the wavelength of the $\Halpha_\mathrm{N}$ component in \cliff, demonstrating the \Halpha emission associated with the satellite has a redshifted offset. Hence, this satellite could correspond to an inflow of pristine gas towards \cliff, and may therefore be associated with the observed low metallicity of this system.}
\label{fig:satellites}
\end{figure*}


Analysis of the \Halpha emission in the vicinity of \cliff reveals a candidate \Halpha-emitting satellite located to the South, as illustrated in the left panel of Fig.~\ref{fig:satellites}. Its \Halpha emission is redshifted by about +380 \kms relative to the narrow \Halpha and \OIIIL components of \cliff (Fig.~\ref{fig:satellites}; right panel). Notably, this satellite lacks detectable \OIII or other emission features within the spectral coverage, indicative of low metallicity gas. The elongated morphology and redshifted velocity of the satellite may trace a tidally stretched, accreting system, potentially associated with the redshifted absorber of \cliff identified in Section~\ref{sec:Fitting_Ha}.

The detection of this potential satellite suggests that \cliff resides in an environment characterised by the accretion of metal-poor gas \citep[e.g. preceding a `compaction' event; see][]{Dekel2023, Mannucci2010}, and this may explain the low metallicity of the galaxy.

However, while the \Halpha identification is the most plausible explanation of the observed feature, we cannot exclude the possibility that it originates from a system at a different redshift which is not physically associated with \cliff. 
Furthermore, the narrow emission lines of \cliff exhibit a remarkably low velocity dispersion ($\sigma_\mathrm{N}\sim38$ \kms), providing no clear evidence for the turbulence typically associated with an ongoing inflow. Higher-sensitivity, deeper follow-up observations are required to robustly confirm the physical association of the satellite with \cliff and further constrain its nature.

\subsection{Dynamical mass}\label{sec:M_dyn}

\begin{figure}
\centering
   \includegraphics[width=\linewidth]{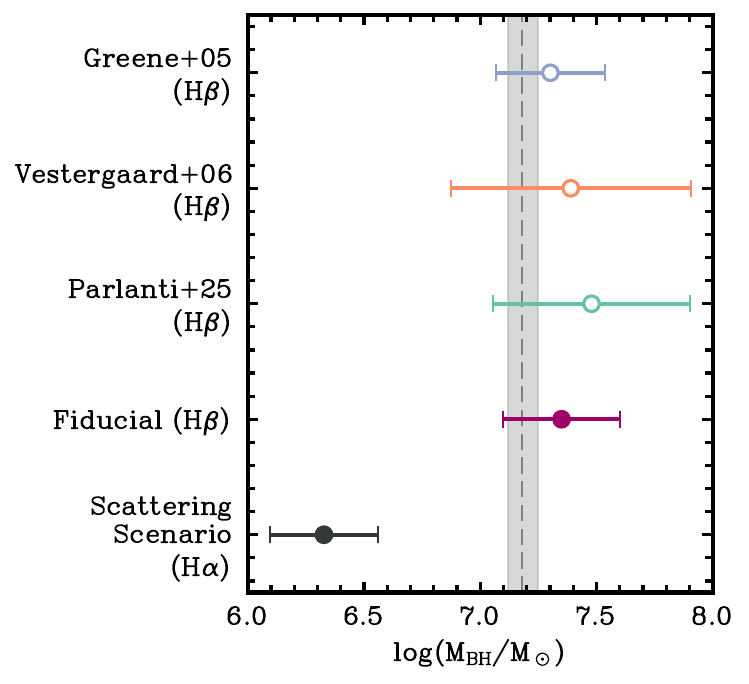}
\caption{Constraints on the BH mass of \cliff from several different calibrations, based on both $\Halpha$ and $\Hbeta$ \citep{Greene2005, Vestergaard2006, Parlanti2025, ReinesVolonteri2015}.
Our \Hbeta BH mass measurements and associated $1\sigma$ uncertainties are shown by the topmost three points; these measurements are combined to derive our fiducial BH mass (the magenta point), which is quoted in Table~\ref{table:galaxyproperties}. 
We also show the `electron scattering' scenario BH mass as determined from our \Halpha fit, following the argument of \citet{Rusakov2026}; this represents a more conservative estimate of the BH mass, and is also quoted in Table~\ref{table:galaxyproperties}.
The dashed gray line and shaded region show the BH mass estimate from \citet{deGraaff2025} along with its $1\sigma$ uncertainty.}
\label{fig:BHmassconstraint}
\end{figure}

In this work, we do not attempt to directly measure the stellar mass of \cliff. 
$M_\ast$ has proven difficult to measure across the wider LRD population, for reasons including but not limited to: degeneracies between AGN and stellar light \citep[e.g.][]{Labbe2024, Greene2024}, non-stellar Balmer breaks \citep[e.g.][]{Ji2025}, and systematic uncertainties in SED fitting \citep[e.g.][]{deGraaff2025, Durodola2025}. 
Additionally, stellar masses inferred from the optical continuum of LRDs may be unreliable, as some LRDs have been found to exhibit a time-variable rest-frame optical continuum \citep{Ji2025, Naidu2025}, suggesting the optical continuum may be AGN- rather than SF-dominated. 
In fact, in the case of \cliff, \cite{deGraaff2025} have already demonstrated the difficulties in measuring $M_\ast$ by fitting the spectral energy distribution (SED) of this object, as SED models favour an implausibly high stellar mass, and stellar emission is insufficient to explain the strength of the observed Balmer break. 
Furthermore, attempts to measure $M_\ast$ of \cliff from its spectral continuum may be affected by contamination from the nearby foreground galaxy (see Fig.~\ref{fig:fov}).
Consequently, in this work we adopt the dynamical mass of \cliff as a strong upper constraint on its stellar mass.

As discussed earlier, the small velocity dispersion inferred from \OIIIL and the narrow component of \Hbeta (see Section~\ref{sec:fitting_Hb_OIII} and Table~\ref{table:fit_properties}) indicates that the dynamical mass of \cliff's host galaxy is likely fairly low. 
To constrain the dynamical mass, we adopt the same approach outlined in other high-redshift studies \citep[e.g.][]{Ji2025, Ubler2023, Maiolino2024_JADES, Ivey2026} of estimating the dynamical mass through the calibration of \cite{VanDerWel22}:
\begin{equation}
M_\mathrm{dyn} = \beta(n)K(q)\frac{\sigma_\ast^2 R_e}{G},
    \label{eq:dynmass}
\end{equation}
where $\beta(n) = 8.87 - 0.831n+0.0241n^2$ with Sérsic index $n$ following \cite{Cappellari2006}, and $K(q) = [0.87+0.38e^{-3.71(1-q)}]^2$ with axis ratio $q$ following \cite{VanDerWel22}. 
Given we lack information about $q$ beyond the fit to the PSF, we adopt $q=1$ as this yields a conservative estimate of the upper limit on the dynamical mass (taking $q = 0$ would reduce $M_\mathrm{dyn}$ by a factor of $\sim2$). 
For the Sérsic index we assume $n=1$ as is the case for the majority of high-$z$ SFGs \citep[][]{Ormerod2024, Danhaive2026}, including AGN-host galaxies \citep{Maiolino2024_JADES,Juodzbalis2026_JADEScensus}. $R_e$ is the effective radius, already estimated by \cite{deGraaff2025} from morphological fitting of \jwst/NIRCam F200W imaging to be $38.6^{+7.4}_{-6.9}\ \mathrm{pc}$, which we adopt here as we have determined \cliff to be spatially unresolved.
Finally, in Equation~\ref{eq:dynmass}, $\sigma_\ast$ is the stellar velocity dispersion. 
To estimate $\sigma_\ast$, we use our constraint on the narrow line width to measure the gas velocity dispersion, 
$\sigma_\mathrm{gas} \approx 38\
 \mathrm{km} \ \mathrm{s}^{-1}$, and then apply a correction of $\Delta \log(\sigma_\mathrm{gas} / [\mathrm{km} \ \mathrm{s}^{-1}]) = +0.1$ following \cite{Ubler2023},
since galaxies with low integrated ionised gas velocity dispersion tend to have a higher integrated stellar velocity dispersion
\citep{Bezanson18}\footnote{Tests of this relation at high-redshift \citep[e.g.][]{Carnall2023,DEugenio2024_Nature, Pascalau2026} indicate that it does appear to hold at high redshift, though this result refers to systems much more massive than \cliff.}. 
By adopting these assumptions, we obtain $2.6\times 10^8 \ M_\odot$ as a conservative upper limit on the dynamical mass of \cliff. 
Hence, we obtain an upper limit on the stellar mass of $M_\ast < 2.6\times 10^8 \ M_\odot$ (Table~\ref{table:galaxyproperties}). We note that dark matter and gas fractions are likely to be high \citep[see e.g.][]{Danhaive2026,McClymont2026_BH}, hence this is a very conservative upper limit.

\cite{InayoshiMaiolino2025} have recently proposed that dense gas clouds in the vicinity of an AGN accretion disc can give rise to strong Balmer series absorption, including both line and bound-free absorption, i.e. the same phenomena happening in the atmosphere of hot stars ($T_\mathrm{eff}\sim10^4\ \mathrm{K}$). 
\cite{Ji2025} have verified that such a model can indeed produce the strong Balmer break and Balmer absorption lines such as those we observe in \cliff and in the case of QSO1, without the need to invoke massive, evolved stellar populations. 
Although QSO1 has a weaker Balmer break than \cliff, a similar argument certainly applies here, meaning a small $M_\ast$ is entirely reasonable.

\begin{figure*}
\centering
   \includegraphics[width=0.85\paperwidth]{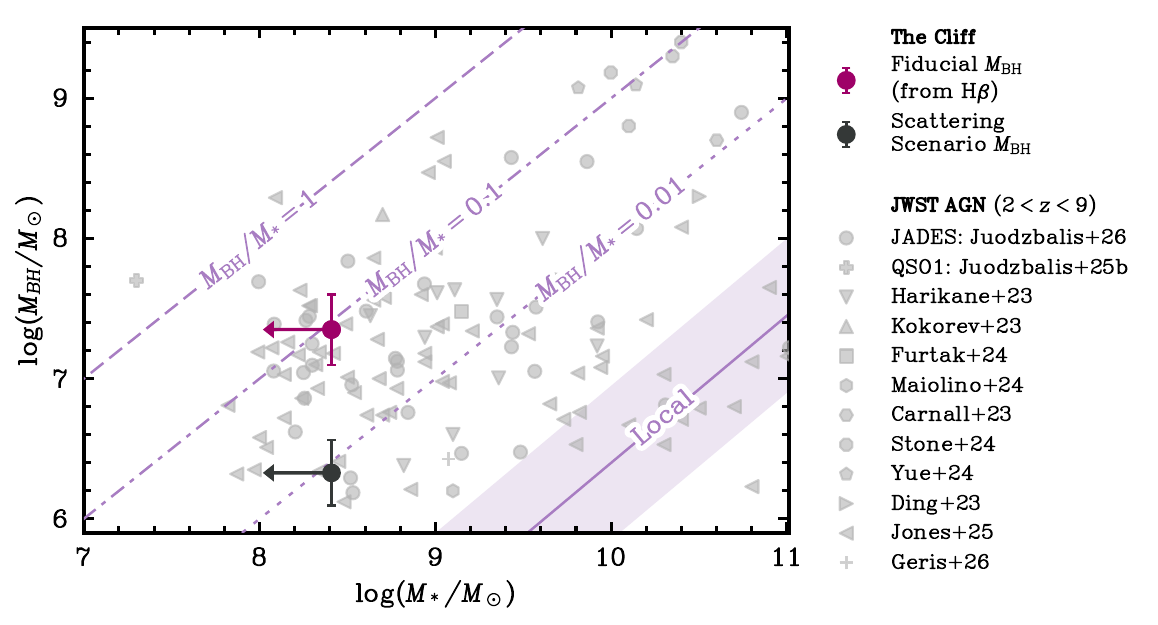}
\caption{The location of \cliff on the $M_\mathrm{BH}$--$M_\ast$ plane, for both the fiducial \Hbeta-measured BH mass (magenta circle) and the scattering scenario BH mass (black circle). 
The grey points correspond to measurements from other \jwst observations of low mass AGN \citep{Juodzbalis2026_JADEScensus, Juodzbalis2025_QSO1, Harikane2023, Kokorev2023, Furtak2024, Maiolino2024_Nature, Carnall2023} and quasars \citep{Stone2024, Yue2024, Ding2023}, along with the sample of 3 < z < 7 broad-line AGN from \citet{JonesKocevski2025} and the mean stacks of low-mass AGN from \citet{Geris2026}. 
The solid purple line shows the local scaling relation from \citet{ReinesVolonteri2015}, with its scatter indicated by the shaded region. 
The other purple lines indicate constant $M_\mathrm{BH}/M_\ast$ ratios. 
Regardless of the BH mass adopted, \cliff is overmassive relative to the local relation, a conclusion made stronger by noting that $M_\ast$ presented here is a conservative upper limit estimated from $M_\mathrm{dyn}$ (see Section~\ref{sec:M_dyn}).}
\label{fig:MstarMBH}
\end{figure*}

\subsection{Black hole mass}\label{sec:BH_mass}

The inferred metallicity provides important constraints for models and simulations, but becomes more powerful when accompanied by measurements of BH mass.

In this work, we estimate the black hole mass of \cliff by assuming the local virial relations between BH mass, and the broad line luminosity \citep[e.g.][]{Greene2005, Vestergaard2006, ReinesVolonteri2015}, which are calibrated either through reverberation mapping or direct measurements. 
It is not obvious that these locally-calibrated relations apply at high redshift. 
In fact, we note that BH mass measurement at high-$z$ remains a contentious topic, with arguments that \jwst-measured BH masses are inaccurately estimated by these calibrations, for example due to super-Eddington accretion \citep[e.g.][]{Marconi2008, Marconi2009, Lambrides2024, Lupi2024} or electron scattering \citep[e.g.][]{Rusakov2026, Naidu2025, Chang2026}.

The black hole mass of \cliff was previously estimated by \cite{deGraaff2025} as $\log(M_\mathrm{BH}/M_\odot) = 7.18^{+0.07}_{-0.06}$, based on the profile of broad \Halpha. 
However, the medium-resolution spectra result in a lack of strong constraints on the narrow line kinematics or the Balmer line absorption component, which have now been revealed by the G235H observations. 

In this work, as we have the best constraints on the narrow and broad components of \Hbeta (compared to \Halpha), we primarily utilise the results of our \Hbeta fit to determine an estimate of $M_\mathrm{BH}$. We take $\mathrm{FWHM}_\mathrm{B}$ as measured from our spectral fit to the \Hbeta-\OIIIL complex, and calculate the luminosity of broad \Hbeta. 
We utilise 3 different \Hbeta-based single epoch calibrations to determine the BH mass: \cite{Greene2005, Vestergaard2006} and \cite{Parlanti2025}. 
For the \cite{Vestergaard2006} and \cite{Parlanti2025} calibrations, when calculating uncertainties we incorporate the quoted intrinsic scatters of $0.43$ and $0.40$ dex, respectively. 

There is a spread of $\sim 0.2 \ \mathrm{dex}$ across the $M_\mathrm{BH}$ values obtained from the various \Hbeta calibrations, but all of the measurements are consistent within $1\sigma$, and this spread is lower than the $\sim0.3-0.4$ dex scatter on the BH mass calibrations overall. 
Hence, to obtain a single \Hbeta-determined value for $M_\mathrm{BH}$, we combine these three measurements with an uncertainty-weighted average and propagate their uncertainties, to obtain the \textit{Mean} BH mass measurement of $\log(M_\mathrm{BH}/M_\odot) = 7.35\pm0.24$. 
This \textit{fiducial} BH mass measurement will be used in discussions and analysis throughout the rest of this work.

Noting that the \Halpha line profile was parameterised by a model compatible with the proposed electron scattering scenario \citep{Rusakov2026}, we obtain an alternative, more conservative estimate of BH mass, by using $\mathrm{FWHM}_\mathrm{B, \Halpha}=480\pm50$ \kms as associated with the intrinsic Gaussian. 
For this calculation, we utilise the single-epoch \Halpha calibrations of \cite{ReinesVolonteri2015}. 
This yields a BH mass an order of magnitude smaller than our fiducial estimate, $\log(M_\mathrm{BH}/M_\odot) = 6.32\pm0.22$. 
However, we emphasise that exponential profiles are not necessarily tracing electron scattering and that simple models of the (virialised) BLR can also naturally result in exponential profiles \citep{Scholtz2026_LRD,Madau2026b}. In particular, \citet{Madau2026b} have recently shown that the black hole masses derived by modelling LRD exponential profiles through BLR stratification, without invoking electron scattering, are fully consistent with standard single epoch estimates.

Finally, one could measure BH mass from \Halpha using the calibration of \cite{ReinesVolonteri2015}, now by using the overall FWHM of the broad component as estimated from its overall profile, $920\pm110$ \kms (which is still significantly lower than the FWHM of $\Hbeta_\mathrm{N}$, $2770\pm210$ \kms). This approach yields a measurement of $\log(M_\mathrm{BH}/M_\odot) = 6.91\pm0.23$, which is consistent with both our fiducial, mean \Hbeta BH mass measurement and the earlier estimate of \cite{deGraaff2025} (within $2\sigma$). For clarity, in this work we will limit further discussion to the \Hbeta-measured and \Halpha `scattering scenario' masses as these correspond to the extreme scenarios and therefore enable a better exploration of the parameter space.

In Fig.~\ref{fig:BHmassconstraint}, we present an overview of the BH masses of \cliff determined in this work (the various coloured points); we also compare to the earlier measurement from \cite{deGraaff2025} (the dotted gray line, with a shaded region illustrating the $1\sigma$ uncertainty).
Our fiducial, mean \Hbeta-estimated BH mass (the magenta circle) is consistent with the value from \cite{deGraaff2025} within $1\sigma$. 
We will discuss the implications of our BH mass estimates in Section~\ref{sec:overmassive}.

\section{Discussion}\label{sec:Discussion}

\subsection{An overmassive black hole}\label{sec:overmassive}

Beyond the challenges for stellar mass measurements discussed in the previous section (Section~\ref{sec:M_dyn}), as, discussed, BH mass measurements for LRDs are also subject to critical systematic uncertainties. 
While virial, locally-calibrated single-epoch SMBH mass estimators are widely adopted to measure BH masses for \jwst-discovered AGN, their reliability is limited by extensive systematic uncertainties in BLR geometry, ionisation and kinematics. 
Furthermore, single-epoch BH mass estimates can vary by up to $\sim0.5$ dex depending on the line used (e.g. \Halpha, \Hbeta) and the adopted calibration \citep[e.g.][]{DallaBonta2025}. 
These systematics are particularly important when considering LRDs, which are effectively a new class of AGN. 
Direct measurements of BH mass are essential for independently validating the single epoch estimators for high redshift AGN and LRDs.
However, while a few direct BH mass measurements exist outside of the local Universe \citep[e.g.][]{Abuter2024, Shen2024, Liao2025, GRAVITYcollab2026}, the only direct $M_\mathrm{BH}$ measurement in an LRD to date is that of QSO1 at $z\approx 7.04$ by \cite{Juodzbalis2025_QSO1}. 
$M_\mathrm{BH}$ of QSO1 was shown to be consistent with that from the single-epoch mass estimates, but this single measurement is insufficient for making any broad conclusions about whether the single epoch estimators are applicable to the wider early AGN population. 

With these caveats on virial BH masses in mind, we compare the upper limit on $M_\ast$ of \cliff (Section~\ref{sec:M_dyn}) to $M_\mathrm{BH}$ inferred from the virial relations (Section~\ref{sec:BH_mass}). 
We interpret this comparison in the context of both local BH-galaxy scaling relations and of other high-redshift \jwst-observed AGN.
Fig.~\ref{fig:MstarMBH} illustrates the relationship between BH mass and host galaxy stellar mass. 
We include \cliff in this figure (the magenta circle), together with other high-$z$ \jwst AGN sources from the literature whose masses were also estimated with single-epoch virial calibrations \citep[or direct dynamical measurement in the case of QSO1;][]{Juodzbalis2025_QSO1}. 
We find that even when adopting the lower of our two BH mass estimates (the BH mass derived assuming the scattering scenario, shown by the black circle), \cliff remains overmassive relative to the local scaling relations. 
If we instead adopt the fiducial BH mass inferred from \Hbeta, \cliff appears even more overmassive ($M_\mathrm{BH}/M_\ast \gtrsim 0.1$), though we note that both measured BH masses render \cliff consistent with the wider population of high-$z$ \jwst AGN shown in this figure.
It is crucial to note that as we have adopted an upper limit on $M_\ast$, the black hole of \cliff is likely even more overmassive than our measurements suggest, as indicated by the arrows on the plot. This argument is further strengthened by recent findings that \Halpha profiles with exponential wings are not direct evidence of electron scattering \citep{Brazzini2025, Brazzini2026}, because they can alternatively be explained by stratification of the BLR \citep{Scholtz2026_LRD, Madau2026b}.

\subsection{Metal-poor LRDs and deviation from the MZR}\label{sec:mass-metal-relation}

\begin{figure*}
\centering
   \includegraphics[width=\linewidth]{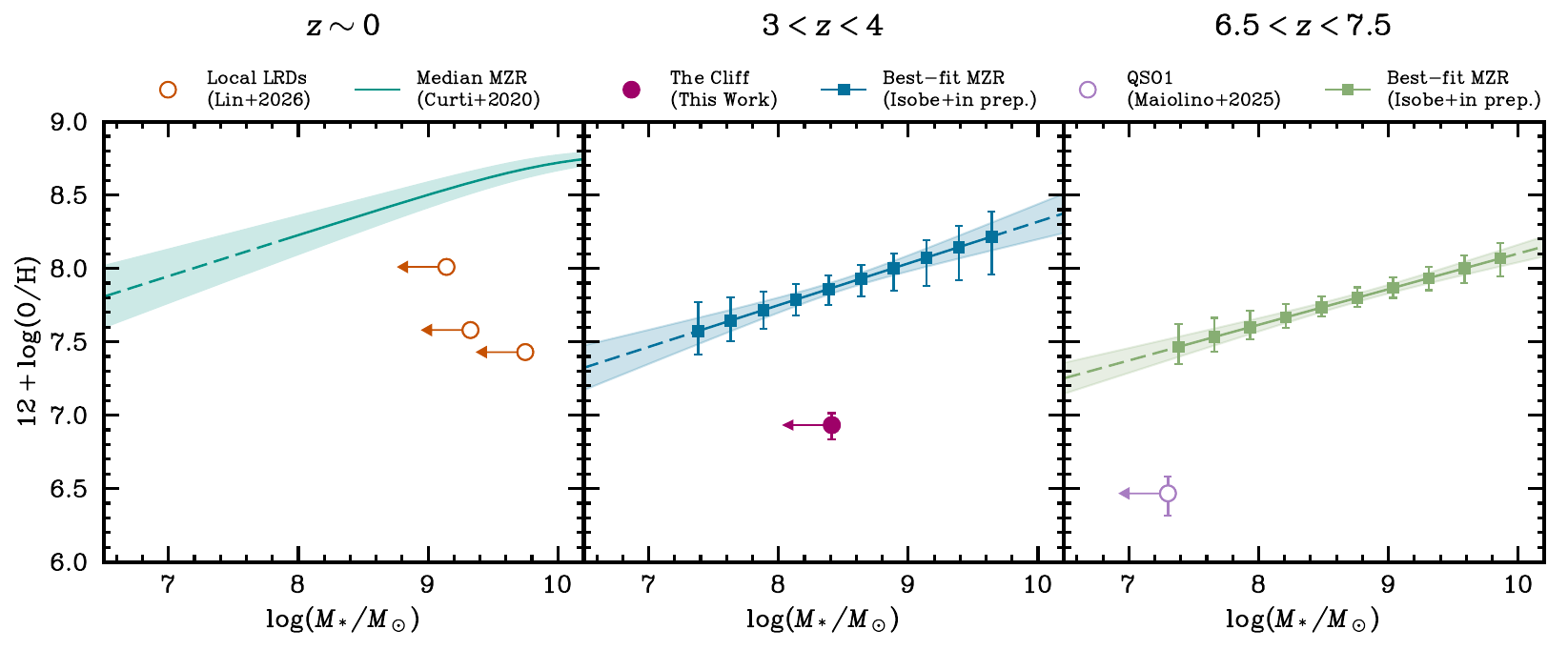}
\caption{
The mass-metallicity relation (MZR) for three different redshift ranges. For each plotted MZR, the dashed segments show the regions for which the relation is extrapolated, and the shaded area indicates the $1\sigma$ uncertainty of the relation.
\textbf{Left panel}: The median MZR for the local Universe (teal line), derived by \citet{Curti2020} based on galaxies from the Sloan Digital Sky Survey (SDSS). 
The unfilled orange circles illustrates the position of the local LRDs \citep{Lin2026} on the mass-metallicity plane, with $M_\mathrm{dyn}$ taken as an upper limit on $M_\ast$, following Equation~\ref{eq:dynmass}. 
Here we plot the direct-$T_e$ method metallicities derived by \citet{Lin2026}; the corresponding uncertainties are $0.01-0.03$ dex, smaller than is visible on the scale of this plot.
\textbf{Center panel}: The MZR for galaxies at $3 < z< 4$ (blue line), as derived by Isobe et al. (in prep.). 
The position of \cliff ($z\sim3.5$) is shown by the filled magenta point, with $1\sigma$ uncertainties on its metallicity and $M_\mathrm{dyn}$ plotted as an upper limit on $M_\ast$.
\textbf{Right panel:} The MZR for galaxies at $6.5 < z< 7.5$ (green line), as derived by Isobe et al. (in prep.).
QSO1 ($z\sim7$) is shown by the unfilled purple point, using the stellar mass upper limit from \citet{Juodzbalis2025_QSO1}, and a metallicity derived from the $\OIIIL/\Hbeta_\mathrm{N}$ flux ratio measured by \citet{Maiolino2025_QSO1} via the Isobe et al. (in prep.) R3 calibration. 
}
\label{fig:massmetallicity}
\end{figure*}

\begin{figure*}
\centering
   \includegraphics[width=\linewidth]{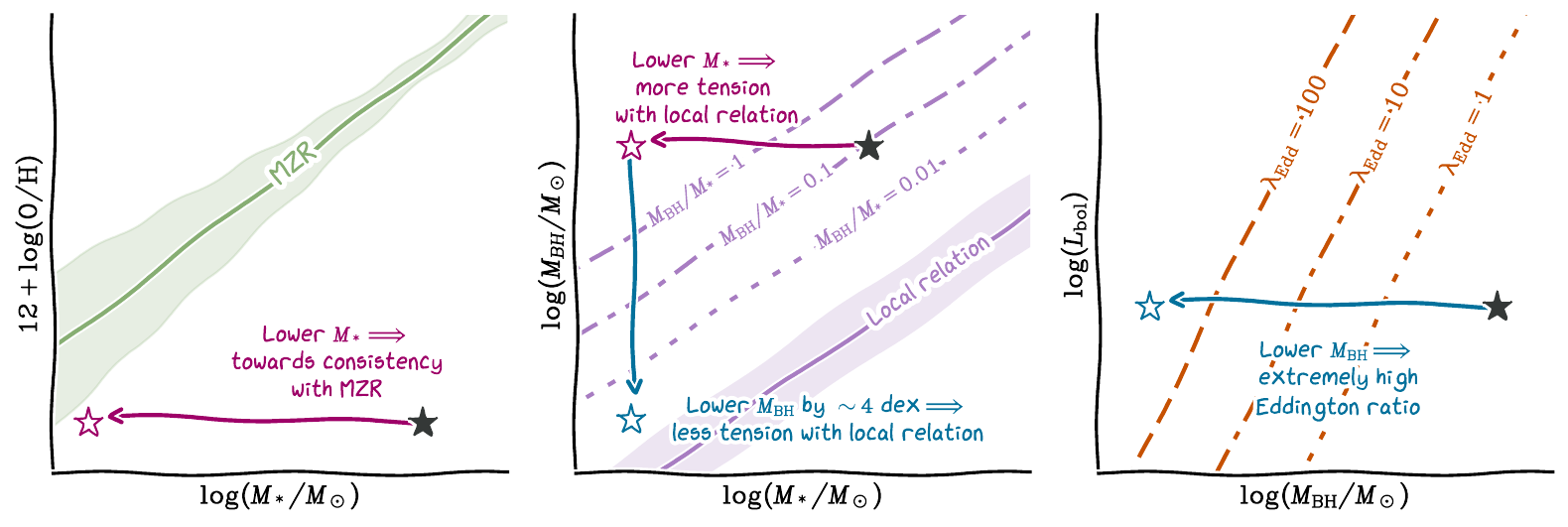}
\caption{
A sketch illustrating the key tensions discussed in this work (see Section~\ref{sec:mass-metal-relation} for a full discussion). The black stars illustrate `real measurements' of observational properties for an LRD such as \cliff, with the hollow stars corresponding to hypothetical scenarios. In brief, from left to right: for an object such as \cliff, using $M_\mathrm{dyn}$ as an upper limit on $M_\ast$ means that if the true $M_\ast$ is substantially lower, the system will be in closer agreement with the corresponding MZR. However, this lower $M_\ast$ would also increase $M_\mathrm{BH}/M_\ast$, so matching the MZR would make the BH even more overmassive relative to the host galaxy. Matching both the MZR and the local $M_\mathrm{BH}\text{--}M_\ast$ relations requires the virial $M_\mathrm{BH}$ to be overestimated by $\sim4$ dex, a factor much larger than can be attributed to even the most extreme cases of electron scattering. A black hole mass this much smaller would increases the inferred Eddington ratio to $\lambda_\mathrm{Edd} \gtrsim 100$, resulting in a system too exotic to obey the typical BH scaling relations, and indicating that systems such as \cliff fundamentally deviate from the local scaling relations.}
\label{fig:key_tensions}
\end{figure*}

As we have noted across the previous sections, \cliff exhibits a number of observational similarities to QSO1. This raises a number of questions: Do the formation pathways producing objects such as \cliff operate across cosmic time? Is low metallicity characteristic of unusual LRDs and therefore an indicator of rare pathways for BH seeding and evolution? 

To further investigate the connection between \cliff and other extreme LRDs, Fig.~\ref{fig:massmetallicity} illustrates comparison of the mass-metallicity relation (MZR) to `unusual' LRDs across three different redshift regimes: $z\sim0$, $3<z<4$ and $6.5<z<7.5$. We note as a caveat for this discussion that this comparison requires extrapolation of MZRs to lower stellar masses. The center panel of Fig.~\ref{fig:massmetallicity}, presents a comparison of \cliff to the $3<z<4$ MZR determined from the calibrations of Isobe et al. (in prep.). Adopting the same calibrations for both the MZR and the strong-line metallicity measurement ensures a self-consistent comparison. 
Visually, \cliff is clearly offset from the MZR at this redshift; its deviation of $\sim11 \sigma$ from the best-fit MZR statistically confirms its status as an outlier.
Indeed, marginal consistency with the MZR (i.e. a $3 \sigma$ deviation) would require a stellar mass low enough to attain $M_\mathrm{BH}/M_\ast \sim 9$, a ratio larger than the lower limit determined for QSO1 ($M_\mathrm{BH}/M_\ast > 2$), which has been identified be a `nearly naked' black hole in an early stage of its evolution \citep{Juodzbalis2025_QSO1}. Even in the case of the electron scattering black hole mass, such a scenario results in $M_\mathrm{BH}/M_\ast \sim 1$. This hints at a fundamental tension between the MZR and $M_\mathrm{BH}-M_\ast$ relation for \cliff.

In the left panel of Fig.~\ref{fig:massmetallicity}, we present a comparison of the so-called `local LRDs' \citep[$z\sim0.1-0.22$][]{Lin2026} to the local MZR from \cite{Curti2020}; these are 3 of a mere handful of local LRDs found to date \citep{Lin2025, Lin2026} and therefore, like \cliff, represent a rare population.
We adopt the direct metallicity measurements of \cite{Lin2026} for this comparison.
We additionally estimate the dynamical masses of the local LRDs of \cite{Lin2026} as an upper limit on their stellar masses using Equation~\ref{eq:dynmass}, adopting the narrow-line FWHMs measured by \cite{Lin2026} and the tightest available constraint on $R_e$ for each LRD \citep[i.e. from][]{Lin2026, Ji2026}.
With these upper limits on stellar mass, even the highest-metallicity of the three local LRDs deviates from the median $z\sim0$ MZR by more than $5\sigma$. The deviation from the MZR is yet more significant for the other two local LRDs.

Finally, in the rightmost panel of Fig.~\ref{fig:massmetallicity}, we present a comparison of QSO1 ($z\sim7$) to the $6.5<z<7.5$ MZR (also from Isobe et al. in prep.).
We re-estimated the metallicity of QSO1 based on the \cite{Maiolino2025_QSO1} measurement of $F(\OIIIL)/F(\Hbeta_\mathrm{N} = 0.55\pm0.16$, now using the Isobe et al. (in prep.) R3 calibration to obtain a self-consistent comparison to the MZR.
At the upper $M_\ast$ limit afforded by $M_\mathrm{dyn}$, we find QSO1 is also a clear outlier from the $6.5<z<7.5$ MZR by $\gtrsim7\sigma$.

Taken together, these comparisons reveal a fundamental tension between extreme LRDs and the established galaxy-black hole scaling relations. The statistical significance with which \cliff, QSO1 and local LRDs deviate from the MZR suggests that standard frameworks for galaxy assembly do not apply to these systems. Indeed, these LRDs may follow a unique evolutionary trajectory, in which the processes governing black hole growth are decoupled from the metal enrichment and stellar assembly of the host galaxy.

\subsection{The \texorpdfstring{$M_\mathrm{BH}$}{MBH}--\texorpdfstring{$M_\ast$}{M*}--\texorpdfstring{$Z$}{Z} tension}

Fig.~\ref{fig:key_tensions} illustrates the key tensions highlighted by the previous sections. The dynamical mass of \cliff represents an upper limit on its stellar mass and a lower $M_\ast$ would therefore align \cliff more closely with the $3<z<4$ MZR (Fig.~\ref{fig:key_tensions}; left panel). However, reducing $M_\ast$ simultaneously pushes \cliff further into the `overmassive' regime on the $M_\mathrm{BH}-M_\ast$ plane. This holds true even when adopting a more conservative BH mass estimate calculated based on an electron scattering scenario, increasing the tension with the local $M_\mathrm{BH}-M_\ast$ relation (Fig.~\ref{fig:key_tensions}; centre panel). Consequently, under standard assumptions, objects like \cliff cannot simultaneously follow both scaling relations. Possible resolutions to this predicament involve invoking extreme density conditions (Section~\ref{sec:Accounting_for_density}), or assuming black hole masses are not measured accurately by the virial relations.

If the virial BH masses are indeed overestimated, the tension with the local $M_\mathrm{BH}-M_\ast$ relation is reduced - however, this would require the BH mass to be overestimated by four orders of magnitude, much higher than then even the most extreme electron scattering scenarios. Additionally, this would require a extremely high Eddington ratios ($\lambda_\mathrm{Edd}\gtrsim 100$; right panel of Fig.~\ref{fig:key_tensions}). Such extreme Eddington ratios are virtually absent even from large-scale quasar surveys, which show that even the most rapidly accreting BHs rarely exceed $\lambda_\mathrm{Edd}\sim1-3$, and ratios significantly above 10 remain extreme outliers \citep[e.g.][]{Kollmeier2006, Jin2012}.

In any case, if one accepts the possibility of such high $\lambda_\mathrm{Edd}$, there is no compelling reason to expect such objects would obey local scaling relations, invalidating the assumptions that led us to this extreme picture. Consequently, our findings suggest that these extreme LRDs, and potentially other high-$z$ \jwst AGN, inherently deviate from the local $M_\mathrm{BH}-M_\ast$ relation. This deviation is likely driven by selection biases which systematically miss lower $M_\mathrm{BH}$ sources \citep[e.g.][]{Geris2026, Ziparo2026}, while simulations indicate that these early overmassive black holes eventually evolve onto the local relation over time \citep{Hu2026}.

As highlighted by Fig.~\ref{fig:massmetallicity}, \cliff is not the only object for which this set of tensions exists, pointing to a formation channel which can produce massive BHs in metal-poor environments across cosmic times. While low metallicity does appear to be characteristic of certain LRDs such as \cliff or QSO1, the evolutionary pathways that produce (or indeed maintain) such metal-poor states remain effectively unconstrained. These systems could have transitioned from an earlier metal-rich phase to their observed metal-poor state due to the efficient ejection of metals by galactic outflows; alternatively, the ISM could have been diluted by significant pristine gas inflows (e.g. in the case of \cliff, by accretion from a metal-poor satellite; see Section~\ref{sec:satellites}). Evaluating the feasibility of such pathways is limited by uncertainties associated with factors including metal transport and outflow efficiencies, for example. Consequently, numerical simulations of black hole formation and evolution provide a vital framework for interpreting the chemical histories and BH seeding pathways of LRDs such as \cliff.

\subsection{Comparison to hydrodynamical simulations}\label{sec:Comparison_to_sims}

\begin{figure*}
\centering
   \includegraphics[width=\linewidth]{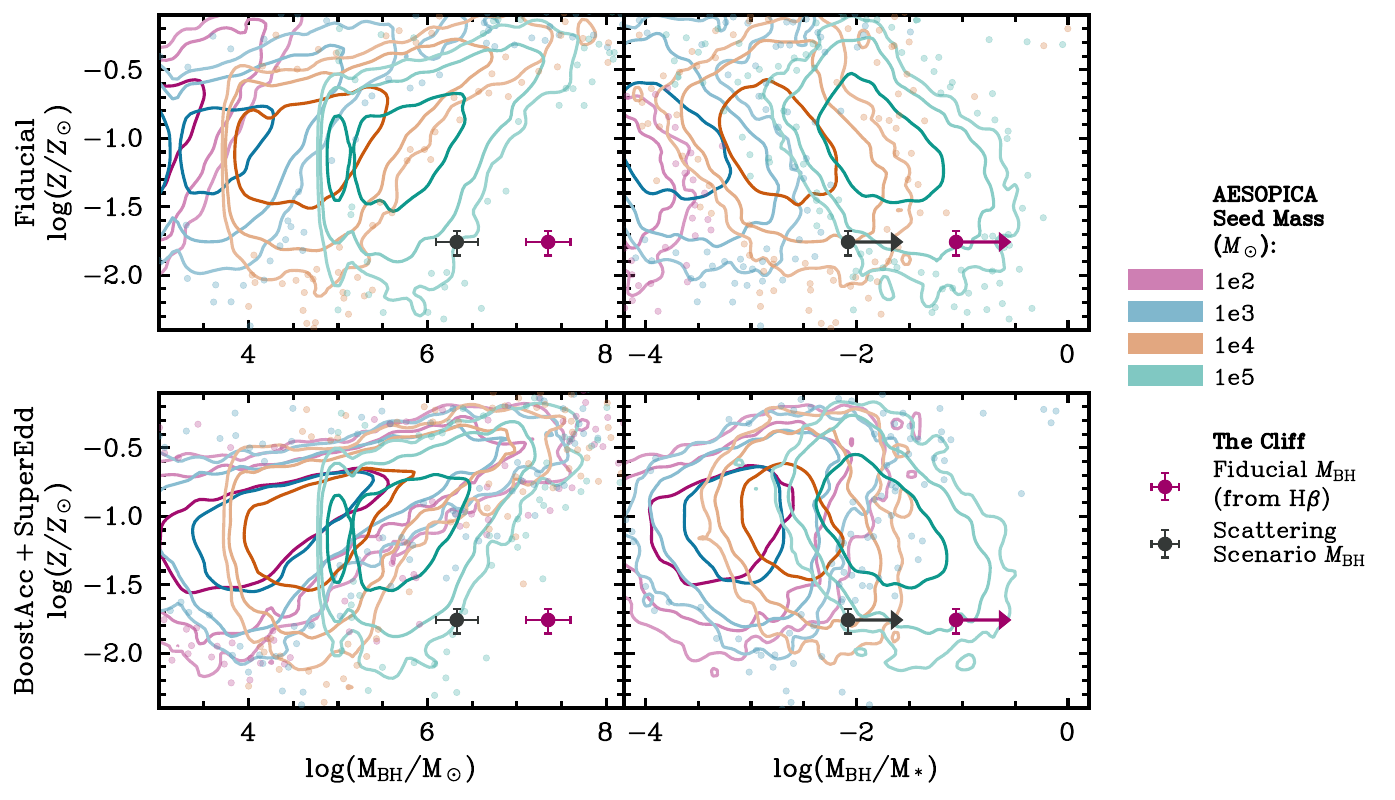}
\caption{Comparison of \cliff to results from \textsc{aesopica} hydrodynamical simulations run down to $z\sim3.5$ on the $M_\mathrm{BH}-Z$ (left) and $M_\mathrm{BH}/M_\ast-Z$ (right) diagrams. 
Further descriptions of the simulations are provided in Section~\ref{sec:Comparison_to_sims} and Appendix~\ref{sec:sims}. 
Contours enclose 68\%, 95\%, and 99\% of the simulations; simulations outside of the 99\% contours are plotted as individual points. 
The large magenta and black circles illustrate the properties of \cliff for our fiducial and scattering scenario BH masses, respectively.}
\label{fig:simcomparison}
\end{figure*}

In many scenarios of BH-galaxy co-evolution, it is unlikely for a BH to grow to a mass $>10^7 \ M_\odot$ in a metal-free environment. 
This follows from the fact that the conditions conducive to rapid black hole growth (i.e abundant cold gas accretion) would also catalyse star formation, resulting in rapid metal enrichment. 
Hence, the findings of this work (and those of \citealt{Maiolino2025_QSO1} and \citealt{Jones2026}, in the case of QSO1) are indicative of a route that could allow a black hole to form and grow within a chemically unevolved stellar system, seemingly at odds with any simple picture of galaxy--black hole co-evolution. 

Due to the complexity and stochasticity of the processes involved in co-evolution, any quantitative approach to the question of whether black holes can grow significantly in pristine environments requires a simulation-based approach. 
In this Section, we compare our findings for \cliff to predictions from the \textsc{Aesopica} simulations (Koudmani et al., in prep.; see Appendix~\ref{sec:sims} for a full description). 
We stress that these simulations incorporate stellar and BH feedback capable of gas and metal removal, as well as merger and gas accretion capable of ISM dilution. 
For the purposes of this work, we utilise two different simulation modes: (i) fiducial Eddington-limited accretion following \textsc{Fable} \citep[\textit{‘Fiducial’};][]{Henden2018} and (ii) boosted accretion with super-Eddington accretion up to ten times the Eddington limit (\textit{‘SuperEdd+BoostAcc’}). 

Fig.~\ref{fig:simcomparison} shows the location of \cliff on the $M_\mathrm{BH}-Z$ plane (first column of panels) and $M_\mathrm{BH}/M_\ast-Z$ plane (second column). 
The large symbols show the location of \cliff as inferred from our analysis: both the single-epoch virial BH mass (magenta circle) and the BH mass inferred in the electron-scattering scenario (black circle), plotted against our measurement of the metallicity. 
Each panel shows the results of four different simulation runs, each associated with a different BH seed mass; the range of seed masses explored in these simulations is $10^2 - 10^5 \ \mathrm{M}_\odot$. 

In the $M_\mathrm{BH}-Z$ plane, even when adopting the lower electron-scattering $M_\mathrm{BH}$ estimate, only simulations utilising the highest seed masses ($10^5 \ M_\odot$) successfully approach the observed parameter space of \cliff; this discrepancy is exacerbated when adopting the fiducial $M_\mathrm{BH}$ estimate.
In particular, the mismatch relative to lower-mass seeds is so large that it is unlikely to be salvaged by earlier seeding.
While a small subset of the highest seed mass simulations recover metallicities and BH masses comparable to \cliff, the simulation runs incorporating boosted accretion exhibit greater metallicity scatter in this plane, suggesting that super-Eddington growth phases may be a key evolutionary component for extreme objects like \cliff. Tracking the specific inflow and outflow histories of the simulated outliers could provide key insights into the mechanisms governing the evolution of \cliff.

The $M_\mathrm{BH}/M_\ast-Z$ plane proves to be less constraining as \cliff overlaps more broadly with the parameter space of the simulation results, although heavy seeds ($10^4-10^5 \ M_\odot)$ remain favoured. 
Crucially, as the stellar mass presented here corresponds to a strict upper limit, the true position of \cliff in this plane likely shifts toward the highest seed mass simulation runs or indeed the extreme outliers of the simulated population, depending on the BH mass estimate adopted.
Collectively, these constraints point toward a rare evolutionary pathway for \cliff, defined by heavy black hole seeding and periods of super-Eddington growth.

To evaluate whether gas inflows are predominantly responsible for the low-$Z$ outliers in the simulations, tracking the evolutionary history of individual outliers is required. Regardless of the underlying driver of these outliers, the left-hand panels of Fig~\ref{fig:simcomparison} demonstrate that \cliff remains $\sim 1$ dex below the simulation $M_\mathrm{BH}-Z$ relation, even for heavy seed simulations runs. Adopting an earlier seeding prescription would not resolve this tension, as the $M_\mathrm{BH}-Z$ relation appears to evolve towards higher metallicities as $M_\mathrm{BH}$ increases.


To gain context for the rareness of objects such as \cliff in the simulations compared to the rareness of \cliff itself, it is key to account for the survey and simulation volumes.  
The \textsc{Aesopica} simulation boxes have side length $60 \ \mathrm{cMpc}$, giving a total simulation volume of $2.16 \times10^5 \ \mathrm{cMpc}^3$ for the snapshot at $z\sim3.5$. 
Adopting the same same cosmology used by \textsc{Aesopica} \citep[see][]{Henden2018} and the survey area of $\sim150 \ \mathrm{arcmin}^2$ \citep{RUBIES2025}, the RUBIES survey volume over the redshift range $3 < z < 4$  is $5.06 \times10^5 \ \mathrm{cMpc}^3$, i.e. about 2.3 times larger than the simulation volume. 
\cliff is an extremely unique object within the RUBIES survey, and seems similarly rare within the simulations, and we conclude it is fair to evaluate the underlying seeding and evolutionary mechanisms as similarly rare. 

One alternative scenario involving primordial black hole seeds (black holes theorised to form shortly after the Big Bang) may naturally produce such rare systems with overmassive BHs in chemically unevolved environments~\citep[for a review, see][]{Carr2020}. Recent hydrodynamical simulations by \citet{Zhang2025_hydrosim_QSO1,Zhang2025_hydrosim} have explored galaxy formation around massive PBH seeds with masses of $\sim 10^{6}-5\times10^{7}\,M_\odot$ (a mass scale motivated by \citealt{Carr2021PDU....3100755C}, for example), which act as deep gravitational potential wells for halo formation and subsequent baryonic inflow. In these models, strong accretion-driven feedback heats the surrounding gas and suppresses or delays stellar assembly. Once star formation begins, feedback-driven outflows expel metal-enriched gas, while continued baryonic inflow replenishes the system with pristine material. Such simulations reproduce several qualitative properties inferred for objects like \cliff\ or QSO1, including extremely low metallicities ($Z \lesssim 0.01\,Z_\odot$), compact morphologies ($R_{\rm e}\lesssim 50~\mathrm{pc}$), and BH-dominated mass budgets ($M_{\rm BH}/M_\ast \gtrsim 0.1$). Although these simulations currently stop at higher redshift, $z\sim7-9$, and are carried out in a small volume of $\sim 1~\mathrm{cMpc}^3$ (targeting a single halo around the PBH), they nonetheless illustrate that PBH seeds could provide a viable evolutionary pathway for near-pristine LRDs.\footnote{For details of the numerical setup and analysis, the reader should refer to \citet{Zhang2025_hydrosim, Zhang2025_hydrosim_QSO1}.}.
In fact, sources like \cliff could present an opportunity to empirically constrain PBH abundance as a fraction of dark matter density, a key cosmological parameter \citep[see e.g.][]{Carr2021} which to date remains poorly understood. For simplicity, if one assumes that the comoving number density of such sources is approximately the inverse survey volume, i.e. $\sim 2\times10^{-6}\,\mathrm{cMpc}^{-3}$, then the corresponding PBH seed density fraction is roughly $f_{\rm PBH, seed} \sim 3.2\times10^{-11} \left(M_{\rm PBH} / 10^{6}\, M_\odot\right)$. This estimate is well below current observational upper limits over the mass range $\sim10^{6}$--$5\times10^{7}\,M_\odot$. If only a fraction of such objects are seeded by PBHs, the implied constraint would become correspondingly tighter.

Due to the high computational costs and variety of relevant physical processes involved in running simulations down to low redshift, very few simulations targeted at BH formation and evolution in the early Universe reach $z\sim3.5$, so we caution that the models presented here represent a limited comparison and are certainly not intended to be exhaustive. 
As an additional caveat for \textsc{Aesopica}, we must stress that the simulations are not necessarily developed with assumptions aimed at reproducing BHs similar to \cliff, nor do they directly model the multi-phase ISM structure which would be crucial to fully unravelling the origin of these pristine galaxies with overmassive BHs.
Finally, it is crucial to note that simulations and observations do not infer metallicities in the same way, which can result in significant systematic differences between observational measurements and measurements derived from simulations. 
For example, with forward modelling of metallicity, \cite{PayyoorVijayan2026} note a discrepancy between simulation-inferred and strong-line metallicities of up to 0.3 dex, which they find is driven by dust attenuation and properties of the underlying stellar population.
Ultimately, future models and simulations will cover a wider parameter space, pushing to higher resolution and lower redshifts, offering a more comprehensive comparison with rare objects such as \cliff and offering better insight into the underlying physics.

\section{Conclusions}\label{sec:Conclusions}
We have presented new, high-resolution \jwst/NIRSpec-IFU observations of \cliff, a remarkable Little Red Dot at $z\sim3.55$. These observations have offered valuable new insights into the ISM properties and chemical enrichment of \cliff. In this work we have focused on the narrow components of the emission lines, \Hbeta and \OIIIall in particular, with the goal of obtaining constraints on BH mass and metallicity which could be combined with simulations to investigate potential BH growth and seeding pathways. The main findings of our analysis are as follows:

\begin{itemize}
    \item The \Hbeta and \OIIIL emission are spatially compact and unresolved beyond the scale of the PSF.
    \item The high-resolution spectral observations reveal deep, redshifted absorption components in both \Halpha and \Hbeta, indicative of inflowing gas. 
    \item We additionally identify a potential satellite south of \cliff detected only in \Halpha, possibly tidally stretched, which could trace pristine gas inflows onto \cliff. However, we note that confirming this satellite and performing a detailed characterisation will require deeper follow-up observations.
    \item In the galaxy-integrated spectrum of \cliff, the \OIIIL[5007] line is weak relative to the narrow component of \Hbeta emission compared to the bulk of the galaxy population at the same redshift, with a flux ratio $F(\OIIIL)/F(\Hbeta_\mathrm{N}) = 1.59^{+0.26}_{-0.24}$.
    \item Adopting the updated strong-line metallicity calibrations of Isobe et al. (in prep.), we estimate the metallicity of \cliff as $Z \approx 0.017\pm0.004 \ Z_\odot$.
    \item This finding is reinforced by the non-detection of key low-ionisation lines (\OIIall, \NIIall and \SIIall). Alternative explanations for the low $\OIIIL/\Hbeta_\mathrm{N}$ ratio, such as collisional saturation or extreme ionisation parameters, are disfavoured. 
    \item We identify \cliff as a significant outlier from the $3<z<4$ MZR (also derived from the calibrations of Isobe et al. in prep.), appearing exceptionally metal-poor for its stellar mass limit ($\log(M_\ast/M_\odot) < 8.41$).
    \item By fitting the Balmer lines, we derive two separate black hole mass measurements: a fiducial virial mass of $\log(M_\mathrm{BH}/M_\odot) = 7.35\pm0.24$, and a scattering scenario mass of $\log(M_\mathrm{BH}/M_\odot) = 6.32\pm0.22$ (from \Halpha). 
    \item In both scenarios, the black hole of \cliff is overmassive relative to the local $M_\mathrm{BH}-M_\ast$ scaling relation. Reconciling \cliff with the MZR would require an even lower stellar mass, further exacerbating its deviation from the $M_\mathrm{BH}-M_\ast$ relation.
    \item The observed tension between the MZR and the $M_\mathrm{BH}-M_\ast$ relation could be resolved by a black hole mass overestimated by four orders of magnitude and an extreme super-Eddington growth phase, but the typical galaxy-black hole scaling relations would not be expected to hold for such a scenario. It is more likely that MZR deviation and overmassive BHs are in some way characteristic of the evolutionary pathway underpinning \cliff, meaning \cliff inherently deviates from these relations.
    \item This tension is not limited to \cliff, but is also shown to apply to the local LRDs \citep[][]{Lin2026} and Abell2744-QSO1 \citep{Maiolino2025_QSO1, Juodzbalis2025_QSO1}, pointing to a potential rare evolutionary pathway shared by these unusual LRDs which operates across cosmic time.
    \item While some simulations of BH growth from heavy seeds ($10^4-10^5 \ M_\odot$) with periods of super-Eddington accretion can capture the high $M_\mathrm{BH}$ and $M_\mathrm{BH}/M_\ast$, or low metallicity, of \cliff, these outcomes are rare. The existence of such a metal-poor, overmassive-BH system at $z\sim3.55$ remains difficult to capture within existing evolutionary frameworks. While a Primordial Black Hole origin for \cliff may be a possibility, this scenario requires further, more quantitative exploration. 
\end{itemize}

The constraints on the metallicity and black hole mass of \cliff, when combined with simulations of BH growth and evolution, suggest a rare BH seeding and growth pathway underpins this system. Without the context of a larger sample of similar systems, it remains unclear whether the MZR deviation and overmassive BH of \cliff are typical of other LRDs. Future simulations incorporating a resolved, multi-phase ISM will be essential to disentangle the mechanisms fuelling the growth of massive black holes within such metal-poor environments. Until then, \cliff represents a critical benchmark for understanding the formation and early evolution of Little Red Dots. 

\section*{Acknowledgements}
LRI, FDE, RM, XJ, GCJ and JS acknowledge support by the Science and Technology Facilities Council (STFC), European Research Council (ERC) Advanced Grant 695671 “QUENCH" and the UKRI Frontier Research grant RISEandFALL. RM also acknowledges funding from Royal Society research professorship. 
YI is supported by JSPS KAKENHI grant number 24KJ0202. 
Ĳ acknowledges support by the Huo Family Foundation through a P.C. Ho PhD Studentship. 
SK is supported by a Junior Research Fellowship from St Catharine’s College, Cambridge and a Research Fellowship from the Royal Commission for the Exhibition of 1851. 
MP acknowledges support through the grants PID2021-127718NB-I00, PID2024-159902NA-I00, and RYC2023-044853-I, funded by the Spain Ministry of Science and Innovation/State Agency of Research MCIN/AEI/10.13039/501100011033 and El Fondo Social Europeo Plus FSE+.
SZ, VB and BL acknowledge the Texas Advanced Computing Center (TACC) for providing HPC resources under allocation AST23026. 
AJB acknowledges funding from the "FirstGalaxies" Advanced Grant from the ERC under the European Union's Horizon 2020 research and innovation prograxm (Grant agreement No. 789056).
SC acknowledges support by EU HE ERC Starting Grant No. 101040227 - WINGS.
KI acknowledges support from the National Natural Science Foundation of China (12573015, 1251101148, 12233001, 12473037), and the China Manned Space Program (CMS-CSST-2025-A09.
BL acknowledges the funding of the Deutsche Forschungsgemeinschaft (DFG, German Research Foundation) under Germany's Excellence Strategy EXC 2181/1 - 390900948 (the Heidelberg STRUCTURES Excellence Cluster).
RP acknowledges funding support from a STFC PhD studentship.
PR and BR acknowledge support from JWST/NIRCam contract to the University of Arizona NAS5-02105.
ST acknowledges support from the Royal Society Research Grant G125142.

This work is based on observations made with the NASA/ESA/CSA James Webb Space Telescope. The data were obtained from the Mikulski Archive for Space Telescopes (MAST) at the Space Telescope Science Institute, which is operated by the Association of Universities for Research in Astronomy, Inc., under NASA contract NAS 5-03127 for JWST. These observations are associated with JWST program ID 9433.

\section*{Data Availability}

The datasets used in this work were derived from a source in the public domain. The \jwst/NIRSpec data can be downloaded from the MAST (\url{https://mast.stsci.edu/portal/Mashup/Clients/Mast/Portal.html}) under PID 9433. Photometry was obtained from PID~1837 and PID~7814.



\bibliographystyle{mnras}
\bibliography{example} 

@ARTICLE{Kido2025,
       author = {{Kido}, Daisaburo and {Ioka}, Kunihito and {Hotokezaka}, Kenta and {Inayoshi}, Kohei and {Irwin}, Christopher M.},
        title = "{Black hole envelopes in Little Red Dots}",
      journal = {\mnras},
         year = 2025,
        month = dec,
       volume = {544},
       number = {4},
        pages = {3407-3416},
          doi = {10.1093/mnras/staf1898},
archivePrefix = {arXiv},
       eprint = {2505.06965},
 primaryClass = {astro-ph.HE},
       adsurl = {https://ui.adsabs.harvard.edu/abs/2025MNRAS.544.3407K}
}

@ARTICLE{Martinez2025,
       author = {{Martinez}, Zorayda and {Berg}, Danielle A. and {James}, Bethan L. and {Arellano-C{\'o}rdova}, Karla Z. and {Stark}, Daniel P. and {Senchyna}, Peter and {Skillman}, Evan D. and {Rogers}, Noah S.~J. and {Chisholm}, John},
        title = "{Under Pressure: Decoding the Effect of High Densities on Derived Nebular Properties}",
      journal = {\apj},
         year = 2025,
        month = dec,
       volume = {995},
       number = {2},
          eid = {204},
        pages = {204},
          doi = {10.3847/1538-4357/ae17c6},
archivePrefix = {arXiv},
       eprint = {2510.21960},
 primaryClass = {astro-ph.GA},
       adsurl = {https://ui.adsabs.harvard.edu/abs/2025ApJ...995..204M}
}

@ARTICLE{Madau2026b,
       author = {{Madau}, Piero and {Maiolino}, Roberto and {Scholtz}, Jan and {D'Eugenio}, Francesco},
        title = "{Wings of little dots: Exponential broad lines from a stratified BLR}",
      journal = {arXiv e-prints},
         year = 2026,
        month = apr,
          eid = {arXiv:2604.04216},
        pages = {arXiv:2604.04216},
archivePrefix = {arXiv},
       eprint = {2604.04216},
 primaryClass = {astro-ph.GA},
       adsurl = {https://ui.adsabs.harvard.edu/abs/2026arXiv260404216M}
}

@ARTICLE{Jeon_LRDs2026,
       author = {{Jeon}, Junehyoung and {Liu}, Boyuan and {Bromm}, Volker and {Fujimoto}, Seiji and {Taylor}, Anthony J. and {Kokorev}, Vasily and {Larson}, Rebecca L. and {Chisholm}, John and {Finkelstein}, Steven L. and {Kocevski}, Dale D.},
        title = "{Little Red Dots and Their Progenitors from Direct Collapse Black Holes}",
      journal = {\apj},
         year = 2026,
        month = feb,
       volume = {998},
       number = {1},
          eid = {148},
        pages = {148},
          doi = {10.3847/1538-4357/ae3725},
archivePrefix = {arXiv},
       eprint = {2508.14155},
 primaryClass = {astro-ph.GA},
       adsurl = {https://ui.adsabs.harvard.edu/abs/2026ApJ...998..148J}
}

@ARTICLE{Inayoshi_rev2020,
       author = {{Inayoshi}, Kohei and {Visbal}, Eli and {Haiman}, Zolt{\'a}n},
        title = "{The Assembly of the First Massive Black Holes}",
      journal = {\araa},
         year = 2020,
        month = aug,
       volume = {58},
        pages = {27-97},
          doi = {10.1146/annurev-astro-120419-014455},
archivePrefix = {arXiv},
       eprint = {1911.05791},
 primaryClass = {astro-ph.GA},
       adsurl = {https://ui.adsabs.harvard.edu/abs/2020ARA&A..58...27I}
}

@ARTICLE{Smith_rev2019,
       author = {{Smith}, Aaron and {Bromm}, Volker},
        title = "{Supermassive black holes in the early universe}",
      journal = {Contemporary Physics},
         year = 2019,
        month = apr,
       volume = {60},
       number = {2},
        pages = {111-126},
          doi = {10.1080/00107514.2019.1615715},
archivePrefix = {arXiv},
       eprint = {1904.12890},
 primaryClass = {astro-ph.GA},
       adsurl = {https://ui.adsabs.harvard.edu/abs/2019ConPh..60..111S}
}

@ARTICLE{Jeon_DCBH2025,
       author = {{Jeon}, Junehyoung and {Bromm}, Volker and {Liu}, Boyuan and {Finkelstein}, Steven L.},
        title = "{Physical Pathways for JWST-observed Supermassive Black Holes in the Early Universe}",
      journal = {\apj},
         year = 2025,
        month = feb,
       volume = {979},
       number = {2},
          eid = {127},
        pages = {127},
          doi = {10.3847/1538-4357/ad9f3a},
archivePrefix = {arXiv},
       eprint = {2402.18773},
 primaryClass = {astro-ph.GA},
       adsurl = {https://ui.adsabs.harvard.edu/abs/2025ApJ...979..127J}
}

@ARTICLE{Zhang_PBH2025,
       author = {{Zhang}, Saiyang and {Liu}, Boyuan and {Bromm}, Volker and {Jeon}, Junehyoung and {Boylan-Kolchin}, Michael and {K{\"u}hnel}, Florian},
        title = "{How do Massive Primordial Black Holes Impact the Formation of the First Stars and Galaxies?}",
      journal = {\apj},
         year = 2025,
        month = jul,
       volume = {987},
       number = {2},
          eid = {185},
        pages = {185},
          doi = {10.3847/1538-4357/adddb4},
archivePrefix = {arXiv},
       eprint = {2503.17585},
 primaryClass = {astro-ph.GA},
       adsurl = {https://ui.adsabs.harvard.edu/abs/2025ApJ...987..185Z}
}

@ARTICLE{Vestergaard2006,
       author = {{Vestergaard}, Marianne and {Peterson}, Bradley M.},
        title = "{Determining Central Black Hole Masses in Distant Active Galaxies and Quasars. II. Improved Optical and UV Scaling Relationships}",
      journal = {\apj},
         year = 2006,
        month = apr,
       volume = {641},
       number = {2},
        pages = {689-709},
          doi = {10.1086/500572},
archivePrefix = {arXiv},
       eprint = {astro-ph/0601303},
 primaryClass = {astro-ph},
       adsurl = {https://ui.adsabs.harvard.edu/abs/2006ApJ...641..689V}
}

@ARTICLE{Inayoshi2025_LRD,
       author = {{Inayoshi}, Kohei},
        title = "{Little Red Dots as the Very First Activity of Black Hole Growth}",
      journal = {\apjl},
         year = 2025,
        month = jul,
       volume = {988},
       number = {1},
          eid = {L22},
        pages = {L22},
          doi = {10.3847/2041-8213/adea66},
archivePrefix = {arXiv},
       eprint = {2503.05537},
 primaryClass = {astro-ph.GA},
       adsurl = {https://ui.adsabs.harvard.edu/abs/2025ApJ...988L..22I}
}

@ARTICLE{Parlanti2025,
       author = {{Parlanti}, Eleonora and {Trefoloni}, Bartolomeo and {Carniani}, Stefano and {D'Eugenio}, Francesco and {Perna}, Michele and {Tozzi}, Giulia and {{\"U}bler}, Hannah and {Venturi}, Giacomo and {Zamora}, Sandra},
        title = "{Doubling NIRSpec/IFS capability to calibrate the single epoch black hole mass relation at high redshift}",
      journal = {arXiv e-prints},
         year = 2025,
        month = dec,
          eid = {arXiv:2512.14844},
        pages = {arXiv:2512.14844},
          doi = {10.48550/arXiv.2512.14844},
archivePrefix = {arXiv},
       eprint = {2512.14844},
 primaryClass = {astro-ph.GA},
       adsurl = {https://ui.adsabs.harvard.edu/abs/2025arXiv251214844P}
}

@ARTICLE{Greene2005,
       author = {{Greene}, Jenny E. and {Ho}, Luis C.},
        title = "{Estimating Black Hole Masses in Active Galaxies Using the H{\ensuremath{\alpha}} Emission Line}",
      journal = {\apj},
         year = 2005,
        month = sep,
       volume = {630},
       number = {1},
        pages = {122-129},
          doi = {10.1086/431897},
archivePrefix = {arXiv},
       eprint = {astro-ph/0508335},
 primaryClass = {astro-ph},
       adsurl = {https://ui.adsabs.harvard.edu/abs/2005ApJ...630..122G}
}

@ARTICLE{deGraaff2025,
       author = {{de Graaff}, Anna and {Rix}, Hans-Walter and {Naidu}, Rohan P. and {Labb{\'e}}, Ivo and {Wang}, Bingjie and {Leja}, Joel and {Matthee}, Jorryt and {Katz}, Harley and {Greene}, Jenny E. and {Hviding}, Raphael E. and {Baggen}, Josephine and {Bezanson}, Rachel and {Boogaard}, Leindert A. and {Brammer}, Gabriel and {Dayal}, Pratika and {van Dokkum}, Pieter and {Goulding}, Andy D. and {Hirschmann}, Michaela and {Maseda}, Michael V. and {McConachie}, Ian and {Miller}, Tim B. and {Nelson}, Erica and {Oesch}, Pascal A. and {Setton}, David J. and {Shivaei}, Irene and {Weibel}, Andrea and {Whitaker}, Katherine E. and {Williams}, Christina C.},
        title = "{A remarkable ruby: Absorption in dense gas, rather than evolved stars, drives the extreme Balmer break of a little red dot at z = 3.5}",
      journal = {\aap},
         year = 2025,
        month = sep,
       volume = {701},
          eid = {A168},
        pages = {A168},
          doi = {10.1051/0004-6361/202554681},
archivePrefix = {arXiv},
       eprint = {2503.16600},
 primaryClass = {astro-ph.GA},
       adsurl = {https://ui.adsabs.harvard.edu/abs/2025A&A...701A.168D}
}

@ARTICLE{InayoshiMaiolino2025,
       author = {{Inayoshi}, Kohei and {Maiolino}, Roberto},
        title = "{Extremely Dense Gas around Little Red Dots and High-redshift Active Galactic Nuclei: A Nonstellar Origin of the Balmer Break and Absorption Features}",
      journal = {\apjl},
         year = 2025,
        month = feb,
       volume = {980},
       number = {2},
          eid = {L27},
        pages = {L27},
          doi = {10.3847/2041-8213/adaebd},
archivePrefix = {arXiv},
       eprint = {2409.07805},
 primaryClass = {astro-ph.GA},
       adsurl = {https://ui.adsabs.harvard.edu/abs/2025ApJ...980L..27I}
}

@ARTICLE{VanDerWel22,
       author = {{van der Wel}, Arjen and {van Houdt}, Josha and {Bezanson}, Rachel and {Franx}, Marijn and {D'Eugenio}, Francesco and {Straatman}, Caroline and {Bell}, Eric F. and {Muzzin}, Adam and {Sobral}, David and {Maseda}, Michael V. and {de Graaff}, Anna and {Holden}, Bradford P.},
        title = "{The Mass Scale of High-redshift Galaxies: Virial Mass Estimates Calibrated with Stellar Dynamical Models from LEGA-C}",
      journal = {\apj},
         year = 2022,
        month = sep,
       volume = {936},
       number = {1},
          eid = {9},
        pages = {9},
          doi = {10.3847/1538-4357/ac83c5},
archivePrefix = {arXiv},
       eprint = {2208.12605},
 primaryClass = {astro-ph.GA},
       adsurl = {https://ui.adsabs.harvard.edu/abs/2022ApJ...936....9V}
}

@ARTICLE{Cappellari2006,
       author = {{Cappellari}, Michele and {Bacon}, R. and {Bureau}, M. and {Damen}, M.~C. and {Davies}, Roger L. and {de Zeeuw}, P.~T. and {Emsellem}, Eric and {Falc{\'o}n-Barroso}, Jes{\'u}s and {Krajnovi{\'c}}, Davor and {Kuntschner}, Harald and {McDermid}, Richard M. and {Peletier}, Reynier F. and {Sarzi}, Marc and {van den Bosch}, Remco C.~E. and {van de Ven}, Glenn},
        title = "{The SAURON project - IV. The mass-to-light ratio, the virial mass estimator and the Fundamental Plane of elliptical and lenticular galaxies}",
      journal = {\mnras},
         year = 2006,
        month = mar,
       volume = {366},
       number = {4},
        pages = {1126-1150},
          doi = {10.1111/j.1365-2966.2005.09981.x},
archivePrefix = {arXiv},
       eprint = {astro-ph/0505042},
 primaryClass = {astro-ph},
       adsurl = {https://ui.adsabs.harvard.edu/abs/2006MNRAS.366.1126C}
}

@ARTICLE{Bezanson18,
       author = {{Bezanson}, Rachel and {van der Wel}, Arjen and {Straatman}, Caroline and {Pacifici}, Camilla and {Wu}, Po-Feng and {Bari{\v{s}}i{\'c}}, Ivana and {Bell}, Eric F. and {Conroy}, Charlie and {D'Eugenio}, Francesco and {Franx}, Marijn and {Gallazzi}, Anna and {van Houdt}, Josha and {Maseda}, Michael V. and {Muzzin}, Adam and {van de Sande}, Jesse and {Sobral}, David and {Spilker}, Justin},
        title = "{1D Kinematics from Stars and Ionized Gas at z {\ensuremath{\sim}} 0.8 from the LEGA-C Spectroscopic Survey of Massive Galaxies}",
      journal = {\apjl},
         year = 2018,
        month = dec,
       volume = {868},
       number = {2},
          eid = {L36},
        pages = {L36},
          doi = {10.3847/2041-8213/aaf16b},
archivePrefix = {arXiv},
       eprint = {1811.07900},
 primaryClass = {astro-ph.GA},
       adsurl = {https://ui.adsabs.harvard.edu/abs/2018ApJ...868L..36B}
}

@ARTICLE{Ji2025,
       author = {{Ji}, Xihan and {Maiolino}, Roberto and {{\"U}bler}, Hannah and {Scholtz}, Jan and {D'Eugenio}, Francesco and {Sun}, Fengwu and {Perna}, Michele and {Turner}, Hannah and {Carniani}, Stefano and {Arribas}, Santiago and {Bennett}, Jake S. and {Bunker}, Andrew and {Charlot}, St{\'e}phane and {Cresci}, Giovanni and {Curti}, Mirko and {Egami}, Eiichi and {Fabian}, Andy and {Inayoshi}, Kohei and {Isobe}, Yuki and {Jones}, Gareth and {Juod{\v{z}}balis}, Ignas and {Kumari}, Nimisha and {Lyu}, Jianwei and {Mazzolari}, Giovanni and {Parlanti}, Eleonora and {Robertson}, Brant and {Rodr{\'\i}guez Del Pino}, Bruno and {Schneider}, Raffaella and {Sijacki}, Debora and {Tacchella}, Sandro and {Trinca}, Alessandro and {Valiante}, Rosa and {Venturi}, Giacomo and {Volonteri}, Marta and {Willott}, Chris and {Witten}, Callum and {Witstok}, Joris},
        title = "{BlackTHUNDER ─ A non-stellar Balmer break in a black hole-dominated little red dot at z = 7.04}",
      journal = {\mnras},
         year = 2025,
        month = dec,
       volume = {544},
       number = {4},
        pages = {3900-3935},
          doi = {10.1093/mnras/staf1867},
archivePrefix = {arXiv},
       eprint = {2501.13082},
 primaryClass = {astro-ph.GA},
       adsurl = {https://ui.adsabs.harvard.edu/abs/2025MNRAS.544.3900J}
}

@ARTICLE{Ziparo2026,
       author = {{Ziparo}, Francesco and {Carniani}, Stefano and {Gallerani}, Simona and {Trefoloni}, Bartolomeo},
        title = "{A Selection Aware View of Black Hole-Galaxy Coevolution at High Redshift}",
      journal = {arXiv e-prints},
         year = 2026,
        month = mar,
          eid = {arXiv:2603.04358},
        pages = {arXiv:2603.04358},
          doi = {10.48550/arXiv.2603.04358},
archivePrefix = {arXiv},
       eprint = {2603.04358},
 primaryClass = {astro-ph.GA},
       adsurl = {https://ui.adsabs.harvard.edu/abs/2026arXiv260304358Z}
}

@ARTICLE{Nicholls2017,
       author = {{Nicholls}, David C. and {Sutherland}, Ralph S. and {Dopita}, Michael A. and {Kewley}, Lisa J. and {Groves}, Brent A.},
        title = "{Abundance scaling in stars, nebulae and galaxies}",
      journal = {\mnras},
         year = 2017,
        month = apr,
       volume = {466},
       number = {4},
        pages = {4403-4422},
          doi = {10.1093/mnras/stw3235},
archivePrefix = {arXiv},
       eprint = {1612.03546},
 primaryClass = {astro-ph.GA},
       adsurl = {https://ui.adsabs.harvard.edu/abs/2017MNRAS.466.4403N}
}

@ARTICLE{Perna2023,
       author = {{Perna}, M. and {Arribas}, S. and {Marshall}, M. and {D'Eugenio}, F. and {{\"U}bler}, H. and {Bunker}, A. and {Charlot}, S. and {Carniani}, S. and {Jakobsen}, P. and {Maiolino}, R. and {Rodr{\'\i}guez Del Pino}, B. and {Willott}, C.~J. and {B{\"o}ker}, T. and {Circosta}, C. and {Cresci}, G. and {Curti}, M. and {Husemann}, B. and {Kumari}, N. and {Lamperti}, I. and {P{\'e}rez-Gonz{\'a}lez}, P.~G. and {Scholtz}, J.},
        title = "{GA-NIFS: The ultra-dense, interacting environment of a dual AGN at z {\ensuremath{\sim}} 3.3 revealed by JWST/NIRSpec IFS}",
      journal = {\aap},
         year = 2023,
        month = nov,
       volume = {679},
          eid = {A89},
        pages = {A89},
          doi = {10.1051/0004-6361/202346649},
archivePrefix = {arXiv},
       eprint = {2304.06756},
 primaryClass = {astro-ph.GA},
       adsurl = {https://ui.adsabs.harvard.edu/abs/2023A&A...679A..89P}
}

@ARTICLE{Reddy2023,
       author = {{Reddy}, Naveen A. and {Topping}, Michael W. and {Sanders}, Ryan L. and {Shapley}, Alice E. and {Brammer}, Gabriel},
        title = "{A JWST/NIRSpec Exploration of the Connection between Ionization Parameter, Electron Density, and Star-formation-rate Surface Density in z = 2.7-6.3 Galaxies}",
      journal = {\apj},
         year = 2023,
        month = aug,
       volume = {952},
       number = {2},
          eid = {167},
        pages = {167},
          doi = {10.3847/1538-4357/acd754},
archivePrefix = {arXiv},
       eprint = {2303.11397},
 primaryClass = {astro-ph.GA},
       adsurl = {https://ui.adsabs.harvard.edu/abs/2023ApJ...952..167R}
}

@ARTICLE{Strom2017,
       author = {{Strom}, Allison L. and {Steidel}, Charles C. and {Rudie}, Gwen C. and {Trainor}, Ryan F. and {Pettini}, Max and {Reddy}, Naveen A.},
        title = "{Nebular Emission Line Ratios in z ≃ 2-3 Star-forming Galaxies with KBSS-MOSFIRE: Exploring the Impact of Ionization, Excitation, and Nitrogen-to-Oxygen Ratio}",
      journal = {\apj},
         year = 2017,
        month = feb,
       volume = {836},
       number = {2},
          eid = {164},
        pages = {164},
          doi = {10.3847/1538-4357/836/2/164},
archivePrefix = {arXiv},
       eprint = {1608.02587},
 primaryClass = {astro-ph.GA},
       adsurl = {https://ui.adsabs.harvard.edu/abs/2017ApJ...836..164S}
}

@ARTICLE{StoreyHummer1995,
       author = {{Storey}, P.~J. and {Hummer}, D.~G.},
        title = "{Recombination line intensities for hydrogenic ions-IV. Total recombination coefficients and machine-readable tables for Z=1 to 8}",
      journal = {\mnras},
         year = 1995,
        month = jan,
       volume = {272},
       number = {1},
        pages = {41-48},
          doi = {10.1093/mnras/272.1.41},
       adsurl = {https://ui.adsabs.harvard.edu/abs/1995MNRAS.272...41S}
}

@ARTICLE{Perna2025,
       author = {{Perna}, Michele and {Arribas}, Santiago and {Lamperti}, Isabella and {Circosta}, Chiara and {Bertola}, Elena and {P{\'e}rez-Gonz{\'a}lez}, Pablo G. and {D'Eugenio}, Francesco and {{\"U}bler}, Hannah and {Cresci}, Giovanni and {Volonteri}, Marta and {Mannucci}, Filippo and {Maiolino}, Roberto and {Rodr{\'\i}guez Del Pino}, Bruno and {B{\"o}ker}, Torsten and {Bunker}, Andrew J. and {Charlot}, St{\'e}phane and {Willott}, Chris J. and {Carniani}, Stefano and {Curti}, Mirko and {Jones}, Gareth C. and {Kumari}, Nimisha and {Marshall}, Madeline A. and {Venturi}, Giacomo and {Saxena}, Aayush and {Scholtz}, Jan and {Witstok}, Joris},
        title = "{GA-NIFS: High number of dual active galactic nuclei at z {\ensuremath{\sim}} 3}",
      journal = {\aap},
         year = 2025,
        month = apr,
       volume = {696},
          eid = {A59},
        pages = {A59},
          doi = {10.1051/0004-6361/202453430},
archivePrefix = {arXiv},
       eprint = {2310.03067},
 primaryClass = {astro-ph.GA},
       adsurl = {https://ui.adsabs.harvard.edu/abs/2025A&A...696A..59P}
}

@ARTICLE{Rauscher2017,
       author = {{Rauscher}, Bernard J. and {Arendt}, Richard G. and {Fixsen}, D.~J. and {Greenhouse}, Matthew A. and {Lander}, Matthew and {Lindler}, Don and {Loose}, Markus and {Moseley}, S.~H. and {Mott}, D. Brent and {Wen}, Yiting and {Wilson}, Donna V. and {Xenophontos}, Christos},
        title = "{Improved Reference Sampling and Subtraction: A Technique for Reducing the Read Noise of Near-infrared Detector Systems}",
      journal = {\pasp},
         year = 2017,
        month = oct,
       volume = {129},
       number = {980},
        pages = {105003},
          doi = {10.1088/1538-3873/aa83fd},
archivePrefix = {arXiv},
       eprint = {1707.09387},
 primaryClass = {astro-ph.IM},
       adsurl = {https://ui.adsabs.harvard.edu/abs/2017PASP..129j5003R}
}

@ARTICLE{Topping2025,
       author = {{Topping}, Michael W. and {Sanders}, Ryan L. and {Shapley}, Alice E. and {Pahl}, Anthony J. and {Reddy}, Naveen A. and {Stark}, Daniel P. and {Berg}, Danielle A. and {Clarke}, Leonardo and {Cullen}, Fergus and {Dunlop}, James S. and {Ellis}, Richard S. and {Schreiber}, N.~M. F{\"o}rster and {Illingworth}, Garth D. and {Jones}, Tucker and {Narayanan}, Desika and {Pettini}, Max and {Schaerer}, Daniel},
        title = "{The AURORA survey: the evolution of multiphase electron densities at high redshift}",
      journal = {\mnras},
         year = 2025,
        month = aug,
       volume = {541},
       number = {2},
        pages = {1707-1721},
          doi = {10.1093/mnras/staf903},
archivePrefix = {arXiv},
       eprint = {2502.08712},
 primaryClass = {astro-ph.GA},
       adsurl = {https://ui.adsabs.harvard.edu/abs/2025MNRAS.541.1707T}
}

@ARTICLE{Isobe2023,
       author = {{Isobe}, Yuki and {Ouchi}, Masami and {Nakajima}, Kimihiko and {Harikane}, Yuichi and {Ono}, Yoshiaki and {Xu}, Yi and {Zhang}, Yechi and {Umeda}, Hiroya},
        title = "{Redshift Evolution of Electron Density in the Interstellar Medium at z   0-9 Uncovered with JWST/NIRSpec Spectra and Line-spread Function Determinations}",
      journal = {\apj},
         year = 2023,
        month = oct,
       volume = {956},
       number = {2},
          eid = {139},
        pages = {139},
          doi = {10.3847/1538-4357/acf376},
archivePrefix = {arXiv},
       eprint = {2301.06811},
 primaryClass = {astro-ph.GA},
       adsurl = {https://ui.adsabs.harvard.edu/abs/2023ApJ...956..139I}
}

@ARTICLE{MadauMaiolino2026,
       author = {{Madau}, Piero and {Maiolino}, Roberto},
        title = "{Little Red Dots as Obscured Little Blue Dots: A Super-Eddington Unification Model}",
      journal = {arXiv e-prints},
         year = 2026,
        month = feb,
          eid = {arXiv:2602.22386},
        pages = {arXiv:2602.22386},
          doi = {10.48550/arXiv.2602.22386},
archivePrefix = {arXiv},
       eprint = {2602.22386},
 primaryClass = {astro-ph.GA},
       adsurl = {https://ui.adsabs.harvard.edu/abs/2026arXiv260222386M}
}

@ARTICLE{Laor2006,
       author = {{Laor}, Ari},
        title = "{Evidence for Line Broadening by Electron Scattering in the Broad-Line Region of NGC 4395}",
      journal = {\apj},
         year = 2006,
        month = may,
       volume = {643},
       number = {1},
        pages = {112-119},
          doi = {10.1086/502798},
archivePrefix = {arXiv},
       eprint = {astro-ph/0601688},
 primaryClass = {astro-ph},
       adsurl = {https://ui.adsabs.harvard.edu/abs/2006ApJ...643..112L}
}

@ARTICLE{Shajib2025,
       author = {{Shajib}, Anowar J. and {Treu}, Tommaso and {Melo}, Alejandra and {Roberts-Borsani}, Guido and {Knabel}, Shawn and {Cappellari}, Michele and {Frieman}, Joshua A.},
        title = "{An accurate measurement of the spectral resolution of the JWST Near Infrared Spectrograph}",
      journal = {\aap},
         year = 2025,
        month = oct,
       volume = {702},
          eid = {L12},
        pages = {L12},
          doi = {10.1051/0004-6361/202556281},
archivePrefix = {arXiv},
       eprint = {2507.03746},
 primaryClass = {astro-ph.IM},
       adsurl = {https://ui.adsabs.harvard.edu/abs/2025A&A...702L..12S}
}

@ARTICLE{RUBIES2025,
       author = {{de Graaff}, Anna and {Brammer}, Gabriel and {Weibel}, Andrea and {Lewis}, Zach and {Maseda}, Michael V. and {Oesch}, Pascal A. and {Bezanson}, Rachel and {Boogaard}, Leindert A. and {Cleri}, Nikko J. and {Cooper}, Olivia R. and {Gottumukkala}, Rashmi and {Greene}, Jenny E. and {Hirschmann}, Michaela and {Hviding}, Raphael E. and {Katz}, Harley and {Labb{\'e}}, Ivo and {Leja}, Joel and {Matthee}, Jorryt and {McConachie}, Ian and {Miller}, Tim B. and {Naidu}, Rohan P. and {Price}, Sedona H. and {Rix}, Hans-Walter and {Setton}, David J. and {Suess}, Katherine A. and {Wang}, Bingjie and {Whitaker}, Katherine E. and {Williams}, Christina C.},
        title = "{RUBIES: A complete census of the bright and red distant Universe with JWST/NIRSpec}",
      journal = {\aap},
         year = 2025,
        month = may,
       volume = {697},
          eid = {A189},
        pages = {A189},
          doi = {10.1051/0004-6361/202452186},
archivePrefix = {arXiv},
       eprint = {2409.05948},
 primaryClass = {astro-ph.GA},
       adsurl = {https://ui.adsabs.harvard.edu/abs/2025A&A...697A.189D}
}

@ARTICLE{DEugenio2025,
       author = {{D'Eugenio}, Francesco and {Maiolino}, Roberto and {Perna}, Michele and {Uebler}, Hannah and {Ji}, Xihan and {McClymont}, William and {Koudmani}, Sophie and {Sijacki}, Debora and {Juod{\v{z}}balis}, Ignas and {Scholtz}, Jan and {Bennett}, Jake and {Bunker}, Andrew J. and {Carniani}, Stefano and {Charlot}, St{\'e}phane and {Cresci}, Giovanni and {Curtis-Lake}, Emma and {Dalla Bont{\`a}}, Elena and {Jones}, Gareth C. and {Lyu}, Jianwei and {Marconi}, Alessandro and {Mazzolari}, Giovanni and {Nelson}, Erica J. and {Parlanti}, Eleonora and {Robertson}, Brant E. and {Schneider}, Raffaella and {Simmonds}, Charlotte and {Tacchella}, Sandro and {Venturi}, Giacomo and {Willott}, Chris and {Witstok}, Joris and {Witten}, Callum},
        title = "{BlackTHUNDER strikes twice: rest-frame Balmer-line absorption and high Eddington accretion rate in a Little Red Dot at $z=7.04$}",
      journal = {arXiv e-prints},
         year = 2025,
        month = mar,
          eid = {arXiv:2503.11752},
        pages = {arXiv:2503.11752},
          doi = {10.48550/arXiv.2503.11752},
archivePrefix = {arXiv},
       eprint = {2503.11752},
 primaryClass = {astro-ph.GA},
       adsurl = {https://ui.adsabs.harvard.edu/abs/2025arXiv250311752D}
}

@ARTICLE{DEugenio2026,
       author = {{D'Eugenio}, Francesco and {Juod{\v{z}}balis}, Ignas and {Ji}, Xihan and {Scholtz}, Jan and {Maiolino}, Roberto and {Carniani}, Stefano and {Perna}, Michele and {Mazzolari}, Giovanni and {{\"U}bler}, Hannah and {Arribas}, Santiago and {Bhatawdekar}, Rachana and {Bunker}, Andrew J. and {Cresci}, Giovanni and {Curtis-Lake}, Emma and {Hainline}, Kevin and {Inayoshi}, Kohei and {Isobe}, Yuki and {Ji}, Zhiyuan and {Johnson}, Benjamin D. and {Jones}, Gareth C. and {Looser}, Tobias J. and {Nelson}, Erica J. and {Parlanti}, Eleonora and {Pusk{\'a}s}, D{\'a}vid and {Rinaldi}, Pierluigi and {Robertson}, Brant and {Rodr{\'\i}guez Del Pino}, Bruno and {Shivaei}, Irene and {Sun}, Fengwu and {Tacchella}, Sandro and {Venturi}, Giacomo and {Volonteri}, Marta and {Williams}, Christina C. and {Willmer}, Christopher N.~A. and {Willott}, Chris and {Witstok}, Joris},
        title = "{JADES and BlackTHUNDER: rest-frame Balmer-line absorption and the local environment in a Little Red Dot at z = 5}",
      journal = {\mnras},
         year = 2026,
        month = jan,
       volume = {545},
       number = {3},
          eid = {staf2117},
        pages = {staf2117},
          doi = {10.1093/mnras/staf2117},
archivePrefix = {arXiv},
       eprint = {2506.14870},
 primaryClass = {astro-ph.GA},
       adsurl = {https://ui.adsabs.harvard.edu/abs/2026MNRAS.545f2117D}
}

@ARTICLE{Nikopoulos2025,
       author = {{Nikopoulos}, G.~P. and {Watson}, D. and {Sneppen}, A. and {Rusakov}, V. and {Heintz}, K.~E. and {Witstok}, J. and {Brammer}, G.},
        title = "{Evidence of violation of Case B recombination in Little Red Dots}",
      journal = {arXiv e-prints},
         year = 2025,
        month = oct,
          eid = {arXiv:2510.06362},
        pages = {arXiv:2510.06362},
          doi = {10.48550/arXiv.2510.06362},
archivePrefix = {arXiv},
       eprint = {2510.06362},
 primaryClass = {astro-ph.GA},
       adsurl = {https://ui.adsabs.harvard.edu/abs/2025arXiv251006362N}
}

@ARTICLE{DEugenio2025_Irony,
       author = {{D'Eugenio}, Francesco and {Nelson}, Erica and {Ji}, Xihan and {Baggen}, Josephine and {Greene}, Jenny and {Labb{\'e}}, Ivo and {Pezzulli}, Gabriele and {Brown}, Vanessa and {Maiolino}, Roberto and {Matthee}, Jorryt and {Terlevich}, Elena and {Terlevich}, Roberto and {Torralba}, Alberto and {Carniani}, Stefano},
        title = "{Irony at z=6.68: a bright AGN with forbidden Fe emission and multi-component Balmer absorption}",
      journal = {arXiv e-prints},
         year = 2025,
        month = sep,
          eid = {arXiv:2510.00101},
        pages = {arXiv:2510.00101},
          doi = {10.48550/arXiv.2510.00101},
archivePrefix = {arXiv},
       eprint = {2510.00101},
 primaryClass = {astro-ph.GA},
       adsurl = {https://ui.adsabs.harvard.edu/abs/2025arXiv251000101D}
}

@ARTICLE{Nagao2006,
       author = {{Nagao}, T. and {Maiolino}, R. and {Marconi}, A.},
        title = "{Gas metallicity diagnostics in star-forming galaxies}",
      journal = {\aap},
         year = 2006,
        month = nov,
       volume = {459},
       number = {1},
        pages = {85-101},
          doi = {10.1051/0004-6361:20065216},
archivePrefix = {arXiv},
       eprint = {astro-ph/0603580},
 primaryClass = {astro-ph},
       adsurl = {https://ui.adsabs.harvard.edu/abs/2006A&A...459...85N}
}

@ARTICLE{Lin2025,
       author = {{Lin}, Ruqiu and {Zheng}, Zhen-Ya and {Jiang}, Chunyan and {Yuan}, Fang-Ting and {Ho}, Luis C. and {Wang}, Junxian and {Jiang}, Linhua and {Rhoads}, James E. and {Malhotra}, Sangeeta and {Barrientos}, L. Felipe and {Wold}, Isak and {Infante}, Leopoldo and {Zhu}, Shuairu and {Ji}, Xiang and {Fu}, Xiaodan},
        title = "{Discovery of Local Analogs to JWST's Little Red Dots}",
      journal = {\apjl},
         year = 2025,
        month = feb,
       volume = {980},
       number = {2},
          eid = {L34},
        pages = {L34},
          doi = {10.3847/2041-8213/adaaf1},
archivePrefix = {arXiv},
       eprint = {2412.08396},
 primaryClass = {astro-ph.GA},
       adsurl = {https://ui.adsabs.harvard.edu/abs/2025ApJ...980L..34L}
}

@ARTICLE{Schwarz1978,
       author = {{Schwarz}, Gideon},
        title = "{Estimating the Dimension of a Model}",
      journal = {Annals of Statistics},
         year = 1978,
        month = jul,
       volume = {6},
       number = {2},
        pages = {461-464},
       adsurl = {https://ui.adsabs.harvard.edu/abs/1978AnSta...6..461S}
}

@ARTICLE{Ferruit2022,
       author = {{Ferruit}, P. and {Jakobsen}, P. and {Giardino}, G. and {Rawle}, T. and {Alves de Oliveira}, C. and {Arribas}, S. and {Beck}, T.~L. and {Birkmann}, S. and {B{\"o}ker}, T. and {Bunker}, A.~J. and {Charlot}, S. and {de Marchi}, G. and {Franx}, M. and {Henry}, A. and {Karakla}, D. and {Kassin}, S.~A. and {Kumari}, N. and {L{\'o}pez-Caniego}, M. and {L{\"u}tzgendorf}, N. and {Maiolino}, R. and {Manjavacas}, E. and {Marston}, A. and {Moseley}, S.~H. and {Muzerolle}, J. and {Pirzkal}, N. and {Rauscher}, B. and {Rix}, H.-W. and {Sabbi}, E. and {Sirianni}, M. and {te Plate}, M. and {Valenti}, J. and {Willott}, C.~J. and {Zeidler}, P.},
        title = "{The Near-Infrared Spectrograph (NIRSpec) on the James Webb Space Telescope. II. Multi-object spectroscopy (MOS)}",
      journal = {\aap},
         year = 2022,
        month = may,
       volume = {661},
          eid = {A81},
        pages = {A81},
          doi = {10.1051/0004-6361/202142673},
archivePrefix = {arXiv},
       eprint = {2202.03306},
 primaryClass = {astro-ph.IM},
       adsurl = {https://ui.adsabs.harvard.edu/abs/2022A&A...661A..81F}
}

@ARTICLE{Carr2020,
       author = {{Carr}, Bernard and {K{\"u}hnel}, Florian},
        title = "{Primordial Black Holes as Dark Matter: Recent Developments}",
      journal = {Annual Review of Nuclear and Particle Science},
         year = 2020,
        month = oct,
       volume = {70},
        pages = {355-394},
          doi = {10.1146/annurev-nucl-050520-125911},
archivePrefix = {arXiv},
       eprint = {2006.02838},
 primaryClass = {astro-ph.CO},
       adsurl = {https://ui.adsabs.harvard.edu/abs/2020ARNPS..70..355C}
}

@ARTICLE{Carr2021,
       author = {{Carr}, Bernard and {Kohri}, Kazunori and {Sendouda}, Yuuiti and {Yokoyama}, Jun'ichi},
        title = "{Constraints on primordial black holes}",
      journal = {Reports on Progress in Physics},
         year = 2021,
        month = nov,
       volume = {84},
       number = {11},
          eid = {116902},
        pages = {116902},
          doi = {10.1088/1361-6633/ac1e31},
archivePrefix = {arXiv},
       eprint = {2002.12778},
 primaryClass = {astro-ph.CO},
       adsurl = {https://ui.adsabs.harvard.edu/abs/2021RPPh...84k6902C}
}

@ARTICLE{Carr2021PDU....3100755C,
       author = {{Carr}, Bernard and {Clesse}, S{\'e}bastien and {Garc{\'\i}a-Bellido}, Juan and {K{\"u}hnel}, Florian},
        title = "{Cosmic conundra explained by thermal history and primordial black holes}",
      journal = {Physics of the Dark Universe},
         year = 2021,
        month = jan,
       volume = {31},
          eid = {100755},
        pages = {100755},
          doi = {10.1016/j.dark.2020.100755},
       adsurl = {https://ui.adsabs.harvard.edu/abs/2021PDU....3100755C}
}

@ARTICLE{Tang2023,
       author = {{Tang}, Mengtao and {Stark}, Daniel P. and {Chen}, Zuyi and {Mason}, Charlotte and {Topping}, Michael and {Endsley}, Ryan and {Senchyna}, Peter and {Plat}, Ad{\`e}le and {Lu}, Ting-Yi and {Whitler}, Lily and {Robertson}, Brant and {Charlot}, St{\'e}phane},
        title = "{JWST/NIRSpec spectroscopy of z = 7-9 star-forming galaxies with CEERS: new insight into bright Ly{\ensuremath{\alpha}} emitters in ionized bubbles}",
      journal = {\mnras},
         year = 2023,
        month = dec,
       volume = {526},
       number = {2},
        pages = {1657-1686},
          doi = {10.1093/mnras/stad2763},
archivePrefix = {arXiv},
       eprint = {2301.07072},
 primaryClass = {astro-ph.GA},
       adsurl = {https://ui.adsabs.harvard.edu/abs/2023MNRAS.526.1657T}
}

@ARTICLE{Sanders2025,
       author = {{Sanders}, Ryan L. and {Shapley}, Alice E. and {Topping}, Michael W. and {Reddy}, Naveen A. and {Berg}, Danielle A. and {Khostovan}, Ali Ahmad and {Bouwens}, Rychard J. and {Brammer}, Gabriel and {Carnall}, Adam C. and {Cullen}, Fergus and {Dav{\'e}}, Romeel and {Dunlop}, James S. and {Ellis}, Richard S. and {F{\"o}rster Schreiber}, N.~M. and {Furlanetto}, Steven R. and {Glazebrook}, Karl and {Illingworth}, Garth D. and {Jones}, Tucker and {Kriek}, Mariska and {McLeod}, Derek J. and {McLure}, Ross J. and {Narayanan}, Desika and {Oesch}, Pascal A. and {Pahl}, Anthony J. and {Pettini}, Max and {Schaerer}, Daniel and {Stark}, Daniel P. and {Steidel}, Charles C. and {Tang}, Mengtao and {Clarke}, Leonardo and {Donnan}, Callum T. and {Kehoe}, Emily},
        title = "{The AURORA Survey: High-Redshift Empirical Metallicity Calibrations from Electron Temperature Measurements at z=2-10}",
      journal = {arXiv e-prints},
         year = 2025,
        month = aug,
          eid = {arXiv:2508.10099},
        pages = {arXiv:2508.10099},
          doi = {10.48550/arXiv.2508.10099},
archivePrefix = {arXiv},
       eprint = {2508.10099},
 primaryClass = {astro-ph.GA},
       adsurl = {https://ui.adsabs.harvard.edu/abs/2025arXiv250810099S}
}

@ARTICLE{Juodzbalis2026_JADEScensus,
       author = {{Juod{\v{z}}balis}, Ignas and {Maiolino}, Roberto and {Baker}, William M. and {Lake}, Emma Curtis and {Scholtz}, Jan and {D'Eugenio}, Francesco and {Trefoloni}, Bartolomeo and {Isobe}, Yuki and {Tacchella}, Sandro and {Bunker}, Andrew J. and {Carniani}, Stefano and {Charlot}, St{\'e}phane and {Jones}, Gareth C. and {Parlanti}, Eleonora and {Perna}, Michele and {Rinaldi}, Pierluigi and {Robertson}, Brant and {{\"U}bler}, Hannah and {Venturi}, Giacomo and {Willott}, Chris},
        title = "{JADES: comprehensive census of broad-line AGN from Reionization to Cosmic Noon revealed by JWST}",
      journal = {\mnras},
         year = 2026,
        month = jan,
          doi = {10.1093/mnras/stag086},
archivePrefix = {arXiv},
       eprint = {2504.03551},
 primaryClass = {astro-ph.GA},
       adsurl = {https://ui.adsabs.harvard.edu/abs/2026MNRAS.tmp...94J}
}

@ARTICLE{Li2025,
       author = {{Li}, Junyao and {Silverman}, John D. and {Shen}, Yue and {Volonteri}, Marta and {Jahnke}, Knud and {Zhuang}, Ming-Yang and {Scoggins}, Matthew T. and {Ding}, Xuheng and {Harikane}, Yuichi and {Onoue}, Masafusa and {Tanaka}, Takumi S.},
        title = "{Tip of the Iceberg: Overmassive Black Holes at 4 < z < 7 Found by JWST Are Not Inconsistent with the Local  Relation}",
      journal = {\apj},
         year = 2025,
        month = mar,
       volume = {981},
       number = {1},
          eid = {19},
        pages = {19},
          doi = {10.3847/1538-4357/ada603},
archivePrefix = {arXiv},
       eprint = {2403.00074},
 primaryClass = {astro-ph.GA},
       adsurl = {https://ui.adsabs.harvard.edu/abs/2025ApJ...981...19L}
}

@ARTICLE{Geris2026,
       author = {{Geris}, Sophia and {Maiolino}, Roberto and {Isobe}, Yuki and {Scholtz}, Jan and {D'Eugenio}, Francesco and {Ji}, Xihan and {Juod{\v{z}}balis}, Ignas and {Simmonds}, Charlotte and {Dayal}, Pratika and {Trinca}, Alessandro and {Schneider}, Raffaella and {Arribas}, Santiago and {Bhatawdekar}, Rachana and {Bunker}, Andrew J. and {Carniani}, Stefano and {Charlot}, St{\'e}phane and {Chevallard}, Jacopo and {Curtis-Lake}, Emma and {Johnson}, Benjamin D. and {Parlanti}, Eleonora and {Rinaldi}, Pierluigi and {Robertson}, Brant and {Tacchella}, Sandro and {{\"U}bler}, Hannah and {Venturi}, Giacomo and {Williams}, Christina C. and {Witstok}, Joris},
        title = "{JADES reveals a large population of low-mass black holes at high redshift}",
      journal = {\mnras},
         year = 2026,
        month = jan,
       volume = {545},
       number = {1},
          eid = {staf1979},
        pages = {staf1979},
          doi = {10.1093/mnras/staf1979},
archivePrefix = {arXiv},
       eprint = {2506.22147},
 primaryClass = {astro-ph.GA},
       adsurl = {https://ui.adsabs.harvard.edu/abs/2026MNRAS.545f1979G}
}

@ARTICLE{Hainline2025,
       author = {{Hainline}, Kevin N. and {Maiolino}, Roberto and {Juod{\v{z}}balis}, Ignas and {Scholtz}, Jan and {{\"U}bler}, Hannah and {D'Eugenio}, Francesco and {Helton}, Jakob M. and {Sun}, Yang and {Sun}, Fengwu and {Robertson}, Brant and {Tacchella}, Sandro and {Bunker}, Andrew J. and {Carniani}, Stefano and {Charlot}, Stephane and {Curtis-Lake}, Emma and {Egami}, Eiichi and {Johnson}, Benjamin D. and {Lin}, Xiaojing and {Lyu}, Jianwei and {P{\'e}rez-Gonz{\'a}lez}, Pablo G. and {Rinaldi}, Pierluigi and {Silcock}, Maddie S. and {Venturi}, Giacomo and {Williams}, Christina C. and {Willmer}, Christopher N.~A. and {Willott}, Chris and {Zhang}, Junyu and {Zhu}, Yongda},
        title = "{An Investigation into the Selection and Colors of Little Red Dots and Active Galactic Nuclei}",
      journal = {\apj},
         year = 2025,
        month = feb,
       volume = {979},
       number = {2},
          eid = {138},
        pages = {138},
          doi = {10.3847/1538-4357/ad9920},
archivePrefix = {arXiv},
       eprint = {2410.00100},
 primaryClass = {astro-ph.GA},
       adsurl = {https://ui.adsabs.harvard.edu/abs/2025ApJ...979..138H}
}

@ARTICLE{Lin2026,
       author = {{Lin}, Xiaojing and {Fan}, Xiaohui and {Cai}, Zheng and {Bian}, Fuyan and {Liu}, Hanpu and {Sun}, Fengwu and {Ma}, Yilun and {Greene}, Jenny E. and {Strauss}, Michael A. and {Green}, Richard and {Lyu}, Jianwei and {Champagne}, Jaclyn B. and {Goulding}, Andy D. and {Inayoshi}, Kohei and {Jin}, Xiangyu and {Leung}, Gene C.~K. and {Li}, Mingyu and {Liu}, Weizhe and {Liu}, Yichen and {Mao}, Junjie and {Pudoka}, Maria Anne and {Tee}, Wei Leong and {Wang}, Ben and {Wang}, Feige and {Wu}, Yunjing and {Yang}, Jinyi and {Zhang}, Haowen and {Zhu}, Yongda},
        title = "{The Discovery of Little Red Dots in the Local Universe: Signatures of Cool Gas Envelopes}",
      journal = {\apj},
         year = 2026,
        month = feb,
       volume = {997},
       number = {2},
          eid = {364},
        pages = {364},
          doi = {10.3847/1538-4357/ae2bdf},
archivePrefix = {arXiv},
       eprint = {2507.10659},
 primaryClass = {astro-ph.GA},
       adsurl = {https://ui.adsabs.harvard.edu/abs/2026ApJ...997..364L}
}

@ARTICLE{Ji2026,
       author = {{Ji}, Xihan and {D'Eugenio}, Francesco and {Juod{\v{z}}balis}, Ignas and {Walton}, Dominic J. and {Fabian}, Andrew C. and {Maiolino}, Roberto and {Ramos Almeida}, Cristina and {Acosta Pulido}, Jose A. and {Belokurov}, Vasily A. and {Isobe}, Yuki and {Jones}, Gareth and {Maraston}, Claudia and {Scholtz}, Jan and {Simmonds}, Charlotte and {Tacchella}, Sandro and {Terlevich}, Elena and {Terlevich}, Roberto},
        title = "{Lord of LRDs: insights into a 'Little Red Dot' with a low-ionization spectrum at z = 0.1}",
      journal = {\mnras},
         year = 2026,
        month = jan,
       volume = {545},
       number = {3},
          eid = {staf2235},
        pages = {staf2235},
          doi = {10.1093/mnras/staf2235},
archivePrefix = {arXiv},
       eprint = {2507.23774},
 primaryClass = {astro-ph.GA},
       adsurl = {https://ui.adsabs.harvard.edu/abs/2026MNRAS.545f2235J}
}

@ARTICLE{Kocevski2023,
       author = {{Kocevski}, Dale D. and {Onoue}, Masafusa and {Inayoshi}, Kohei and {Trump}, Jonathan R. and {Arrabal Haro}, Pablo and {Grazian}, Andrea and {Dickinson}, Mark and {Finkelstein}, Steven L. and {Kartaltepe}, Jeyhan S. and {Hirschmann}, Michaela and {Aird}, James and {Holwerda}, Benne W. and {Fujimoto}, Seiji and {Juneau}, St{\'e}phanie and {Amor{\'\i}n}, Ricardo O. and {Backhaus}, Bren E. and {Bagley}, Micaela B. and {Barro}, Guillermo and {Bell}, Eric F. and {Bisigello}, Laura and {Calabr{\`o}}, Antonello and {Cleri}, Nikko J. and {Cooper}, M.~C. and {Ding}, Xuheng and {Grogin}, Norman A. and {Ho}, Luis C. and {Hutchison}, Taylor A. and {Inoue}, Akio K. and {Jiang}, Linhua and {Jones}, Brenda and {Koekemoer}, Anton M. and {Li}, Wenxiu and {Li}, Zhengrong and {McGrath}, Elizabeth J. and {Molina}, Juan and {Papovich}, Casey and {P{\'e}rez-Gonz{\'a}lez}, Pablo G. and {Pirzkal}, Nor and {Wilkins}, Stephen M. and {Yang}, Guang and {Yung}, L.~Y. Aaron},
        title = "{Hidden Little Monsters: Spectroscopic Identification of Low-mass, Broad-line AGNs at z > 5 with CEERS}",
      journal = {\apjl},
         year = 2023,
        month = sep,
       volume = {954},
       number = {1},
          eid = {L4},
        pages = {L4},
          doi = {10.3847/2041-8213/ace5a0},
archivePrefix = {arXiv},
       eprint = {2302.00012},
 primaryClass = {astro-ph.GA},
       adsurl = {https://ui.adsabs.harvard.edu/abs/2023ApJ...954L...4K}
}

@ARTICLE{Ubler2023,
       author = {{{\"U}bler}, Hannah and {Maiolino}, Roberto and {Curtis-Lake}, Emma and {P{\'e}rez-Gonz{\'a}lez}, Pablo G. and {Curti}, Mirko and {Perna}, Michele and {Arribas}, Santiago and {Charlot}, St{\'e}phane and {Marshall}, Madeline A. and {D'Eugenio}, Francesco and {Scholtz}, Jan and {Bunker}, Andrew and {Carniani}, Stefano and {Ferruit}, Pierre and {Jakobsen}, Peter and {Rix}, Hans-Walter and {Rodr{\'\i}guez Del Pino}, Bruno and {Willott}, Chris J. and {Boeker}, Torsten and {Cresci}, Giovanni and {Jones}, Gareth C. and {Kumari}, Nimisha and {Rawle}, Tim},
        title = "{GA-NIFS: A massive black hole in a low-metallicity AGN at z {\ensuremath{\sim}} 5.55 revealed by JWST/NIRSpec IFS}",
      journal = {\aap},
         year = 2023,
        month = sep,
       volume = {677},
          eid = {A145},
        pages = {A145},
          doi = {10.1051/0004-6361/202346137},
archivePrefix = {arXiv},
       eprint = {2302.06647},
 primaryClass = {astro-ph.GA},
       adsurl = {https://ui.adsabs.harvard.edu/abs/2023A&A...677A.145U}
}

@ARTICLE{Harikane2023,
       author = {{Harikane}, Yuichi and {Zhang}, Yechi and {Nakajima}, Kimihiko and {Ouchi}, Masami and {Isobe}, Yuki and {Ono}, Yoshiaki and {Hatano}, Shun and {Xu}, Yi and {Umeda}, Hiroya},
        title = "{A JWST/NIRSpec First Census of Broad-line AGNs at z = 4-7: Detection of 10 Faint AGNs with M $_{BH}$ {}10$^{6}$-{}10$^{8}$ M $_{{\ensuremath{\odot}}}$ and Their Host Galaxy Properties}",
      journal = {\apj},
         year = 2023,
        month = dec,
       volume = {959},
       number = {1},
          eid = {39},
        pages = {39},
          doi = {10.3847/1538-4357/ad029e},
archivePrefix = {arXiv},
       eprint = {2303.11946},
 primaryClass = {astro-ph.GA},
       adsurl = {https://ui.adsabs.harvard.edu/abs/2023ApJ...959...39H}
}

@ARTICLE{Matthee2024,
       author = {{Matthee}, Jorryt and {Naidu}, Rohan P. and {Brammer}, Gabriel and {Chisholm}, John and {Eilers}, Anna-Christina and {Goulding}, Andy and {Greene}, Jenny and {Kashino}, Daichi and {Labbe}, Ivo and {Lilly}, Simon J. and {Mackenzie}, Ruari and {Oesch}, Pascal A. and {Weibel}, Andrea and {Wuyts}, Stijn and {Xiao}, Mengyuan and {Bordoloi}, Rongmon and {Bouwens}, Rychard and {van Dokkum}, Pieter and {Illingworth}, Garth and {Kramarenko}, Ivan and {Maseda}, Michael V. and {Mason}, Charlotte and {Meyer}, Romain A. and {Nelson}, Erica J. and {Reddy}, Naveen A. and {Shivaei}, Irene and {Simcoe}, Robert A. and {Yue}, Minghao},
        title = "{Little Red Dots: An Abundant Population of Faint Active Galactic Nuclei at z {\ensuremath{\sim}} 5 Revealed by the EIGER and FRESCO JWST Surveys}",
      journal = {\apj},
         year = 2024,
        month = mar,
       volume = {963},
       number = {2},
          eid = {129},
        pages = {129},
          doi = {10.3847/1538-4357/ad2345},
archivePrefix = {arXiv},
       eprint = {2306.05448},
 primaryClass = {astro-ph.GA},
       adsurl = {https://ui.adsabs.harvard.edu/abs/2024ApJ...963..129M}
}

@ARTICLE{Maiolino2024_JADES,
       author = {{Maiolino}, Roberto and {Scholtz}, Jan and {Curtis-Lake}, Emma and {Carniani}, Stefano and {Baker}, William and {de Graaff}, Anna and {Tacchella}, Sandro and {{\"U}bler}, Hannah and {D'Eugenio}, Francesco and {Witstok}, Joris and {Curti}, Mirko and {Arribas}, Santiago and {Bunker}, Andrew J. and {Charlot}, St{\'e}phane and {Chevallard}, Jacopo and {Eisenstein}, Daniel J. and {Egami}, Eiichi and {Ji}, Zhiyuan and {Jones}, Gareth C. and {Lyu}, Jianwei and {Rawle}, Tim and {Robertson}, Brant and {Rujopakarn}, Wiphu and {Perna}, Michele and {Sun}, Fengwu and {Venturi}, Giacomo and {Williams}, Christina C. and {Willott}, Chris},
        title = "{JADES: The diverse population of infant black holes at 4 < z < 11: Merging, tiny, poor, but mighty}",
      journal = {\aap},
         year = 2024,
        month = nov,
       volume = {691},
          eid = {A145},
        pages = {A145},
          doi = {10.1051/0004-6361/202347640},
archivePrefix = {arXiv},
       eprint = {2308.01230},
 primaryClass = {astro-ph.GA},
       adsurl = {https://ui.adsabs.harvard.edu/abs/2024A&A...691A.145M}
}

@ARTICLE{Ma2025_CountingLRD,
       author = {{Ma}, Yilun and {Greene}, Jenny E. and {Setton}, David J. and {Goulding}, Andy D. and {Annunziatella}, Marianna and {Fan}, Xiaohui and {Kokorev}, Vasily and {Labbe}, Ivo and {Li}, Jiaxuan and {Lin}, Xiaojing and {Marchesini}, Danilo and {Matthee}, Jorryt and {Robbins}, Luke and {Sajina}, Anna and {Sawicki}, Marcin and {Telford}, O. Grace},
        title = "{Counting Little Red Dots at $z<4$ with Ground-based Surveys and Spectroscopic Follow-up}",
      journal = {arXiv e-prints},
         year = 2025,
        month = apr,
          eid = {arXiv:2504.08032},
        pages = {arXiv:2504.08032},
          doi = {10.48550/arXiv.2504.08032},
archivePrefix = {arXiv},
       eprint = {2504.08032},
 primaryClass = {astro-ph.GA},
       adsurl = {https://ui.adsabs.harvard.edu/abs/2025arXiv250408032M}
}

@BOOK{Draine2011,
       author = {{Draine}, Bruce T.},
        title = "{Physics of the Interstellar and Intergalactic Medium}",
        publisher = "{Princeton University Press}",
         year = 2011,
       adsurl = {https://ui.adsabs.harvard.edu/abs/2011piim.book.....D}
}

@ARTICLE{Maiolino2024_Nature,
       author = {{Maiolino}, Roberto and {Scholtz}, Jan and {Witstok}, Joris and {Carniani}, Stefano and {D'Eugenio}, Francesco and {de Graaff}, Anna and {{\"U}bler}, Hannah and {Tacchella}, Sandro and {Curtis-Lake}, Emma and {Arribas}, Santiago and {Bunker}, Andrew and {Charlot}, St{\'e}phane and {Chevallard}, Jacopo and {Curti}, Mirko and {Looser}, Tobias J. and {Maseda}, Michael V. and {Rawle}, Timothy D. and {Rodr{\'\i}guez del Pino}, Bruno and {Willott}, Chris J. and {Egami}, Eiichi and {Eisenstein}, Daniel J. and {Hainline}, Kevin N. and {Robertson}, Brant and {Williams}, Christina C. and {Willmer}, Christopher N.~A. and {Baker}, William M. and {Boyett}, Kristan and {DeCoursey}, Christa and {Fabian}, Andrew C. and {Helton}, Jakob M. and {Ji}, Zhiyuan and {Jones}, Gareth C. and {Kumari}, Nimisha and {Laporte}, Nicolas and {Nelson}, Erica J. and {Perna}, Michele and {Sandles}, Lester and {Shivaei}, Irene and {Sun}, Fengwu},
        title = "{A small and vigorous black hole in the early Universe}",
      journal = {\nat},
         year = 2024,
        month = mar,
       volume = {627},
       number = {8002},
        pages = {59-63},
          doi = {10.1038/s41586-024-07052-5},
archivePrefix = {arXiv},
       eprint = {2305.12492},
 primaryClass = {astro-ph.GA},
       adsurl = {https://ui.adsabs.harvard.edu/abs/2024Natur.627...59M}
}

@ARTICLE{Kokorev2023,
       author = {{Kokorev}, Vasily and {Fujimoto}, Seiji and {Labbe}, Ivo and {Greene}, Jenny E. and {Bezanson}, Rachel and {Dayal}, Pratika and {Nelson}, Erica J. and {Atek}, Hakim and {Brammer}, Gabriel and {Caputi}, Karina I. and {Chemerynska}, Iryna and {Cutler}, Sam E. and {Feldmann}, Robert and {Fudamoto}, Yoshinobu and {Furtak}, Lukas J. and {Goulding}, Andy D. and {de Graaff}, Anna and {Leja}, Joel and {Marchesini}, Danilo and {Miller}, Tim B. and {Nanayakkara}, Themiya and {Oesch}, Pascal A. and {Pan}, Richard and {Price}, Sedona H. and {Setton}, David J. and {Smit}, Renske and {Stefanon}, Mauro and {Wang}, Bingjie and {Weaver}, John R. and {Whitaker}, Katherine E. and {Williams}, Christina C. and {Zitrin}, Adi},
        title = "{UNCOVER: A NIRSpec Identification of a Broad-line AGN at z = 8.50}",
      journal = {\apjl},
         year = 2023,
        month = nov,
       volume = {957},
       number = {1},
          eid = {L7},
        pages = {L7},
          doi = {10.3847/2041-8213/ad037a},
archivePrefix = {arXiv},
       eprint = {2308.11610},
 primaryClass = {astro-ph.GA},
       adsurl = {https://ui.adsabs.harvard.edu/abs/2023ApJ...957L...7K}
}

@ARTICLE{Furtak2024,
       author = {{Furtak}, Lukas J. and {Labb{\'e}}, Ivo and {Zitrin}, Adi and {Greene}, Jenny E. and {Dayal}, Pratika and {Chemerynska}, Iryna and {Kokorev}, Vasily and {Miller}, Tim B. and {Goulding}, Andy D. and {de Graaff}, Anna and {Bezanson}, Rachel and {Brammer}, Gabriel B. and {Cutler}, Sam E. and {Leja}, Joel and {Pan}, Richard and {Price}, Sedona H. and {Wang}, Bingjie and {Weaver}, John R. and {Whitaker}, Katherine E. and {Atek}, Hakim and {Bogd{\'a}n}, {\'A}kos and {Charlot}, St{\'e}phane and {Curtis-Lake}, Emma and {van Dokkum}, Pieter and {Endsley}, Ryan and {Feldmann}, Robert and {Fudamoto}, Yoshinobu and {Fujimoto}, Seiji and {Glazebrook}, Karl and {Juneau}, St{\'e}phanie and {Marchesini}, Danilo and {Maseda}, Micheal V. and {Nelson}, Erica and {Oesch}, Pascal A. and {Plat}, Ad{\`e}le and {Setton}, David J. and {Stark}, Daniel P. and {Williams}, Christina C.},
        title = "{A high black-hole-to-host mass ratio in a lensed AGN in the early Universe}",
      journal = {\nat},
         year = 2024,
        month = apr,
       volume = {628},
       number = {8006},
        pages = {57-61},
          doi = {10.1038/s41586-024-07184-8},
archivePrefix = {arXiv},
       eprint = {2308.05735},
 primaryClass = {astro-ph.GA},
       adsurl = {https://ui.adsabs.harvard.edu/abs/2024Natur.628...57F}
}

@ARTICLE{Greene2024,
       author = {{Greene}, Jenny E. and {Labbe}, Ivo and {Goulding}, Andy D. and {Furtak}, Lukas J. and {Chemerynska}, Iryna and {Kokorev}, Vasily and {Dayal}, Pratika and {Volonteri}, Marta and {Williams}, Christina C. and {Wang}, Bingjie and {Setton}, David J. and {Burgasser}, Adam J. and {Bezanson}, Rachel and {Atek}, Hakim and {Brammer}, Gabriel and {Cutler}, Sam E. and {Feldmann}, Robert and {Fujimoto}, Seiji and {Glazebrook}, Karl and {de Graaff}, Anna and {Khullar}, Gourav and {Leja}, Joel and {Marchesini}, Danilo and {Maseda}, Michael V. and {Matthee}, Jorryt and {Miller}, Tim B. and {Naidu}, Rohan P. and {Nanayakkara}, Themiya and {Oesch}, Pascal A. and {Pan}, Richard and {Papovich}, Casey and {Price}, Sedona H. and {van Dokkum}, Pieter and {Weaver}, John R. and {Whitaker}, Katherine E. and {Zitrin}, Adi},
        title = "{UNCOVER Spectroscopy Confirms the Surprising Ubiquity of Active Galactic Nuclei in Red Sources at z > 5}",
      journal = {\apj},
         year = 2024,
        month = mar,
       volume = {964},
       number = {1},
          eid = {39},
        pages = {39},
          doi = {10.3847/1538-4357/ad1e5f},
archivePrefix = {arXiv},
       eprint = {2309.05714},
 primaryClass = {astro-ph.GA},
       adsurl = {https://ui.adsabs.harvard.edu/abs/2024ApJ...964...39G}
}

@ARTICLE{Taylor2025,
       author = {{Taylor}, Anthony J. and {Finkelstein}, Steven L. and {Kocevski}, Dale D. and {Jeon}, Junehyoung and {Bromm}, Volker and {Amor{\'\i}n}, Ricardo O. and {Arrabal Haro}, Pablo and {Backhaus}, Bren E. and {Bagley}, Micaela B. and {Banados}, Eduardo and {Bhatawdekar}, Rachana and {Brooks}, Madisyn and {Calabr{\`o}}, Antonello and {Ch{\'a}vez Ortiz}, {\'O}scar A. and {Cheng}, Yingjie and {Cleri}, Nikko J. and {Cole}, Justin W. and {Davis}, Kelcey and {Dickinson}, Mark and {Donnan}, Callum and {Dunlop}, James S. and {Ellis}, Richard S. and {Fern{\'a}ndez}, Vital and {Fontana}, Adriano and {Fujimoto}, Seiji and {Giavalisco}, Mauro and {Grazian}, Andrea and {Guo}, Jingsong and {Hathi}, Nimish P. and {Holwerda}, Benne W. and {Hirschmann}, Michaela and {Inayoshi}, Kohei and {Kartaltepe}, Jeyhan S. and {Khusanova}, Yana and {Koekemoer}, Anton M. and {Kokorev}, Vasily and {Larson}, Rebecca L. and {Leung}, Gene C.~K. and {Lucas}, Ray A. and {McLeod}, Derek J. and {Napolitano}, Lorenzo and {Onoue}, Masafusa and {Pacucci}, Fabio and {Papovich}, Casey and {P{\'e}rez-Gonz{\'a}lez}, Pablo G. and {Pirzkal}, Nor and {Somerville}, Rachel S. and {Trump}, Jonathan R. and {Wilkins}, Stephen M. and {Yung}, L.~Y. Aaron and {Zhang}, Haowen},
        title = "{Broad-line AGNs at 3.5 < z < 6: The Black Hole Mass Function and a Connection with Little Red Dots}",
      journal = {\apj},
         year = 2025,
        month = jun,
       volume = {986},
       number = {2},
          eid = {165},
        pages = {165},
          doi = {10.3847/1538-4357/add15b},
archivePrefix = {arXiv},
       eprint = {2409.06772},
 primaryClass = {astro-ph.GA},
       adsurl = {https://ui.adsabs.harvard.edu/abs/2025ApJ...986..165T}
}

@ARTICLE{Juodzbalis2024_Nature,
       author = {{Juod{\v{z}}balis}, Ignas and {Maiolino}, Roberto and {Baker}, William M. and {Tacchella}, Sandro and {Scholtz}, Jan and {D'Eugenio}, Francesco and {Witstok}, Joris and {Schneider}, Raffaella and {Trinca}, Alessandro and {Valiante}, Rosa and {DeCoursey}, Christa and {Curti}, Mirko and {Carniani}, Stefano and {Chevallard}, Jacopo and {de Graaff}, Anna and {Arribas}, Santiago and {Bennett}, Jake S. and {Bourne}, Martin A. and {Bunker}, Andrew J. and {Charlot}, St{\'e}phane and {Jiang}, Brian and {Koudmani}, Sophie and {Perna}, Michele and {Robertson}, Brant and {Sijacki}, Debora and {{\"U}bler}, Hannah and {Williams}, Christina C. and {Willott}, Chris},
        title = "{A dormant overmassive black hole in the early Universe}",
      journal = {\nat},
         year = 2024,
        month = dec,
       volume = {636},
       number = {8043},
        pages = {594-597},
          doi = {10.1038/s41586-024-08210-5},
archivePrefix = {arXiv},
       eprint = {2403.03872},
 primaryClass = {astro-ph.GA},
       adsurl = {https://ui.adsabs.harvard.edu/abs/2024Natur.636..594J}
}

@ARTICLE{Scholtz2025,
       author = {{Scholtz}, Jan and {Maiolino}, Roberto and {D'Eugenio}, Francesco and {Curtis-Lake}, Emma and {Carniani}, Stefano and {Charlot}, Stephane and {Curti}, Mirko and {Silcock}, Maddie S. and {Arribas}, Santiago and {Baker}, William and {Bhatawdekar}, Rachana and {Boyett}, Kristan and {Bunker}, Andrew J. and {Chevallard}, Jacopo and {Circosta}, Chiara and {Eisenstein}, Daniel J. and {Hainline}, Kevin and {Hausen}, Ryan and {Ji}, Xihan and {Ji}, Zhiyuan and {Johnson}, Benjamin D. and {Kumari}, Nimisha and {Looser}, Tobias J. and {Lyu}, Jianwei and {Maseda}, Michael V. and {Parlanti}, Eleonora and {Perna}, Michele and {Rieke}, Marcia and {Robertson}, Brant and {Del Pino}, Bruno Rodr{\'\i}guez and {Sun}, Fengwu and {Tacchella}, Sandro and {{\"U}bler}, Hannah and {Venturi}, Giacomo and {Williams}, Christina C. and {Willmer}, Christopher N.~A. and {Willott}, Chris and {Witstok}, Joris},
        title = "{JADES: A large population of obscured, narrow-line active galactic nuclei at high redshift}",
      journal = {\aap},
         year = 2025,
        month = may,
       volume = {697},
          eid = {A175},
        pages = {A175},
          doi = {10.1051/0004-6361/202348804},
archivePrefix = {arXiv},
       eprint = {2311.18731},
 primaryClass = {astro-ph.GA},
       adsurl = {https://ui.adsabs.harvard.edu/abs/2025A&A...697A.175S}
}

@ARTICLE{Taylor2025_CAPERS,
       author = {{Taylor}, Anthony J. and {Kokorev}, Vasily and {Kocevski}, Dale D. and {Akins}, Hollis B. and {Cullen}, Fergus and {Dickinson}, Mark and {Finkelstein}, Steven L. and {Arrabal Haro}, Pablo and {Bromm}, Volker and {Giavalisco}, Mauro and {Inayoshi}, Kohei and {Juneau}, St{\'e}phanie and {Leung}, Gene C.~K. and {P{\'e}rez-Gonz{\'a}lez}, Pablo G. and {Somerville}, Rachel S. and {Trump}, Jonathan R. and {Amor{\'\i}n}, Ricardo O. and {Barro}, Guillermo and {Burgarella}, Denis and {Brooks}, Madisyn and {Carnall}, Adam C. and {Casey}, Caitlin M. and {Cheng}, Yingjie and {Chisholm}, John and {Chworowsky}, Katherine and {Davis}, Kelcey and {Donnan}, Callum T. and {Dunlop}, James S. and {Ellis}, Richard S. and {Fern{\'a}ndez}, Vital and {Fujimoto}, Seiji and {Grogin}, Norman A. and {Gupta}, Ansh R. and {Hathi}, Nimish P. and {Jung}, Intae and {Hirschmann}, Michaela and {Kartaltepe}, Jeyhan S. and {Koekemoer}, Anton M. and {Larson}, Rebecca L. and {Leung}, Ho-Hin and {Llerena}, Mario and {Lucas}, Ray A. and {McLeod}, Derek J. and {McLure}, Ross and {Napolitano}, Lorenzo and {Papovich}, Casey and {Stanton}, Thomas M. and {Tripodi}, Roberta and {Wang}, Xin and {Wilkins}, Stephen M. and {Yung}, L.~Y. Aaron and {Zavala}, Jorge A.},
        title = "{CAPERS-LRD-z9: A Gas-enshrouded Little Red Dot Hosting a Broad-line Active Galactic Nucleus at z = 9.288}",
      journal = {\apjl},
         year = 2025,
        month = aug,
       volume = {989},
       number = {1},
          eid = {L7},
        pages = {L7},
          doi = {10.3847/2041-8213/ade789},
archivePrefix = {arXiv},
       eprint = {2505.04609},
 primaryClass = {astro-ph.GA},
       adsurl = {https://ui.adsabs.harvard.edu/abs/2025ApJ...989L...7T}
}

@ARTICLE{Mazzolari2025,
       author = {{Mazzolari}, Giovanni and {Scholtz}, Jan and {Maiolino}, Roberto and {Gilli}, Roberto and {Traina}, Alberto and {L{\'o}pez}, Ivan E. and {{\"U}bler}, Hannah and {Trefoloni}, Bartolomeo and {D'Eugenio}, Francesco and {Ji}, Xihan and {Mignoli}, Marco and {Vito}, Fabio and {Vignali}, Cristian and {Brusa}, Marcella},
        title = "{Narrow-line AGN selection in CEERS: Spectroscopic selection, physical properties, and X-ray and radio analysis}",
      journal = {\aap},
         year = 2025,
        month = aug,
       volume = {700},
          eid = {A12},
        pages = {A12},
          doi = {10.1051/0004-6361/202451860},
archivePrefix = {arXiv},
       eprint = {2408.15615},
 primaryClass = {astro-ph.GA},
       adsurl = {https://ui.adsabs.harvard.edu/abs/2025A&A...700A..12M}
}

@ARTICLE{Chisholm2024,
       author = {{Chisholm}, J. and {Berg}, D.~A. and {Endsley}, R. and {Gazagnes}, S. and {Richardson}, C.~T. and {Lambrides}, E. and {Greene}, J. and {Finkelstein}, S. and {Flury}, S. and {Guseva}, N.~G. and {Henry}, A. and {Hutchison}, T.~A. and {Izotov}, Y.~I. and {Marques-Chaves}, R. and {Oesch}, P. and {Papovich}, C. and {Saldana-Lopez}, A. and {Schaerer}, D. and {Stephenson}, M.~G.},
        title = "{[Ne v] emission from a faint epoch of reionization-era galaxy: evidence for a narrow-line intermediate-mass black hole}",
      journal = {\mnras},
         year = 2024,
        month = nov,
       volume = {534},
       number = {3},
        pages = {2633-2652},
          doi = {10.1093/mnras/stae2199},
archivePrefix = {arXiv},
       eprint = {2402.18643},
 primaryClass = {astro-ph.GA},
       adsurl = {https://ui.adsabs.harvard.edu/abs/2024MNRAS.534.2633C}
}

@ARTICLE{Maiolino2025_Chandra,
       author = {{Maiolino}, Roberto and {Risaliti}, Guido and {Signorini}, Matilde and {Trefoloni}, Bartolomeo and {Juod{\v{z}}balis}, Ignas and {Scholtz}, Jan and {{\"U}bler}, Hannah and {D'Eugenio}, Francesco and {Carniani}, Stefano and {Fabian}, Andy and {Ji}, Xihan and {Mazzolari}, Giovanni and {Bertola}, Elena and {Brusa}, Marcella and {Bunker}, Andrew J. and {Charlot}, Stephane and {Comastri}, Andrea and {Cresci}, Giovanni and {DeCoursey}, Christa Noel and {Egami}, Eiichi and {Fiore}, Fabrizio and {Gilli}, Roberto and {Perna}, Michele and {Tacchella}, Sandro and {Venturi}, Giacomo},
        title = "{JWST meets Chandra: a large population of Compton thick, feedback-free, and intrinsically X-ray weak AGN, with a sprinkle of SNe}",
      journal = {\mnras},
         year = 2025,
        month = apr,
       volume = {538},
       number = {3},
        pages = {1921-1943},
          doi = {10.1093/mnras/staf359},
archivePrefix = {arXiv},
       eprint = {2405.00504},
 primaryClass = {astro-ph.GA},
       adsurl = {https://ui.adsabs.harvard.edu/abs/2025MNRAS.538.1921M}
}

@ARTICLE{BaronNetzer2019,
       author = {{Baron}, Dalya and {Netzer}, Hagai},
        title = "{Discovering AGN-driven winds through their infrared emission - II. Mass outflow rate and energetics}",
      journal = {\mnras},
         year = 2019,
        month = jul,
       volume = {486},
       number = {3},
        pages = {4290-4303},
          doi = {10.1093/mnras/stz1070},
archivePrefix = {arXiv},
       eprint = {1903.11076},
 primaryClass = {astro-ph.GA},
       adsurl = {https://ui.adsabs.harvard.edu/abs/2019MNRAS.486.4290B}
}

@ARTICLE{Danhaive2026,
       author = {{Danhaive}, A. Lola and {Tacchella}, Sandro and {McClymont}, William and {Robertson}, Brant and {Carniani}, Stefano and {Carreira}, Courtney and {Egami}, Eiichi and {Bunker}, Andrew J. and {Curtis-Lake}, Emma and {Eisenstein}, Daniel J. and {Ji}, Zhiyuan and {Johnson}, Benjamin D. and {Rieke}, Marcia and {Villanueva}, Natalia C. and {Willmer}, Christopher N.~A. and {Willot}, Chris and {Wu}, Zihao and {Zhu}, Yongda},
        title = "{Beyond the stars: Linking H{\ensuremath{\alpha}} sizes, kinematics, and star formation in galaxies at z ≍ 4 - 6 with JWST grism surveys and GEKO}",
      journal = {\mnras},
         year = 2026,
        month = mar,
          doi = {10.1093/mnras/stag437},
archivePrefix = {arXiv},
       eprint = {2510.06315},
 primaryClass = {astro-ph.GA},
       adsurl = {https://ui.adsabs.harvard.edu/abs/2026MNRAS.tmp..413D}
}

@ARTICLE{Pacucci2023,
       author = {{Pacucci}, Fabio and {Nguyen}, Bao and {Carniani}, Stefano and {Maiolino}, Roberto and {Fan}, Xiaohui},
        title = "{JWST CEERS and JADES Active Galaxies at z = 4-7 Violate the Local M $_{{\textbullet}}$-M $_{{\ensuremath{\star}}}$ Relation at >3{\ensuremath{\sigma}}: Implications for Low-mass Black Holes and Seeding Models}",
      journal = {\apjl},
         year = 2023,
        month = nov,
       volume = {957},
       number = {1},
          eid = {L3},
        pages = {L3},
          doi = {10.3847/2041-8213/ad0158},
archivePrefix = {arXiv},
       eprint = {2308.12331},
 primaryClass = {astro-ph.GA},
       adsurl = {https://ui.adsabs.harvard.edu/abs/2023ApJ...957L...3P}
}

@ARTICLE{Mezcua2024,
       author = {{Mezcua}, Mar and {Pacucci}, Fabio and {Suh}, Hyewon and {Siudek}, Malgorzata and {Natarajan}, Priyamvada},
        title = "{Overmassive Black Holes at Cosmic Noon: Linking the Local and the High-redshift Universe}",
      journal = {\apjl},
         year = 2024,
        month = may,
       volume = {966},
       number = {2},
          eid = {L30},
        pages = {L30},
          doi = {10.3847/2041-8213/ad3c2a},
archivePrefix = {arXiv},
       eprint = {2404.05793},
 primaryClass = {astro-ph.GA},
       adsurl = {https://ui.adsabs.harvard.edu/abs/2024ApJ...966L..30M}
}

@ARTICLE{Danhaive2026_GEKO,
       author = {{Danhaive}, A. Lola and {Tacchella}, Sandro and {Bunker}, Andrew J. and {Curtis-Lake}, Emma and {de Graaff}, Anna and {D'Eugenio}, Francesco and {Duan}, Qiao and {Egami}, Eiichi and {Eisenstein}, Daniel J. and {Johnson}, Benjamin D. and {Maiolino}, Roberto and {McClymont}, William and {Rieke}, Marcia and {Robertson}, Brant and {Sun}, Fengwu and {Willmer}, Christopher N.~A. and {Wu}, Zihao and {Zhu}, Yongda},
        title = "{The dark side of early galaxies: GEKO uncovers dark-matter fractions at z {\ensuremath{\sim}} 4 {\ensuremath{-}} 6}",
      journal = {\mnras},
         year = 2026,
        month = mar,
       volume = {546},
       number = {3},
          eid = {stag119},
        pages = {stag119},
          doi = {10.1093/mnras/stag119},
archivePrefix = {arXiv},
       eprint = {2510.14779},
 primaryClass = {astro-ph.GA},
       adsurl = {https://ui.adsabs.harvard.edu/abs/2026MNRAS.546ag119D}
}

@ARTICLE{Muzzin2025,
       author = {{Muzzin}, Adam and {Suess}, Katherine A. and {Marchesini}, Danilo and {Robbins}, Luke and {Willott}, Chris J. and {Alberts}, Stacey and {Antwi-Danso}, Jacqueline and {Asada}, Yoshihisa and {Brammer}, Gabriel and {Cutler}, Sam E. and {Iyer}, Kartheik G. and {Labbe}, Ivo and {Martis}, Nicholas S. and {Miller}, Tim B. and {Mitsuhashi}, Ikki and {Pope}, Alexandra and {Sajina}, Anna and {Sarrouh}, Ghassan T.~E. and {Sharma}, Monu and {Stefanon}, Mauro and {Whitaker}, Katherine E. and {Abraham}, Roberto and {Atek}, Hakim and {Bradac}, Marusa and {Berek}, Samantha and {Bezanson}, Rachel and {Brown}, Westley and {Burgasser}, Adam J. and {Chicoine}, Nathalie and {Cloonan}, Aidan P. and {Cooper}, Olivia R. and {Dayal}, Pratika and {de Graaff}, Anna and {Desprez}, Guillaume and {Feldmann}, Robert and {Forrest}, Ben and {Franx}, Marijn and {Fudamoto}, Yoshinobu and {Fujimoto}, Seiji and {Furtak}, Lukas J. and {Glazebrook}, Karl and {Goovaerts}, Ilias and {Greene}, Jenny E. and {Jagga}, Naadiyah and {Jarvis}, William W.~H. and {Kriek}, Mariska and {Khullar}, Gourav and {La Torre}, Valentina and {Leja}, Joel and {Lin}, Jamie and {Lorenz}, Brian and {Lyon}, Daniel and {Markov}, Vladan and {Maseda}, Michael V. and {McConachie}, Ian and {Merchant}, Maya and {Merida}, Rosa M. and {Mowla}, Lamiya and {Myers}, Katherine and {Naidu}, Rohan P. and {Nanayakkara}, Themiya and {Nelson}, Erica J. and {Noirot}, Gael and {Oesch}, Pascal A. and {Omori}, Kiyoaki C. and {Pan}, Richard and {Porraz Barrera}, Natalia and {Price}, Sedona H. and {Ravindranath}, Swara and {Sawicki}, Marcin and {Setton}, David J. and {Smit}, Renske and {Sok}, Visal and {Speagle}, Joshua S. and {Taylor}, Edward N. and {Tan}, Vivian Yun Yan and {Tripodi}, Roberta and {van der Wel}, Arjen and {Perez Vidal}, Edgar and {Wang}, Bingjie and {Weaver}, John R. and {Williams}, Christina C. and {Withers}, Sunna and {Zaidi}, Kumail},
        title = "{MINERVA: A NIRCam Medium Band and MIRI Imaging Survey to Unlock the Hidden Gems of the Distant Universe}",
      journal = {arXiv e-prints},
         year = 2025,
        month = jul,
          eid = {arXiv:2507.19706},
        pages = {arXiv:2507.19706},
          doi = {10.48550/arXiv.2507.19706},
archivePrefix = {arXiv},
       eprint = {2507.19706},
 primaryClass = {astro-ph.GA},
       adsurl = {https://ui.adsabs.harvard.edu/abs/2025arXiv250719706M}
}

@ARTICLE{Hu2026,
       author = {{Hu}, Haojie and {Yanagisawa}, Hiroto and {Nishigaki}, Moka and {Kiyota}, Tomokazu and {Ishiyama}, Tomoaki and {Ohsuga}, Ken},
        title = "{Back to Normal Again: Possible Destinies of JWST overmassive SMBHs and ``Little Red Dots'' in the View of Shin-Uchuu Simulation}",
      journal = {arXiv e-prints},
         year = 2026,
        month = feb,
          eid = {arXiv:2602.14496},
        pages = {arXiv:2602.14496},
          doi = {10.48550/arXiv.2602.14496},
archivePrefix = {arXiv},
       eprint = {2602.14496},
 primaryClass = {astro-ph.GA},
       adsurl = {https://ui.adsabs.harvard.edu/abs/2026arXiv260214496H}
}

@ARTICLE{Kollmeier2006,
       author = {{Kollmeier}, Juna A. and {Onken}, Christopher A. and {Kochanek}, Christopher S. and {Gould}, Andrew and {Weinberg}, David H. and {Dietrich}, Matthias and {Cool}, Richard and {Dey}, Arjun and {Eisenstein}, Daniel J. and {Jannuzi}, Buell T. and {Le Floc'h}, Emeric and {Stern}, Daniel},
        title = "{Black Hole Masses and Eddington Ratios at 0.3 < z < 4}",
      journal = {\apj},
         year = 2006,
        month = sep,
       volume = {648},
       number = {1},
        pages = {128-139},
          doi = {10.1086/505646},
archivePrefix = {arXiv},
       eprint = {astro-ph/0508657},
 primaryClass = {astro-ph},
       adsurl = {https://ui.adsabs.harvard.edu/abs/2006ApJ...648..128K}
}

@ARTICLE{Mannucci2010,
       author = {{Mannucci}, F. and {Cresci}, G. and {Maiolino}, R. and {Marconi}, A. and {Gnerucci}, A.},
        title = "{A fundamental relation between mass, star formation rate and metallicity in local and high-redshift galaxies}",
      journal = {\mnras},
         year = 2010,
        month = nov,
       volume = {408},
       number = {4},
        pages = {2115-2127},
          doi = {10.1111/j.1365-2966.2010.17291.x},
archivePrefix = {arXiv},
       eprint = {1005.0006},
 primaryClass = {astro-ph.CO},
       adsurl = {https://ui.adsabs.harvard.edu/abs/2010MNRAS.408.2115M}
}

@ARTICLE{Begelman2006,
       author = {{Begelman}, Mitchell C. and {Volonteri}, Marta and {Rees}, Martin J.},
        title = "{Formation of supermassive black holes by direct collapse in pre-galactic haloes}",
      journal = {\mnras},
         year = 2006,
        month = jul,
       volume = {370},
       number = {1},
        pages = {289-298},
          doi = {10.1111/j.1365-2966.2006.10467.x},
archivePrefix = {arXiv},
       eprint = {astro-ph/0602363},
 primaryClass = {astro-ph},
       adsurl = {https://ui.adsabs.harvard.edu/abs/2006MNRAS.370..289B}
}

@ARTICLE{Brazzini2025,
       author = {{Brazzini}, Matilde and {D'Eugenio}, Francesco and {Maiolino}, Roberto and {Juod{\v{z}}balis}, Ignas and {Ji}, Xihan and {Scholtz}, Jan and {Chang}, Seok-Jun},
        title = "{Ruling out dominant electron scattering in Little Red Dots' Rosetta Stone using multiple hydrogen lines}",
      journal = {\mnras},
         year = 2025,
        month = nov,
       volume = {544},
       number = {1},
        pages = {L167-L173},
          doi = {10.1093/mnrasl/slaf116},
archivePrefix = {arXiv},
       eprint = {2507.08929},
 primaryClass = {astro-ph.GA},
       adsurl = {https://ui.adsabs.harvard.edu/abs/2025MNRAS.544L.167B}
}

@ARTICLE{Adamo2025,
       author = {{Adamo}, Angela and {Atek}, Hakim and {Bagley}, Micaela B. and {Ba{\~n}ados}, Eduardo and {Barrow}, Kirk S.~S. and {Berg}, Danielle A. and {Bezanson}, Rachel and {Brada{\v{c}}}, Maru{\v{s}}a and {Brammer}, Gabriel and {Carnall}, Adam C. and {Chisholm}, John and {Coe}, Dan and {Dayal}, Pratika and {Eisenstein}, Daniel J. and {Eldridge}, Jan J. and {Ferrara}, Andrea and {Fujimoto}, Seiji and {Graaff}, Anna de and {Habouzit}, Melanie and {Hutchison}, Taylor A. and {Kartaltepe}, Jeyhan S. and {Kassin}, Susan A. and {Kriek}, Mariska and {Labb{\'e}}, Ivo and {Maiolino}, Roberto and {Marques-Chaves}, Rui and {Maseda}, Michael V. and {Mason}, Charlotte and {Matthee}, Jorryt and {McQuinn}, Kristen B.~W. and {Meynet}, Georges and {Naidu}, Rohan P. and {Oesch}, Pascal A. and {Pentericci}, Laura and {P{\'e}rez-Gonz{\'a}lez}, Pablo G. and {Rigby}, Jane R. and {Roberts-Borsani}, Guido and {Schaerer}, Daniel and {Shapley}, Alice E. and {Stark}, Daniel P. and {Stiavelli}, Massimo and {Strom}, Allison L. and {Vanzella}, Eros and {Wang}, Feige and {Wilkins}, Stephen M. and {Williams}, Christina C. and {Willott}, Chris J. and {Wylezalek}, Dominika and {Nota}, Antonella},
        title = "{The first billion years according to JWST}",
      journal = {Nature Astronomy},
         year = 2025,
        month = aug,
       volume = {9},
        pages = {1134-1147},
          doi = {10.1038/s41550-025-02624-5},
archivePrefix = {arXiv},
       eprint = {2405.21054},
 primaryClass = {astro-ph.GA},
       adsurl = {https://ui.adsabs.harvard.edu/abs/2025NatAs...9.1134A}
}

@ARTICLE{Habouzit2025,
       author = {{Habouzit}, Melanie},
        title = "{Is the JWST detecting too many AGN candidates?}",
      journal = {\mnras},
         year = 2025,
        month = mar,
       volume = {537},
       number = {3},
        pages = {2323-2333},
          doi = {10.1093/mnras/staf167},
archivePrefix = {arXiv},
       eprint = {2405.05319},
 primaryClass = {astro-ph.GA},
       adsurl = {https://ui.adsabs.harvard.edu/abs/2025MNRAS.537.2323H}
}

@ARTICLE{Jin2012,
       author = {{Jin}, Chichuan and {Ward}, Martin and {Done}, Chris},
        title = "{A combined optical and X-ray study of unobscured type 1 active galactic nuclei - III. Broad-band SED properties}",
      journal = {\mnras},
         year = 2012,
        month = sep,
       volume = {425},
       number = {2},
        pages = {907-929},
          doi = {10.1111/j.1365-2966.2012.21272.x},
archivePrefix = {arXiv},
       eprint = {1205.1846},
 primaryClass = {astro-ph.HE},
       adsurl = {https://ui.adsabs.harvard.edu/abs/2012MNRAS.425..907J}
}

@ARTICLE{Koekemoer2011,
       author = {{Koekemoer}, Anton M. and {Faber}, S.~M. and {Ferguson}, Henry C. and {Grogin}, Norman A. and {Kocevski}, Dale D. and {Koo}, David C. and {Lai}, Kamson and {Lotz}, Jennifer M. and {Lucas}, Ray A. and {McGrath}, Elizabeth J. and {Ogaz}, Sara and {Rajan}, Abhijith and {Riess}, Adam G. and {Rodney}, Steve A. and {Strolger}, Louis and {Casertano}, Stefano and {Castellano}, Marco and {Dahlen}, Tomas and {Dickinson}, Mark and {Dolch}, Timothy and {Fontana}, Adriano and {Giavalisco}, Mauro and {Grazian}, Andrea and {Guo}, Yicheng and {Hathi}, Nimish P. and {Huang}, Kuang-Han and {van der Wel}, Arjen and {Yan}, Hao-Jing and {Acquaviva}, Viviana and {Alexander}, David M. and {Almaini}, Omar and {Ashby}, Matthew L.~N. and {Barden}, Marco and {Bell}, Eric F. and {Bournaud}, Fr{\'e}d{\'e}ric and {Brown}, Thomas M. and {Caputi}, Karina I. and {Cassata}, Paolo and {Challis}, Peter J. and {Chary}, Ranga-Ram and {Cheung}, Edmond and {Cirasuolo}, Michele and {Conselice}, Christopher J. and {Roshan Cooray}, Asantha and {Croton}, Darren J. and {Daddi}, Emanuele and {Dav{\'e}}, Romeel and {de Mello}, Duilia F. and {de Ravel}, Loic and {Dekel}, Avishai and {Donley}, Jennifer L. and {Dunlop}, James S. and {Dutton}, Aaron A. and {Elbaz}, David and {Fazio}, Giovanni G. and {Filippenko}, Alexei V. and {Finkelstein}, Steven L. and {Frazer}, Chris and {Gardner}, Jonathan P. and {Garnavich}, Peter M. and {Gawiser}, Eric and {Gruetzbauch}, Ruth and {Hartley}, Will G. and {H{\"a}ussler}, Boris and {Herrington}, Jessica and {Hopkins}, Philip F. and {Huang}, Jia-Sheng and {Jha}, Saurabh W. and {Johnson}, Andrew and {Kartaltepe}, Jeyhan S. and {Khostovan}, Ali A. and {Kirshner}, Robert P. and {Lani}, Caterina and {Lee}, Kyoung-Soo and {Li}, Weidong and {Madau}, Piero and {McCarthy}, Patrick J. and {McIntosh}, Daniel H. and {McLure}, Ross J. and {McPartland}, Conor and {Mobasher}, Bahram and {Moreira}, Heidi and {Mortlock}, Alice and {Moustakas}, Leonidas A. and {Mozena}, Mark and {Nandra}, Kirpal and {Newman}, Jeffrey A. and {Nielsen}, Jennifer L. and {Niemi}, Sami and {Noeske}, Kai G. and {Papovich}, Casey J. and {Pentericci}, Laura and {Pope}, Alexandra and {Primack}, Joel R. and {Ravindranath}, Swara and {Reddy}, Naveen A. and {Renzini}, Alvio and {Rix}, Hans-Walter and {Robaina}, Aday R. and {Rosario}, David J. and {Rosati}, Piero and {Salimbeni}, Sara and {Scarlata}, Claudia and {Siana}, Brian and {Simard}, Luc and {Smidt}, Joseph and {Snyder}, Diana and {Somerville}, Rachel S. and {Spinrad}, Hyron and {Straughn}, Amber N. and {Telford}, Olivia and {Teplitz}, Harry I. and {Trump}, Jonathan R. and {Vargas}, Carlos and {Villforth}, Carolin and {Wagner}, Cory R. and {Wandro}, Pat and {Wechsler}, Risa H. and {Weiner}, Benjamin J. and {Wiklind}, Tommy and {Wild}, Vivienne and {Wilson}, Grant and {Wuyts}, Stijn and {Yun}, Min S.},
        title = "{CANDELS: The Cosmic Assembly Near-infrared Deep Extragalactic Legacy Survey{\textemdash}The Hubble Space Telescope Observations, Imaging Data Products, and Mosaics}",
      journal = {\apjs},
         year = 2011,
        month = dec,
       volume = {197},
       number = {2},
          eid = {36},
        pages = {36},
          doi = {10.1088/0067-0049/197/2/36},
archivePrefix = {arXiv},
       eprint = {1105.3754},
 primaryClass = {astro-ph.CO},
       adsurl = {https://ui.adsabs.harvard.edu/abs/2011ApJS..197...36K}
}

@ARTICLE{Grogin2011,
       author = {{Grogin}, Norman A. and {Kocevski}, Dale D. and {Faber}, S.~M. and {Ferguson}, Henry C. and {Koekemoer}, Anton M. and {Riess}, Adam G. and {Acquaviva}, Viviana and {Alexander}, David M. and {Almaini}, Omar and {Ashby}, Matthew L.~N. and {Barden}, Marco and {Bell}, Eric F. and {Bournaud}, Fr{\'e}d{\'e}ric and {Brown}, Thomas M. and {Caputi}, Karina I. and {Casertano}, Stefano and {Cassata}, Paolo and {Castellano}, Marco and {Challis}, Peter and {Chary}, Ranga-Ram and {Cheung}, Edmond and {Cirasuolo}, Michele and {Conselice}, Christopher J. and {Roshan Cooray}, Asantha and {Croton}, Darren J. and {Daddi}, Emanuele and {Dahlen}, Tomas and {Dav{\'e}}, Romeel and {de Mello}, Du{\'\i}lia F. and {Dekel}, Avishai and {Dickinson}, Mark and {Dolch}, Timothy and {Donley}, Jennifer L. and {Dunlop}, James S. and {Dutton}, Aaron A. and {Elbaz}, David and {Fazio}, Giovanni G. and {Filippenko}, Alexei V. and {Finkelstein}, Steven L. and {Fontana}, Adriano and {Gardner}, Jonathan P. and {Garnavich}, Peter M. and {Gawiser}, Eric and {Giavalisco}, Mauro and {Grazian}, Andrea and {Guo}, Yicheng and {Hathi}, Nimish P. and {H{\"a}ussler}, Boris and {Hopkins}, Philip F. and {Huang}, Jia-Sheng and {Huang}, Kuang-Han and {Jha}, Saurabh W. and {Kartaltepe}, Jeyhan S. and {Kirshner}, Robert P. and {Koo}, David C. and {Lai}, Kamson and {Lee}, Kyoung-Soo and {Li}, Weidong and {Lotz}, Jennifer M. and {Lucas}, Ray A. and {Madau}, Piero and {McCarthy}, Patrick J. and {McGrath}, Elizabeth J. and {McIntosh}, Daniel H. and {McLure}, Ross J. and {Mobasher}, Bahram and {Moustakas}, Leonidas A. and {Mozena}, Mark and {Nandra}, Kirpal and {Newman}, Jeffrey A. and {Niemi}, Sami-Matias and {Noeske}, Kai G. and {Papovich}, Casey J. and {Pentericci}, Laura and {Pope}, Alexandra and {Primack}, Joel R. and {Rajan}, Abhijith and {Ravindranath}, Swara and {Reddy}, Naveen A. and {Renzini}, Alvio and {Rix}, Hans-Walter and {Robaina}, Aday R. and {Rodney}, Steven A. and {Rosario}, David J. and {Rosati}, Piero and {Salimbeni}, Sara and {Scarlata}, Claudia and {Siana}, Brian and {Simard}, Luc and {Smidt}, Joseph and {Somerville}, Rachel S. and {Spinrad}, Hyron and {Straughn}, Amber N. and {Strolger}, Louis-Gregory and {Telford}, Olivia and {Teplitz}, Harry I. and {Trump}, Jonathan R. and {van der Wel}, Arjen and {Villforth}, Carolin and {Wechsler}, Risa H. and {Weiner}, Benjamin J. and {Wiklind}, Tommy and {Wild}, Vivienne and {Wilson}, Grant and {Wuyts}, Stijn and {Yan}, Hao-Jing and {Yun}, Min S.},
        title = "{CANDELS: The Cosmic Assembly Near-infrared Deep Extragalactic Legacy Survey}",
      journal = {\apjs},
         year = 2011,
        month = dec,
       volume = {197},
       number = {2},
          eid = {35},
        pages = {35},
          doi = {10.1088/0067-0049/197/2/35},
archivePrefix = {arXiv},
       eprint = {1105.3753},
 primaryClass = {astro-ph.CO},
       adsurl = {https://ui.adsabs.harvard.edu/abs/2011ApJS..197...35G}
}

@ARTICLE{PayyoorVijayan2026,
       author = {{Payyoor Vijayan}, Aswin and {Yates}, Robert M. and {Lovell}, Christopher C. and {Roper}, WIlliam J. and {Wilkins}, Stephen M. and {Algera}, Hiddo S.~B. and {Liao}, Shihong and {Punyasheel}, Paurush and {Rowland}, Lucie E. and {Seeyave}, Louise T.~C.},
        title = "{Interpreting nebular emission lines in the high-redshift Universe}",
      journal = {The Open Journal of Astrophysics},
         year = 2026,
        month = feb,
       volume = {9},
        pages = {57554},
          doi = {10.33232/001c.157554},
archivePrefix = {arXiv},
       eprint = {2507.20190},
 primaryClass = {astro-ph.GA},
       adsurl = {https://ui.adsabs.harvard.edu/abs/2026OJAp....957554P}
}

@ARTICLE{Pascalau2026_gradient,
       author = {{Pascalau}, Robert G. and {D'Eugenio}, Francesco and {Maiolino}, Roberto and {Duan}, Qiao and {Isobe}, Yuki and {Arribas}, Santiago and {Bunker}, Andrew J. and {Charlot}, St{\'e}phane and {Perna}, Michele and {Rodriguez Del Pino}, Bruno and {Ubler}, Hannah and {Bertola}, Elena and {Boker}, Torsten and {Carniani}, Stefano and {Coe}, Dan and {Cresci}, Giovanni and {Curti}, Mirko and {Hsiao}, Tiger Y.~Y. and {Ivey}, Lucy R. and {Jones}, Gareth C. and {Lamperti}, Isabella and {Parlanti}, Eleonora and {Scholtz}, Jan and {Tacchella}, Sandro and {Ulivi}, Lorenzo and {Venturi}, Giacomo and {Witstok}, Joris and {Zamora}, Sandra},
        title = "{GA-NIFS: Dissecting The Alchemised: NIRSpec/IFU reveals turbulent gas inflows in a complex system at $z=10.17$}",
      journal = {arXiv e-prints},
         year = 2026,
        month = feb,
          eid = {arXiv:2603.00232},
        pages = {arXiv:2603.00232},
          doi = {10.48550/arXiv.2603.00232},
archivePrefix = {arXiv},
       eprint = {2603.00232},
 primaryClass = {astro-ph.GA},
       adsurl = {https://ui.adsabs.harvard.edu/abs/2026arXiv260300232P}
}

@ARTICLE{Rinaldi2025,
       author = {{Rinaldi}, Pierluigi and {Rieke}, George H. and {Wu}, Zihao and {Gilbert}, Carys J.~E. and {Pacucci}, Fabio and {Barchiesi}, Luigi and {Alberts}, Stacey and {Carniani}, Stefano and {Bunker}, Andrew J. and {Bhatawdekar}, Rachana and {D'Eugenio}, Francesco and {Ji}, Zhiyuan and {Johnson}, Benjamin D. and {Hainline}, Kevin and {Kokorev}, Vasily and {Kumari}, Nimisha and {Iani}, Edoardo and {Lyu}, Jianwei and {Maiolino}, Roberto and {Parlanti}, Eleonora and {Robertson}, Brant E. and {Sun}, Yang and {Vignali}, Cristian and {Williams}, Christina C. and {Willmer}, Christopher N.~A. and {Zhu}, Yongda},
        title = "{Beyond the Dot: an LRD-like nucleus at the Heart of an IR-Bright Galaxy and its implications for high-redshift LRDs}",
      journal = {arXiv e-prints},
         year = 2025,
        month = jul,
          eid = {arXiv:2507.17738},
        pages = {arXiv:2507.17738},
          doi = {10.48550/arXiv.2507.17738},
archivePrefix = {arXiv},
       eprint = {2507.17738},
 primaryClass = {astro-ph.GA},
       adsurl = {https://ui.adsabs.harvard.edu/abs/2025arXiv250717738R}
}

@ARTICLE{Ananna2024,
       author = {{Ananna}, Tonima Tasnim and {Bogd{\'a}n}, {\'A}kos and {Kov{\'a}cs}, Orsolya E. and {Natarajan}, Priyamvada and {Hickox}, Ryan C.},
        title = "{X-Ray View of Little Red Dots: Do They Host Supermassive Black Holes?}",
      journal = {\apjl},
         year = 2024,
        month = jul,
       volume = {969},
       number = {1},
          eid = {L18},
        pages = {L18},
          doi = {10.3847/2041-8213/ad5669},
archivePrefix = {arXiv},
       eprint = {2404.19010},
 primaryClass = {astro-ph.GA},
       adsurl = {https://ui.adsabs.harvard.edu/abs/2024ApJ...969L..18A}
}

@ARTICLE{Yue2024_Stack,
       author = {{Yue}, Minghao and {Eilers}, Anna-Christina and {Ananna}, Tonima Tasnim and {Panagiotou}, Christos and {Kara}, Erin and {Miyaji}, Takamitsu},
        title = "{Stacking X-Ray Observations of ``Little Red Dots'': Implications for Their Active Galactic Nucleus Properties}",
      journal = {\apjl},
         year = 2024,
        month = oct,
       volume = {974},
       number = {2},
          eid = {L26},
        pages = {L26},
          doi = {10.3847/2041-8213/ad7eba},
archivePrefix = {arXiv},
       eprint = {2404.13290},
 primaryClass = {astro-ph.GA},
       adsurl = {https://ui.adsabs.harvard.edu/abs/2024ApJ...974L..26Y}
}

@ARTICLE{Mazzolari2024,
       author = {{Mazzolari}, G. and {Gilli}, R. and {Maiolino}, R. and {Prandoni}, I. and {Delvecchio}, I. and {Norman}, C. and {Jimenez-Andrade}, E.~F. and {Belladitta}, S. and {Vito}, F. and {Momjian}, E. and {Chiaberge}, M. and {Trefoloni}, B. and {Signorini}, M. and {Ji}, X. and {D'Amato}, Q. and {Risaliti}, G. and {Baldi}, R.~D. and {Fabian}, A. and {{\"U}bler}, H. and {D'Eugenio}, F. and {Scholtz}, J. and {Juod{\v{z}}balis}, I. and {Mignoli}, M. and {Brusa}, M. and {Murphy}, E. and {Muxlow}, T.~W.~B.},
        title = "{The radio properties of the JWST-discovered AGN}",
      journal = {arXiv e-prints},
         year = 2024,
        month = dec,
          eid = {arXiv:2412.04224},
        pages = {arXiv:2412.04224},
          doi = {10.48550/arXiv.2412.04224},
archivePrefix = {arXiv},
       eprint = {2412.04224},
 primaryClass = {astro-ph.GA},
       adsurl = {https://ui.adsabs.harvard.edu/abs/2024arXiv241204224M}
}

@ARTICLE{KokuboHarikane2025,
       author = {{Kokubo}, Mitsuru and {Harikane}, Yuichi},
        title = "{Challenging the Active Galactic Nucleus Scenario for JWST/NIRSpec Little Red Dot and Non─Little Red Dot Broad H{\ensuremath{\alpha}} Emitters in Light of Nondetection of NIRCam Photometric Variability and X-Ray}",
      journal = {\apj},
         year = 2025,
        month = dec,
       volume = {995},
       number = {1},
          eid = {24},
        pages = {24},
          doi = {10.3847/1538-4357/ae119e},
archivePrefix = {arXiv},
       eprint = {2407.04777},
 primaryClass = {astro-ph.GA},
       adsurl = {https://ui.adsabs.harvard.edu/abs/2025ApJ...995...24K}
}

@ARTICLE{Zhang2025,
       author = {{Zhang}, Zijian and {Jiang}, Linhua and {Liu}, Weiyang and {Ho}, Luis C.},
        title = "{Analysis of Multi-epoch JWST Images of {\ensuremath{\sim}}300 Little Red Dots: Tentative Detection of Variability in a Minority of Sources}",
      journal = {\apj},
         year = 2025,
        month = may,
       volume = {985},
       number = {1},
          eid = {119},
        pages = {119},
          doi = {10.3847/1538-4357/adcb3e},
archivePrefix = {arXiv},
       eprint = {2411.02729},
 primaryClass = {astro-ph.GA},
       adsurl = {https://ui.adsabs.harvard.edu/abs/2025ApJ...985..119Z}
}

@ARTICLE{Furtak2025,
       author = {{Furtak}, Lukas J. and {Secunda}, Amy R. and {Greene}, Jenny E. and {Zitrin}, Adi and {Labb{\'e}}, Ivo and {Golubchik}, Miriam and {Bezanson}, Rachel and {Kokorev}, Vasily and {Atek}, Hakim and {Brammer}, Gabriel B. and {Chemerynska}, Iryna and {Cutler}, Sam E. and {Dayal}, Pratika and {Feldmann}, Robert and {Fujimoto}, Seiji and {Glazebrook}, Karl and {Leja}, Joel and {Ma}, Yilun and {Matthee}, Jorryt and {Naidu}, Rohan P. and {Nelson}, Erica J. and {Oesch}, Pascal A. and {Pan}, Richard and {Price}, Sedona H. and {Suess}, Katherine A. and {Wang}, Bingjie and {Weaver}, John R. and {Whitaker}, Katherine E.},
        title = "{Investigating photometric and spectroscopic variability in the multiply imaged little red dot A2744-QSO1}",
      journal = {\aap},
         year = 2025,
        month = jun,
       volume = {698},
          eid = {A227},
        pages = {A227},
          doi = {10.1051/0004-6361/202554110},
archivePrefix = {arXiv},
       eprint = {2502.07875},
 primaryClass = {astro-ph.GA},
       adsurl = {https://ui.adsabs.harvard.edu/abs/2025A&A...698A.227F}
}

@ARTICLE{Naidu2025,
       author = {{Naidu}, Rohan P. and {Matthee}, Jorryt and {Katz}, Harley and {de Graaff}, Anna and {Oesch}, Pascal and {Smith}, Aaron and {Greene}, Jenny E. and {Brammer}, Gabriel and {Weibel}, Andrea and {Hviding}, Raphael and {Chisholm}, John and {Labb\textbackslash'e}, Ivo and {Simcoe}, Robert A. and {Witten}, Callum and {Atek}, Hakim and {Baggen}, Josephine F.~W. and {Belli}, Sirio and {Bezanson}, Rachel and {Boogaard}, Leindert A. and {Bose}, Sownak and {Covelo-Paz}, Alba and {Dayal}, Pratika and {Fudamoto}, Yoshinobu and {Furtak}, Lukas J. and {Giovinazzo}, Emma and {Goulding}, Andy and {Gronke}, Max and {Heintz}, Kasper E. and {Hirschmann}, Michaela and {Illingworth}, Garth and {Inoue}, Akio K. and {Johnson}, Benjamin D. and {Leja}, Joel and {Leonova}, Ecaterina and {McConachie}, Ian and {Maseda}, Michael V. and {Natarajan}, Priyamvada and {Nelson}, Erica and {Setton}, David J. and {Shivaei}, Irene and {Sobral}, David and {Stefanon}, Mauro and {Tacchella}, Sandro and {Toft}, Sune and {Torralba}, Alberto and {van Dokkum}, Pieter and {van der Wel}, Arjen and {Volonteri}, Marta and {Walter}, Fabian and {Wang}, Bingjie and {Watson}, Darach},
        title = "{A ``Black Hole Star'' Reveals the Remarkable Gas-Enshrouded Hearts of the Little Red Dots}",
      journal = {arXiv e-prints},
         year = 2025,
        month = mar,
          eid = {arXiv:2503.16596},
        pages = {arXiv:2503.16596},
          doi = {10.48550/arXiv.2503.16596},
archivePrefix = {arXiv},
       eprint = {2503.16596},
 primaryClass = {astro-ph.GA},
       adsurl = {https://ui.adsabs.harvard.edu/abs/2025arXiv250316596N}
}

@ARTICLE{Chen2025,
       author = {{Chen}, Chang-Hao and {Ho}, Luis C. and {Li}, Ruancun and {Zhuang}, Ming-Yang},
        title = "{The Host Galaxy (If Any) of the Little Red Dots}",
      journal = {\apj},
         year = 2025,
        month = apr,
       volume = {983},
       number = {1},
          eid = {60},
        pages = {60},
          doi = {10.3847/1538-4357/ada93a},
archivePrefix = {arXiv},
       eprint = {2411.04446},
 primaryClass = {astro-ph.GA},
       adsurl = {https://ui.adsabs.harvard.edu/abs/2025ApJ...983...60C}
}

@ARTICLE{Isobe2025,
       author = {{Isobe}, Yuki and {Maiolino}, Roberto and {D'Eugenio}, Francesco and {Curti}, Mirko and {Ji}, Xihan and {Juod{\v{z}}balis}, Ignas and {Scholtz}, Jan and {Feltre}, Anne and {Charlot}, St{\'e}phane and {{\"U}bler}, Hannah and {J. Bunker}, Andrew and {Carniani}, Stefano and {Curtis-Lake}, Emma and {Ji}, Zhiyuan and {Kumari}, Nimisha and {Rinaldi}, Pierluigi and {Robertson}, Brant and {Willott}, Chris and {Witstok}, Joris},
        title = "{JADES: nitrogen enhancement in high-redshift broad-line active galactic nuclei}",
      journal = {\mnras},
         year = 2025,
        month = jul,
       volume = {541},
       number = {1},
        pages = {L71-L79},
          doi = {10.1093/mnrasl/slaf056},
archivePrefix = {arXiv},
       eprint = {2502.12091},
 primaryClass = {astro-ph.GA},
       adsurl = {https://ui.adsabs.harvard.edu/abs/2025MNRAS.541L..71I}
}

@ARTICLE{Ferrara2014,
       author = {{Ferrara}, A. and {Salvadori}, S. and {Yue}, B. and {Schleicher}, D.},
        title = "{Initial mass function of intermediate-mass black hole seeds}",
      journal = {\mnras},
         year = 2014,
        month = sep,
       volume = {443},
       number = {3},
        pages = {2410-2425},
          doi = {10.1093/mnras/stu1280},
archivePrefix = {arXiv},
       eprint = {1406.6685},
 primaryClass = {astro-ph.GA},
       adsurl = {https://ui.adsabs.harvard.edu/abs/2014MNRAS.443.2410F}
}

@ARTICLE{Latif2016,
       author = {{Latif}, Muhammad A. and {Ferrara}, Andrea},
        title = "{Formation of Supermassive Black Hole Seeds}",
      journal = {\pasa},
         year = 2016,
        month = oct,
       volume = {33},
          eid = {e051},
        pages = {e051},
          doi = {10.1017/pasa.2016.41},
archivePrefix = {arXiv},
       eprint = {1605.07391},
 primaryClass = {astro-ph.GA},
       adsurl = {https://ui.adsabs.harvard.edu/abs/2016PASA...33...51L}
}

@ARTICLE{Lin2026_variability,
       author = {{Lin}, Ruqiu and {Zheng}, Zhen-Ya and {Wang}, Junxian and {Ho}, Luis C. and {Zavala}, Jorge A. and {Zhang}, Zijian and {Jiang}, Chunyan and {Lin}, Jiaqi and {Yuan}, Fang-Ting and {Jiang}, Linhua and {Wang}, Tinggui and {Zhang}, Xiaer},
        title = "{Local Analogs of Little Red Dots: Optical Variability and Evidence for an AGN Origin}",
      journal = {arXiv e-prints},
         year = 2026,
        month = mar,
          eid = {arXiv:2603.01473},
        pages = {arXiv:2603.01473},
          doi = {10.48550/arXiv.2603.01473},
archivePrefix = {arXiv},
       eprint = {2603.01473},
 primaryClass = {astro-ph.GA},
       adsurl = {https://ui.adsabs.harvard.edu/abs/2026arXiv260301473L}
}

@ARTICLE{Scholtz2026_LRD,
       author = {{Scholtz}, J. and {D'Eugenio}, F. and {Maiolino}, R. and {Brazzini}, M. and {{\"U}bler}, H. and {Ji}, X. and {Perna}, M. and {Sun}, F. and {Brocchi}, G. and {Carniani}, S. and {Cresci}, G. and {Ivey}, L.~R. and {Juod{\v{z}}balis}, I. and {Marconi}, A. and {Mazzolari}, G. and {Risaliti}, G. and {Trefoloni}, B.},
        title = "{Little Red and Blue Dots: simply stratified Broad Line Regions}",
      journal = {arXiv e-prints},
         year = 2026,
        month = mar,
          eid = {arXiv:2603.22277},
        pages = {arXiv:2603.22277},
          doi = {10.48550/arXiv.2603.22277},
archivePrefix = {arXiv},
       eprint = {2603.22277},
 primaryClass = {astro-ph.GA},
       adsurl = {https://ui.adsabs.harvard.edu/abs/2026arXiv260322277S}
}

@ARTICLE{Lambrides2024,
       author = {{Lambrides}, Erini and {Garofali}, Kristen and {Larson}, Rebecca and {Ptak}, Andrew and {Chiaberge}, Marco and {Long}, Arianna S. and {Hutchison}, Taylor A. and {Norman}, Colin and {McKinney}, Jed and {Akins}, Hollis B. and {Berg}, Danielle A. and {Chisholm}, John and {Civano}, Francesca and {Cloonan}, Aidan P. and {Endsley}, Ryan and {Faisst}, Andreas L. and {Gilli}, Roberto and {Gillman}, Steven and {Hirschmann}, Michaela and {Kartaltepe}, Jeyhan S. and {Kocevski}, Dale D. and {Kokorev}, Vasily and {Pacucci}, Fabio and {Richardson}, Chris T. and {Stiavelli}, Massimo and {Whalen}, Kelly E.},
        title = "{The Case for Super-Eddington Accretion: Connecting Weak X-ray and UV Line Emission in JWST Broad-Line AGN During the First Gyr of Cosmic Time}",
      journal = {arXiv e-prints},
         year = 2024,
        month = sep,
          eid = {arXiv:2409.13047},
        pages = {arXiv:2409.13047},
          doi = {10.48550/arXiv.2409.13047},
archivePrefix = {arXiv},
       eprint = {2409.13047},
 primaryClass = {astro-ph.HE},
       adsurl = {https://ui.adsabs.harvard.edu/abs/2024arXiv240913047L}
}

@ARTICLE{Trefoloni2025,
       author = {{Trefoloni}, Bartolomeo and {Ji}, Xihan and {Maiolino}, Roberto and {D'Eugenio}, Francesco and {{\"U}bler}, Hannah and {Scholtz}, Jan and {Marconi}, Alessandro and {Marconcini}, Cosimo and {Mazzolari}, Giovanni},
        title = "{The missing Fe II bump in faint JWST active galactic nuclei: Possible evidence of metal-poor broad-line regions at early cosmic times}",
      journal = {\aap},
         year = 2025,
        month = aug,
       volume = {700},
          eid = {A203},
        pages = {A203},
          doi = {10.1051/0004-6361/202452795},
archivePrefix = {arXiv},
       eprint = {2410.21867},
 primaryClass = {astro-ph.GA},
       adsurl = {https://ui.adsabs.harvard.edu/abs/2025A&A...700A.203T}
}

@ARTICLE{PacucciLoeb2024,
       author = {{Pacucci}, Fabio and {Loeb}, Abraham},
        title = "{The Redshift Evolution of the M $_{{\textbullet}}${\textendash}M $_{{\ensuremath{\star}}}$ Relation for JWST's Supermassive Black Holes at z > 4}",
      journal = {\apj},
         year = 2024,
        month = apr,
       volume = {964},
       number = {2},
          eid = {154},
        pages = {154},
          doi = {10.3847/1538-4357/ad3044},
archivePrefix = {arXiv},
       eprint = {2401.04159},
 primaryClass = {astro-ph.GA},
       adsurl = {https://ui.adsabs.harvard.edu/abs/2024ApJ...964..154P}
}

@ARTICLE{JonesKocevski2025,
       author = {{Jones}, Brenda L. and {Kocevski}, Dale D. and {Pacucci}, Fabio and {Taylor}, Anthony J. and {Finkelstein}, Steven L. and {Buchner}, Johannes and {Trump}, Jonathan R. and {Somerville}, Rachel S. and {Hirschmann}, Michaela and {Yung}, L.~Y. Aaron and {Barro}, Guillermo and {Bell}, Eric F. and {Bisigello}, Laura and {Calabro}, Antonello and {Cleri}, Nikko J. and {Dekel}, Avishai and {Dickinson}, Mark and {Gandolfi}, Giovanni and {Giavalisco}, Mauro and {Grogin}, Norman A. and {Inayoshi}, Kohei and {Kartaltepe}, Jeyhan S. and {Koekemoer}, Anton M. and {Napolitano}, Lorenzo and {Onoue}, Masafusa and {Ravindranath}, Swara and {Rodighiero}, Giulia and {Wilkins}, Stephen M.},
        title = "{The $M_{\rm BH}-M_{*}$ Relationship at $3<z<7$: Big Black Holes in Little Red Dots}",
      journal = {arXiv e-prints},
         year = 2025,
        month = oct,
          eid = {arXiv:2510.07376},
        pages = {arXiv:2510.07376},
          doi = {10.48550/arXiv.2510.07376},
archivePrefix = {arXiv},
       eprint = {2510.07376},
 primaryClass = {astro-ph.GA},
       adsurl = {https://ui.adsabs.harvard.edu/abs/2025arXiv251007376J}
}

@ARTICLE{Pacucci2026,
       author = {{Pacucci}, Fabio and {Ferrara}, Andrea and {Kocevski}, Dale D.},
        title = "{The Little Red Dots Are Direct Collapse Black Holes}",
      journal = {arXiv e-prints},
         year = 2026,
        month = jan,
          eid = {arXiv:2601.14368},
        pages = {arXiv:2601.14368},
          doi = {10.48550/arXiv.2601.14368},
archivePrefix = {arXiv},
       eprint = {2601.14368},
 primaryClass = {astro-ph.GA},
       adsurl = {https://ui.adsabs.harvard.edu/abs/2026arXiv260114368P}
}

@ARTICLE{Natarajan2024,
       author = {{Natarajan}, Priyamvada and {Pacucci}, Fabio and {Ricarte}, Angelo and {Bogd{\'a}n}, {\'A}kos and {Goulding}, Andy D. and {Cappelluti}, Nico},
        title = "{First Detection of an Overmassive Black Hole Galaxy UHZ1: Evidence for Heavy Black Hole Seed Formation from Direct Collapse}",
      journal = {\apjl},
         year = 2024,
        month = jan,
       volume = {960},
       number = {1},
          eid = {L1},
        pages = {L1},
          doi = {10.3847/2041-8213/ad0e76},
archivePrefix = {arXiv},
       eprint = {2308.02654},
 primaryClass = {astro-ph.HE},
       adsurl = {https://ui.adsabs.harvard.edu/abs/2024ApJ...960L...1N}
}

@ARTICLE{Partmann2025,
       author = {{Partmann}, Christian and {Naab}, Thorsten and {Lah{\'e}n}, Natalia and {Rantala}, Antti and {Hirschmann}, Michaela and {Hislop}, Jessica M. and {Petersson}, Jonathan and {Johansson}, Peter H.},
        title = "{The importance of nuclear star clusters for massive black hole growth and nuclear star formation in simulated low-mass galaxies}",
      journal = {\mnras},
         year = 2025,
        month = feb,
       volume = {537},
       number = {2},
        pages = {956-977},
          doi = {10.1093/mnras/staf002},
archivePrefix = {arXiv},
       eprint = {2409.18096},
 primaryClass = {astro-ph.GA},
       adsurl = {https://ui.adsabs.harvard.edu/abs/2025MNRAS.537..956P}
}

@ARTICLE{Rantala2025,
       author = {{Rantala}, Antti and {Lah{\'e}n}, Natalia and {Naab}, Thorsten and {Escobar}, Gast{\'o}n J. and {Iorio}, Giuliano},
        title = "{FROST-CLUSTERS ─ II. Massive stars, binaries, and triples boost supermassive black hole seed formation in assembling star clusters}",
      journal = {\mnras},
         year = 2025,
        month = nov,
       volume = {543},
       number = {3},
        pages = {2130-2158},
          doi = {10.1093/mnras/staf1519},
archivePrefix = {arXiv},
       eprint = {2506.04330},
 primaryClass = {astro-ph.GA},
       adsurl = {https://ui.adsabs.harvard.edu/abs/2025MNRAS.543.2130R}
}

@ARTICLE{Martin1993,
       author = {{Martin}, W.~C. and {Kaufman}, Victor and {Musgrove}, Arlene},
        title = "{A Compilation of Energy Levels and Wavelengths for the Spectrum of Singly-Ionized Oxygen (O ii)}",
      journal = {Journal of Physical and Chemical Reference Data},
         year = 1993,
        month = sep,
       volume = {22},
       number = {5},
        pages = {1179-1212},
          doi = {10.1063/1.555928},
       adsurl = {https://ui.adsabs.harvard.edu/abs/1993JPCRD..22.1179M}
}

@ARTICLE{Moreschini2026,
       author = {{Moreschini}, Bianca and {Belfiore}, Francesco and {Marconi}, Alessandro and {Cataldi}, Elisa and {Curti}, Mirko and {Amiri}, Amirnezam and {Feltre}, Anna and {Mannucci}, Filippo and {Bertola}, Elena and {Bracci}, Caterina and {Ceci}, Matteo and {Chakraborty}, Avinanda and {Cresci}, Giovanni and {D'Amato}, Quirino and {di Teodoro}, Enrico and {Ginolfi}, Michele and {Lamperti}, Isabella and {Marconcini}, Cosimo and {Scialpi}, Martina and {Ulivi}, Lorenzo and {Vittoria Zanchettin}, Maria},
        title = "{One cloud is not enough: extreme conditions bias chemical abundances in high-redshift galaxies}",
      journal = {arXiv e-prints},
         year = 2026,
        month = jan,
          eid = {arXiv:2601.08939},
        pages = {arXiv:2601.08939},
          doi = {10.48550/arXiv.2601.08939},
archivePrefix = {arXiv},
       eprint = {2601.08939},
 primaryClass = {astro-ph.GA},
       adsurl = {https://ui.adsabs.harvard.edu/abs/2026arXiv260108939M}
}

@ARTICLE{Gordon2003,
       author = {{Gordon}, Karl D. and {Clayton}, Geoffrey C. and {Misselt}, K.~A. and {Landolt}, Arlo U. and {Wolff}, Michael J.},
        title = "{A Quantitative Comparison of the Small Magellanic Cloud, Large Magellanic Cloud, and Milky Way Ultraviolet to Near-Infrared Extinction Curves}",
      journal = {\apj},
         year = 2003,
        month = sep,
       volume = {594},
       number = {1},
        pages = {279-293},
          doi = {10.1086/376774},
archivePrefix = {arXiv},
       eprint = {astro-ph/0305257},
 primaryClass = {astro-ph},
       adsurl = {https://ui.adsabs.harvard.edu/abs/2003ApJ...594..279G}
}

@ARTICLE{Zu2025,
       author = {{Yan}, Zu and {Inayoshi}, Kohei and {Chen}, Kejian and {Guo}, Jingsong},
        title = "{Balmer Transition Signatures from Gas-Enshrouded, Dust-Poor Active Galactic Nuclei}",
      journal = {arXiv e-prints},
         year = 2025,
        month = dec,
          eid = {arXiv:2512.11050},
        pages = {arXiv:2512.11050},
          doi = {10.48550/arXiv.2512.11050},
archivePrefix = {arXiv},
       eprint = {2512.11050},
 primaryClass = {astro-ph.GA},
       adsurl = {https://ui.adsabs.harvard.edu/abs/2025arXiv251211050Y}
}

@ARTICLE{Wenaker1990,
       author = {{Wenaker}, Ingvar},
        title = "{The spectrum of singly ionized oxygen, O II}",
      journal = {\physscr},
         year = 1990,
        month = dec,
       volume = {42},
       number = {6},
        pages = {667-684},
          doi = {10.1088/0031-8949/42/6/008},
       adsurl = {https://ui.adsabs.harvard.edu/abs/1990PhyS...42..667W}
}

@ARTICLE{Belfiore2017,
       author = {{Belfiore}, Francesco and {Maiolino}, Roberto and {Tremonti}, Christy and {S{\'a}nchez}, Sebastian F. and {Bundy}, Kevin and {Bershady}, Matthew and {Westfall}, Kyle and {Lin}, Lihwai and {Drory}, Niv and {Boquien}, M{\'e}d{\'e}ric and {Thomas}, Daniel and {Brinkmann}, Jonathan},
        title = "{SDSS IV MaNGA - metallicity and nitrogen abundance gradients in local galaxies}",
      journal = {\mnras},
         year = 2017,
        month = jul,
       volume = {469},
       number = {1},
        pages = {151-170},
          doi = {10.1093/mnras/stx789},
archivePrefix = {arXiv},
       eprint = {1703.03813},
 primaryClass = {astro-ph.GA},
       adsurl = {https://ui.adsabs.harvard.edu/abs/2017MNRAS.469..151B}
}

@ARTICLE{Katz2022,
       author = {{Katz}, Harley and {Liu}, Shenghua and {Kimm}, Taysun and {Rey}, Martin P. and {Andersson}, Eric P. and {Cameron}, Alex J. and {Rodriguez-Montero}, Francisco and {Agertz}, Oscar and {Devriendt}, Julien and {Slyz}, Adrianne},
        title = "{PRISM: A Non-Equilibrium, Multiphase Interstellar Medium Model for Radiation Hydrodynamics Simulations of Galaxies}",
      journal = {arXiv e-prints},
         year = 2022,
        month = nov,
          eid = {arXiv:2211.04626},
        pages = {arXiv:2211.04626},
          doi = {10.48550/arXiv.2211.04626},
archivePrefix = {arXiv},
       eprint = {2211.04626},
 primaryClass = {astro-ph.GA},
       adsurl = {https://ui.adsabs.harvard.edu/abs/2022arXiv221104626K}
}

@ARTICLE{Dekel2023,
       author = {{Dekel}, Avishai and {Sarkar}, Kartick C. and {Birnboim}, Yuval and {Mandelker}, Nir and {Li}, Zhaozhou},
        title = "{Efficient formation of massive galaxies at cosmic dawn by feedback-free starbursts}",
      journal = {\mnras},
         year = 2023,
        month = aug,
       volume = {523},
       number = {3},
        pages = {3201-3218},
          doi = {10.1093/mnras/stad1557},
archivePrefix = {arXiv},
       eprint = {2303.04827},
 primaryClass = {astro-ph.GA},
       adsurl = {https://ui.adsabs.harvard.edu/abs/2023MNRAS.523.3201D}
}

@ARTICLE{Dekel2025_LRD,
       author = {{Dekel}, Avishai and {Dutta Chowdhury}, Dhruba and {Lapiner}, Sharon and {Yao}, Zhiyuan and {Gilbaum}, Shmuel and {Ceverino}, Daniel and {Primack}, Joel and {Somerville}, Rachel and {Teyssier}, Romain},
        title = "{From Feedback-Free Star Clusters to Little Red Dots via Compaction}",
      journal = {arXiv e-prints},
         year = 2025,
        month = nov,
          eid = {arXiv:2511.07578},
        pages = {arXiv:2511.07578},
          doi = {10.48550/arXiv.2511.07578},
archivePrefix = {arXiv},
       eprint = {2511.07578},
 primaryClass = {astro-ph.GA},
       adsurl = {https://ui.adsabs.harvard.edu/abs/2025arXiv251107578D}
}

@ARTICLE{Dekel2025,
       author = {{Dekel}, Avishai and {Stone}, Nicholas C. and {Chowdhury}, Dhruba Dutta and {Gilbaum}, Shmuel and {Li}, Zhaozhou and {Mandelker}, Nir and {van den Bosch}, Frank C.},
        title = "{Growth of massive black holes in FFB galaxies at cosmic dawn}",
      journal = {\aap},
         year = 2025,
        month = mar,
       volume = {695},
          eid = {A97},
        pages = {A97},
          doi = {10.1051/0004-6361/202452393},
archivePrefix = {arXiv},
       eprint = {2409.18605},
 primaryClass = {astro-ph.GA},
       adsurl = {https://ui.adsabs.harvard.edu/abs/2025A&A...695A..97D}
}

@ARTICLE{Nandal2025,
       author = {{Nandal}, Devesh and {Buldgen}, Ga{\"e}l and {Whalen}, Daniel J. and {Regan}, John and {Woods}, Tyrone E. and {Tan}, Jonathan C.},
        title = "{Rotating supermassive Pop III stars on the main sequence}",
      journal = {\aap},
         year = 2025,
        month = sep,
       volume = {701},
          eid = {A262},
        pages = {A262},
          doi = {10.1051/0004-6361/202555878},
archivePrefix = {arXiv},
       eprint = {2506.08268},
 primaryClass = {astro-ph.SR},
       adsurl = {https://ui.adsabs.harvard.edu/abs/2025A&A...701A.262N}
}

@ARTICLE{Prole2025,
       author = {{Prole}, Lewis R. and {Regan}, John A. and {Mehta}, Daxal and {Pakmor}, Rudiger and {Koudmani}, Sophie and {Bourne}, Martin A. and {Glover}, Simon C.~O. and {Wise}, John H. and {Klessen}, Ralf S. and {Tremmel}, Michael and {Sijacki}, Debora and {Beckmann}, Ricarda S. and {Haehnelt}, Martin G. and {Brennan}, John and {van de Bor}, Pelle and {Clark}, Paul C.},
        title = "{The SEEDZ Simulations: Methodology and First Results on Massive Black Hole Seeding and Early Galaxy Growth}",
      journal = {arXiv e-prints},
         year = 2025,
        month = nov,
          eid = {arXiv:2511.09640},
        pages = {arXiv:2511.09640},
          doi = {10.48550/arXiv.2511.09640},
archivePrefix = {arXiv},
       eprint = {2511.09640},
 primaryClass = {astro-ph.GA},
       adsurl = {https://ui.adsabs.harvard.edu/abs/2025arXiv251109640P}
}

@ARTICLE{Sanati2025,
       author = {{Sanati}, Mahsa and {Tan}, Jonathan C. and {Devriendt}, Julien and {Slyz}, Adrianne and {Martin-Alvarez}, Sergio and {la Torre}, Matteo and {Keller}, Benjamin and {Petkova}, Maya A. and {Monaco}, Pierluigi and {Cammelli}, Vieri and {Singh}, Jasbir and {Hayes}, Matthew},
        title = "{The emergence and ionizing feedback of Pop III.1 stars as progenitors for supermassive black holes}",
      journal = {\mnras},
         year = 2025,
        month = sep,
       volume = {542},
       number = {2},
        pages = {1532-1543},
          doi = {10.1093/mnras/staf1313},
archivePrefix = {arXiv},
       eprint = {2507.23004},
 primaryClass = {astro-ph.GA},
       adsurl = {https://ui.adsabs.harvard.edu/abs/2025MNRAS.542.1532S}
}

@ARTICLE{Cammelli2025,
       author = {{Cammelli}, Vieri and {Monaco}, Pierluigi and {Tan}, Jonathan C. and {Singh}, Jasbir and {Fontanot}, Fabio and {De Lucia}, Gabriella and {Hirschmann}, Michaela and {Xie}, Lizhi},
        title = "{The formation of supermassive black holes from Population III.1 seeds. III. Galaxy evolution and black hole growth from semi-analytic modelling}",
      journal = {\mnras},
         year = 2025,
        month = jan,
       volume = {536},
       number = {1},
        pages = {851-870},
          doi = {10.1093/mnras/stae2663},
archivePrefix = {arXiv},
       eprint = {2407.09949},
 primaryClass = {astro-ph.GA},
       adsurl = {https://ui.adsabs.harvard.edu/abs/2025MNRAS.536..851C}
}

@ARTICLE{Zhang2025_hydrosim_QSO1,
       author = {{Zhang}, Saiyang and {Liu}, Boyuan and {Bromm}, Volker and {K{\"u}hnel}, Florian},
        title = "{Primordial Black Holes as Seeds for Extremely Overmassive AGN Observed by JWST}",
      journal = {arXiv e-prints},
         year = 2025,
        month = dec,
          eid = {arXiv:2512.14066},
        pages = {arXiv:2512.14066},
          doi = {10.48550/arXiv.2512.14066},
archivePrefix = {arXiv},
       eprint = {2512.14066},
 primaryClass = {astro-ph.GA},
       adsurl = {https://ui.adsabs.harvard.edu/abs/2025arXiv251214066Z}
}

@ARTICLE{Zhang2025_hydrosim,
       author = {{Zhang}, Saiyang and {Liu}, Boyuan and {Bromm}, Volker and {Jeon}, Junehyoung and {Boylan-Kolchin}, Michael and {K{\"u}hnel}, Florian},
        title = "{How do Massive Primordial Black Holes Impact the Formation of the First Stars and Galaxies?}",
      journal = {\apj},
         year = 2025,
        month = jul,
       volume = {987},
       number = {2},
          eid = {185},
        pages = {185},
          doi = {10.3847/1538-4357/adddb4},
archivePrefix = {arXiv},
       eprint = {2503.17585},
 primaryClass = {astro-ph.GA},
       adsurl = {https://ui.adsabs.harvard.edu/abs/2025ApJ...987..185Z}
}

@ARTICLE{Dayal2024,
       author = {{Dayal}, Pratika},
        title = "{Exploring a primordial solution for early black holes detected with JWST}",
      journal = {\aap},
         year = 2024,
        month = oct,
       volume = {690},
          eid = {A182},
        pages = {A182},
          doi = {10.1051/0004-6361/202451481},
archivePrefix = {arXiv},
       eprint = {2407.07162},
 primaryClass = {astro-ph.GA},
       adsurl = {https://ui.adsabs.harvard.edu/abs/2024A&A...690A.182D}
}

@ARTICLE{DayalMaiolino2026,
       author = {{Dayal}, Pratika and {Maiolino}, Roberto},
        title = "{The properties of primordially-seeded black holes and their hosts in the first billion years: implications for JWST}",
      journal = {\aap},
         year = 2026,
        month = feb,
       volume = {706},
          eid = {A72},
        pages = {A72},
          doi = {10.1051/0004-6361/202555959},
archivePrefix = {arXiv},
       eprint = {2506.08116},
 primaryClass = {astro-ph.GA},
       adsurl = {https://ui.adsabs.harvard.edu/abs/2026A&A...706A..72D}
}

@ARTICLE{Ziparo2025,
       author = {{Ziparo}, F. and {Gallerani}, S. and {Ferrara}, A.},
        title = "{Primordial black holes as supermassive black hole seeds}",
      journal = {\jcap},
         year = 2025,
        month = apr,
       volume = {2025},
       number = {4},
          eid = {040},
        pages = {040},
          doi = {10.1088/1475-7516/2025/04/040},
archivePrefix = {arXiv},
       eprint = {2411.03448},
 primaryClass = {astro-ph.CO},
       adsurl = {https://ui.adsabs.harvard.edu/abs/2025JCAP...04..040Z}
}

@ARTICLE{Hviding2025,
       author = {{Hviding}, Raphael E. and {de Graaff}, Anna and {Miller}, Tim B. and {Setton}, David J. and {Greene}, Jenny E. and {Labb{\'e}}, Ivo and {Brammer}, Gabriel and {Bezanson}, Rachel and {Boogaard}, Leindert A. and {Cleri}, Nikko J. and {Leja}, Joel and {Maseda}, Michael V. and {McConachie}, Ian and {Matthee}, Jorryt and {Naidu}, Rohan P. and {Oesch}, Pascal A. and {Wang}, Bingjie and {Whitaker}, Katherine E. and {Williams}, Christina C.},
        title = "{RUBIES: A spectroscopic census of little red dots: All point sources with v-shaped continua have broad lines}",
      journal = {\aap},
         year = 2025,
        month = oct,
       volume = {702},
          eid = {A57},
        pages = {A57},
          doi = {10.1051/0004-6361/202555816},
archivePrefix = {arXiv},
       eprint = {2506.05459},
 primaryClass = {astro-ph.GA},
       adsurl = {https://ui.adsabs.harvard.edu/abs/2025A&A...702A..57H}
}

@ARTICLE{Setton2025,
       author = {{Setton}, David J. and {Greene}, Jenny E. and {de Graaff}, Anna and {Ma}, Yilun and {Leja}, Joel and {Matthee}, Jorryt and {Bezanson}, Rachel and {Boogaard}, Leindert A. and {Cleri}, Nikko J. and {Katz}, Harley and {Labbe}, Ivo and {Maseda}, Michael V. and {McConachie}, Ian and {Miller}, Tim B. and {Price}, Sedona H. and {Suess}, Katherine A. and {van Dokkum}, Pieter and {Wang}, Bingjie and {Weibel}, Andrea and {Whitaker}, Katherine E. and {Williams}, Christina C.},
        title = "{Little Red Dots at an Inflection Point: Ubiquitous V-shaped Turnover Consistently Occurs at the Balmer Limit}",
      journal = {\apj},
         year = 2025,
        month = dec,
       volume = {995},
       number = {1},
          eid = {118},
        pages = {118},
          doi = {10.3847/1538-4357/ae1500},
archivePrefix = {arXiv},
       eprint = {2411.03424},
 primaryClass = {astro-ph.GA},
       adsurl = {https://ui.adsabs.harvard.edu/abs/2025ApJ...995..118S}
}

@ARTICLE{Rusakov2026,
       author = {{Rusakov}, V. and {Watson}, D. and {Nikopoulos}, G.~P. and {Brammer}, G. and {Gottumukkala}, R. and {Harvey}, T. and {Heintz}, K.~E. and {Damgaard}, R. and {Sim}, S.~A. and {Sneppen}, A. and {Vijayan}, A.~P. and {Adams}, N. and {Austin}, D. and {Conselice}, C.~J. and {Goolsby}, C.~M. and {Toft}, S. and {Witstok}, J.},
        title = "{Little red dots as young supermassive black holes in dense ionized cocoons}",
      journal = {\nat},
         year = 2026,
        month = jan,
       volume = {649},
       number = {8097},
        pages = {574-579},
          doi = {10.1038/s41586-025-09900-4},
archivePrefix = {arXiv},
       eprint = {2503.16595},
 primaryClass = {astro-ph.GA},
       adsurl = {https://ui.adsabs.harvard.edu/abs/2026Natur.649..574R}
}

@ARTICLE{Juodzbalis2024_Rosetta,
       author = {{Juod{\v{z}}balis}, Ignas and {Ji}, Xihan and {Maiolino}, Roberto and {D'Eugenio}, Francesco and {Scholtz}, Jan and {Risaliti}, Guido and {Fabian}, Andrew C. and {Mazzolari}, Giovanni and {Gilli}, Roberto and {Prandoni}, Isabella and {Arribas}, Santiago and {Bunker}, Andrew J. and {Carniani}, Stefano and {Charlot}, St{\'e}phane and {Curtis-Lake}, Emma and {de Graaff}, Anna and {Hainline}, Kevin and {Parlanti}, Eleonora and {Perna}, Michele and {P{\'e}rez-Gonz{\'a}lez}, Pablo G. and {Robertson}, Brant and {Tacchella}, Sandro and {{\"U}bler}, Hannah and {Williams}, Christina C. and {Willott}, Chris and {Witstok}, Joris},
        title = "{JADES - the Rosetta stone of JWST-discovered AGN: deciphering the intriguing nature of early AGN}",
      journal = {\mnras},
         year = 2024,
        month = nov,
       volume = {535},
       number = {1},
        pages = {853-873},
          doi = {10.1093/mnras/stae2367},
archivePrefix = {arXiv},
       eprint = {2407.08643},
 primaryClass = {astro-ph.GA},
       adsurl = {https://ui.adsabs.harvard.edu/abs/2024MNRAS.535..853J}
}

@ARTICLE{Lin2026_COSMOS,
       author = {{Lin}, Xiaojing and {Fan}, Xiaohui and {Wang}, Feige and {Sun}, Fengwu and {Champagne}, Jaclyn B. and {Egami}, Eiichi and {Kakiichi}, Koki and {Lyu}, Jianwei and {Tee}, Wei Leong and {Yang}, Jinyi and {Bian}, Fuyan and {Bosman}, Sarah E.~I. and {Cai}, Zheng and {Casey}, Caitlin M. and {Decarli}, Roberto and {Faisst}, Andreas L. and {Finkelstein}, Steven L. and {Fujimoto}, Seiji and {Harish}, Santosh and {Ilbert}, Olivier and {Inoue}, Akio K. and {Jin}, Xiangyu and {Kartaltepe}, Jeyhan S. and {Kocevski}, Dale D. and {Li}, Mingyu and {Liu}, Weizhe and {Liu}, Yichen and {Schindler}, Jan-Torge and {Shuntov}, Marko and {Tanaka}, Takumi S. and {Vestergaard}, Marianne and {Wu}, Yunjing and {Zhang}, Haowen and {Zhang}, Zijian},
        title = "{Bridging Quasars and Little Red Dots: Insights into Broad-line Active Galactic Nuclei at z = 5─8 from the First JWST COSMOS-3D Dataset}",
      journal = {\apj},
         year = 2026,
        month = jan,
       volume = {996},
       number = {1},
          eid = {93},
        pages = {93},
          doi = {10.3847/1538-4357/ae1b9b},
archivePrefix = {arXiv},
       eprint = {2504.08039},
 primaryClass = {astro-ph.GA},
       adsurl = {https://ui.adsabs.harvard.edu/abs/2026ApJ...996...93L}
}

@ARTICLE{Zhang2025_LRDvariability,
       author = {{Zhang}, Zijian and {Li}, Mingyu and {Oguri}, Masamune and {Lin}, Xiaojing and {Inayoshi}, Kohei and {Cerny}, Catherine and {Coe}, Dan and {Diego}, Jose M. and {Fujimoto}, Seiji and {Jiang}, Linhua and {Mahler}, Guillaume and {Matthee}, Jorryt and {Naidu}, Rohan P. and {Sharon}, Keren and {Shen}, Yue and {Zitrin}, Adi and {Abdurro'uf} and {Akins}, Hollis and {Allingham}, Joseph F.~V. and {Amor{\'\i}n}, Ricardo and {Asada}, Yoshihisa and {Atek}, Hakim and {Bauer}, Franz E. and {Brada{\v{c}}}, Maru{\v{s}}a and {Bradley}, Larry D. and {Cai}, Zheng and {Cantalupo}, Sebastiano and {Conselice}, Christopher and {Dai}, Liang and {Dayal}, Pratika and {Egami}, Eiichi and {Eisenstein}, Daniel J. and {Faisst}, Andreas L. and {Fan}, Xiaohui and {Fei}, Qinyue and {Frye}, Brenda L. and {Fudamoto}, Yoshinobu and {Furtak}, Lukas J. and {Golubchik}, Miriam and {Gonz{\'a}lez-Otero}, Mauro and {Harikane}, Yuichi and {Hsiao}, Tiger Yu-Yang and {Jim{\'e}nez-Teja}, Yolanda and {Kartaltepe}, Jeyhan S. and {Kiyota}, Tomokazu and {Koekemoer}, Anton M. and {Kohno}, Kotaro and {Kokorev}, Vasily and {Kumari}, Nimisha and {Labbe}, Ivo and {Lagos}, Claudia D.~P. and {Larison}, Conor and {Liang}, Yongming and {Lucas}, Ray A. and {Lyu}, Jianwei and {Martis}, Nicholas S. and {Magdis}, Georgios E. and {Messa}, Matteo and {Nakane}, Minami and {Noirot}, Ga{\"e}l and {Ortiz}, III, Rafael and {Ouchi}, Masami and {Pierel}, Justin D.~R. and {Postman}, Marc and {Reddy}, Naveen and {Ricotti}, Massimo and {Schaerer}, Daniel and {Schneider}, Raffaella and {Steidel}, Charles C. and {Tee}, Wei Leong and {Tripodi}, Roberta and {Trussler}, James A.~A. and {Umeda}, Hiroya and {Valentino}, Francesco and {Vanzella}, Eros and {Wang}, Feige and {Windhorst}, Rogier and {Wu}, Yunjing and {Wu}, Zihao and {Yanagisawa}, Hiroto and {Yang}, Jinyi and {Sun}, Fengwu},
        title = "{Little red dot variability over a century reveals black hole envelope via a giant Einstein cross}",
      journal = {arXiv e-prints},
         year = 2025,
        month = dec,
          eid = {arXiv:2512.05180},
        pages = {arXiv:2512.05180},
          doi = {10.48550/arXiv.2512.05180},
archivePrefix = {arXiv},
       eprint = {2512.05180},
 primaryClass = {astro-ph.GA},
       adsurl = {https://ui.adsabs.harvard.edu/abs/2025arXiv251205180Z}
}

@ARTICLE{Baggen2024,
       author = {{Baggen}, Josephine F.~W. and {van Dokkum}, Pieter and {Brammer}, Gabriel and {de Graaff}, Anna and {Franx}, Marijn and {Greene}, Jenny and {Labb{\'e}}, Ivo and {Leja}, Joel and {Maseda}, Michael V. and {Nelson}, Erica J. and {Rix}, Hans-Walter and {Wang}, Bingjie and {Weibel}, Andrea},
        title = "{The Small Sizes and High Implied Densities of ``Little Red Dots'' with Balmer Breaks Could Explain Their Broad Emission Lines without an Active Galactic Nucleus}",
      journal = {\apjl},
         year = 2024,
        month = dec,
       volume = {977},
       number = {1},
          eid = {L13},
        pages = {L13},
          doi = {10.3847/2041-8213/ad90b8},
archivePrefix = {arXiv},
       eprint = {2408.07745},
 primaryClass = {astro-ph.GA},
       adsurl = {https://ui.adsabs.harvard.edu/abs/2024ApJ...977L..13B}
}

@ARTICLE{Chang2026,
       author = {{Chang}, Seok-Jun and {Gronke}, Max and {Matthee}, Jorryt and {Mason}, Charlotte},
        title = "{Impact of resonance, Raman, and Thomson scattering on hydrogen line formation in Little Red Dots}",
      journal = {\mnras},
         year = 2026,
        month = feb,
       volume = {545},
       number = {4},
          eid = {staf2131},
        pages = {staf2131},
          doi = {10.1093/mnras/staf2131},
archivePrefix = {arXiv},
       eprint = {2508.08768},
 primaryClass = {astro-ph.GA},
       adsurl = {https://ui.adsabs.harvard.edu/abs/2026MNRAS.545f2131C}
}

@ARTICLE{Wang2025_HeII,
       author = {{Wang}, Bingjie and {Leja}, Joel and {Katz}, Harley and {Inayoshi}, Kohei and {Cleri}, Nikko J. and {de Graaff}, Anna and {Hviding}, Raphael E. and {van Dokkum}, Pieter and {Greene}, Jenny E. and {Labb{\'e}}, Ivo and {Matthee}, Jorryt and {McConachie}, Ian and {Naidu}, Rohan P. and {Nelson}, Erica J.},
        title = "{The Missing Hard Photons of Little Red Dots: Their Incident Ionizing Spectra Resemble Massive Stars}",
      journal = {arXiv e-prints},
         year = 2025,
        month = aug,
          eid = {arXiv:2508.18358},
        pages = {arXiv:2508.18358},
          doi = {10.48550/arXiv.2508.18358},
archivePrefix = {arXiv},
       eprint = {2508.18358},
 primaryClass = {astro-ph.GA},
       adsurl = {https://ui.adsabs.harvard.edu/abs/2025arXiv250818358W}
}

@ARTICLE{Cameron2023,
       author = {{Cameron}, Alex J. and {Saxena}, Aayush and {Bunker}, Andrew J. and {D'Eugenio}, Francesco and {Carniani}, Stefano and {Maiolino}, Roberto and {Curtis-Lake}, Emma and {Ferruit}, Pierre and {Jakobsen}, Peter and {Arribas}, Santiago and {Bonaventura}, Nina and {Charlot}, Stephane and {Chevallard}, Jacopo and {Curti}, Mirko and {Looser}, Tobias J. and {Maseda}, Michael V. and {Rawle}, Tim and {Rodr{\'\i}guez Del Pino}, Bruno and {Smit}, Renske and {{\"U}bler}, Hannah and {Willott}, Chris and {Witstok}, Joris and {Egami}, Eiichi and {Eisenstein}, Daniel J. and {Johnson}, Benjamin D. and {Hainline}, Kevin and {Rieke}, Marcia and {Robertson}, Brant E. and {Stark}, Daniel P. and {Tacchella}, Sandro and {Williams}, Christina C. and {Willmer}, Christopher N.~A. and {Bhatawdekar}, Rachana and {Bowler}, Rebecca and {Boyett}, Kristan and {Circosta}, Chiara and {Helton}, Jakob M. and {Jones}, Gareth C. and {Kumari}, Nimisha and {Ji}, Zhiyuan and {Nelson}, Erica and {Parlanti}, Eleonora and {Sandles}, Lester and {Scholtz}, Jan and {Sun}, Fengwu},
        title = "{JADES: Probing interstellar medium conditions at z {\ensuremath{\sim}} 5.5-9.5 with ultra-deep JWST/NIRSpec spectroscopy}",
      journal = {\aap},
         year = 2023,
        month = sep,
       volume = {677},
          eid = {A115},
        pages = {A115},
          doi = {10.1051/0004-6361/202346107},
archivePrefix = {arXiv},
       eprint = {2302.04298},
 primaryClass = {astro-ph.GA},
       adsurl = {https://ui.adsabs.harvard.edu/abs/2023A&A...677A.115C}
}

@ARTICLE{Wang2025,
       author = {{Wang}, Bingjie and {de Graaff}, Anna and {Davies}, Rebecca L. and {Greene}, Jenny E. and {Leja}, Joel and {Brammer}, Gabriel B. and {Goulding}, Andy D. and {Miller}, Tim B. and {Suess}, Katherine A. and {Weibel}, Andrea and {Williams}, Christina C. and {Bezanson}, Rachel and {Boogaard}, Leindert A. and {Cleri}, Nikko J. and {Hirschmann}, Michaela and {Katz}, Harley and {Labb{\'e}}, Ivo and {Maseda}, Michael V. and {Matthee}, Jorryt and {McConachie}, Ian and {Naidu}, Rohan P. and {Oesch}, Pascal A. and {Rix}, Hans-Walter and {Setton}, David J. and {Whitaker}, Katherine E.},
        title = "{RUBIES: JWST/NIRSpec Confirmation of an Infrared-luminous, Broad-line Little Red Dot with an Ionized Outflow}",
      journal = {\apj},
         year = 2025,
        month = may,
       volume = {984},
       number = {2},
          eid = {121},
        pages = {121},
          doi = {10.3847/1538-4357/adc1ca},
archivePrefix = {arXiv},
       eprint = {2403.02304},
 primaryClass = {astro-ph.GA},
       adsurl = {https://ui.adsabs.harvard.edu/abs/2025ApJ...984..121W}
}

@ARTICLE{McClymont2026_BH,
       author = {{McClymont}, William and {Tacchella}, Sandro and {Ji}, Xihan and {Kannan}, Rahul and {Maiolino}, Roberto and {Simmonds}, Charlotte and {Smith}, Aaron and {Puchwein}, Ewald and {Garaldi}, Enrico and {Vogelsberger}, Mark and {D'Eugenio}, Francesco and {Keating}, Laura and {Shen}, Xuejian and {Trefoloni}, Bartolomeo and {Zier}, Oliver},
        title = "{Overmassive black holes in the early Universe can be explained by gas-rich, dark matter-dominated galaxies}",
      journal = {\mnras},
         year = 2026,
        month = jan,
       volume = {545},
       number = {1},
          eid = {staf2092},
        pages = {staf2092},
          doi = {10.1093/mnras/staf2092},
archivePrefix = {arXiv},
       eprint = {2506.13852},
 primaryClass = {astro-ph.GA},
       adsurl = {https://ui.adsabs.harvard.edu/abs/2026MNRAS.545f2092M}
}

@ARTICLE{Loiacono2025,
       author = {{Loiacono}, Federica and {Gilli}, Roberto and {Mignoli}, Marco and {Mazzolari}, Giovanni and {Decarli}, Roberto and {Brusa}, Marcella and {Calura}, Francesco and {Chiaberge}, Marco and {Comastri}, Andrea and {D'Amato}, Quirino and {Iwasawa}, Kazushi and {Juod{\v{z}}balis}, Ignas and {Lanzuisi}, Giorgio and {Maiolino}, Roberto and {Marchesi}, Stefano and {Norman}, Colin and {Peca}, Alessandro and {Prandoni}, Isabella and {Sapori}, Matteo and {Signorini}, Matilde and {Tozzi}, Paolo and {Vanzella}, Eros and {Vignali}, Cristian and {Vito}, Fabio and {Zamorani}, Gianni},
        title = "{A big red dot at cosmic noon}",
      journal = {\aap},
         year = 2025,
        month = oct,
       volume = {703},
          eid = {A36},
        pages = {A36},
          doi = {10.1051/0004-6361/202555946},
archivePrefix = {arXiv},
       eprint = {2506.12141},
 primaryClass = {astro-ph.GA},
       adsurl = {https://ui.adsabs.harvard.edu/abs/2025A&A...703A..36L}
}

@ARTICLE{EuclidCollaboration2025,
       author = {{Euclid Collaboration} and {Bisigello}, L. and {Rodighiero}, G. and {Fotopoulou}, S. and {Ricci}, F. and {Jahnke}, K. and {Feltre}, A. and {Allevato}, V. and {Shankar}, F. and {Cassata}, P. and {Dalla Bont{\`a}}, E. and {Gandolfi}, G. and {Girardi}, G. and {Giulietti}, M. and {Grazian}, A. and {Lovell}, C.~C. and {Maiolino}, R. and {Matamoro Zatarain}, T. and {Mezcua}, M. and {Prandoni}, I. and {Roberts}, D. and {Roster}, W. and {Salvato}, M. and {Siudek}, M. and {Tarsitano}, F. and {Toba}, Y. and {Vietri}, A. and {Wang}, L. and {Zamorani}, G. and {Baes}, M. and {Belladitta}, S. and {Nersesian}, A. and {Spinoglio}, L. and {Lopez Lopez}, X. and {Aghanim}, N. and {Altieri}, B. and {Amara}, A. and {Andreon}, S. and {Auricchio}, N. and {Aussel}, H. and {Baccigalupi}, C. and {Baldi}, M. and {Balestra}, A. and {Bardelli}, S. and {Basset}, A. and {Battaglia}, P. and {Bender}, R. and {Biviano}, A. and {Bonchi}, A. and {Branchini}, E. and {Brescia}, M. and {Brinchmann}, J. and {Camera}, S. and {Ca{\~n}as-Herrera}, G. and {Capobianco}, V. and {Carbone}, C. and {Carretero}, J. and {Casas}, S. and {Castellano}, M. and {Castignani}, G. and {Cavuoti}, S. and {Chambers}, K.~C. and {Cimatti}, A. and {Colodro-Conde}, C. and {Congedo}, G. and {Conselice}, C.~J. and {Conversi}, L. and {Copin}, Y. and {Courbin}, F. and {Courtois}, H.~M. and {Cropper}, M. and {Da Silva}, A. and {Degaudenzi}, H. and {De Lucia}, G. and {Di Giorgio}, A.~M. and {Dolding}, C. and {Dole}, H. and {Dubath}, F. and {Duncan}, C.~A.~J. and {Dupac}, X. and {Dusini}, S. and {Ealet}, A. and {Escoffier}, S. and {Farina}, M. and {Farinelli}, R. and {Faustini}, F. and {Ferriol}, S. and {Finelli}, F. and {Frailis}, M. and {Franceschi}, E. and {Galeotta}, S. and {George}, K. and {Gillard}, W. and {Gillis}, B. and {Giocoli}, C. and {G{\'o}mez-Alvarez}, P. and {Gracia-Carpio}, J. and {Granett}, B.~R. and {Grupp}, F. and {Gwyn}, S. and {Haugan}, S.~V.~H. and {Hoekstra}, H. and {Holmes}, W. and {Hook}, I.~M. and {Hormuth}, F. and {Hornstrup}, A. and {Hudelot}, P. and {Jhabvala}, M. and {Keih{\"a}nen}, E. and {Kermiche}, S. and {Kiessling}, A. and {Kubik}, B. and {K{\"u}mmel}, M. and {Kunz}, M. and {Kurki-Suonio}, H. and {Le Boulc'h}, Q. and {Le Brun}, A.~M.~C. and {Le Mignant}, D. and {Liebing}, P. and {Ligori}, S. and {Lilje}, P.~B. and {Lindholm}, V. and {Lloro}, I. and {Mainetti}, G. and {Maino}, D. and {Maiorano}, E. and {Mansutti}, O. and {Marcin}, S. and {Marggraf}, O. and {Martinelli}, M. and {Martinet}, N. and {Marulli}, F. and {Massey}, R. and {Maurogordato}, S. and {Medinaceli}, E. and {Mei}, S. and {Melchior}, M. and {Mellier}, Y. and {Meneghetti}, M. and {Merlin}, E. and {Meylan}, G. and {Mora}, A. and {Moresco}, M. and {Moscardini}, L. and {Nakajima}, R. and {Neissner}, C. and {Niemi}, S.-M. and {Nightingale}, J.~W. and {Padilla}, C. and {Paltani}, S. and {Pasian}, F. and {Pedersen}, K. and {Percival}, W.~J. and {Pettorino}, V. and {Pires}, S. and {Polenta}, G. and {Poncet}, M. and {Popa}, L.~A. and {Pozzetti}, L. and {Raison}, F. and {Rebolo}, R. and {Renzi}, A. and {Rhodes}, J. and {Riccio}, G. and {Romelli}, E. and {Roncarelli}, M. and {Rossetti}, E. and {Rottgering}, H.~J.~A. and {Rusholme}, B. and {Saglia}, R. and {Sakr}, Z. and {Sapone}, D. and {Sartoris}, B. and {Schewtschenko}, J.~A. and {Schirmer}, M. and {Schneider}, P. and {Schrabback}, T. and {Scodeggio}, M. and {Secroun}, A. and {Seidel}, G. and {Serrano}, S. and {Simon}, P. and {Sirignano}, C. and {Sirri}, G. and {Stanco}, L. and {Steinwagner}, J. and {Tallada-Cresp{\'\i}}, P. and {Taylor}, A.~N. and {Teplitz}, H.~I. and {Tereno}, I. and {Toft}, S. and {Toledo-Moreo}, R. and {Torradeflot}, F. and {Tutusaus}, I. and {Valenziano}, L. and {Valiviita}, J. and {Vassallo}, T. and {Verdoes Kleijn}, G. and {Veropalumbo}, A. and {Wang}, Y.},
        title = "{Euclid Quick Data Release (Q1). Extending the quest for little red dots to z<4}",
      journal = {arXiv e-prints},
         year = 2025,
        month = mar,
          eid = {arXiv:2503.15323},
        pages = {arXiv:2503.15323},
          doi = {10.48550/arXiv.2503.15323},
archivePrefix = {arXiv},
       eprint = {2503.15323},
 primaryClass = {astro-ph.GA},
       adsurl = {https://ui.adsabs.harvard.edu/abs/2025arXiv250315323E}
}

@ARTICLE{Juodzbalis2025_QSO1,
       author = {{Juod{\v{z}}balis}, Ignas and {Marconcini}, Cosimo and {D'Eugenio}, Francesco and {Maiolino}, Roberto and {Marconi}, Alessandro and {{\"U}bler}, Hannah and {Scholtz}, Jan and {Ji}, Xihan and {Arribas}, Santiago and {Bennett}, Jake S. and {Bromm}, Volker and {Bunker}, Andrew J. and {Carniani}, Stefano and {Charlot}, St{\'e}phane and {Cresci}, Giovanni and {Dayal}, Pratika and {Egami}, Eiichi and {Fabian}, Andrew and {Inayoshi}, Kohei and {Isobe}, Yuki and {Ivey}, Lucy and {Jones}, Gareth C. and {Koudmani}, Sophie and {Laporte}, Nicolas and {Liu}, Boyuan and {Lyu}, Jianwei and {Mazzolari}, Giovanni and {Monty}, Stephanie and {Parlanti}, Eleonora and {P{\'e}rez-Gonz{\'a}lez}, Pablo G. and {Perna}, Michele and {Robertson}, Brant and {Schneider}, Raffaella and {Sijacki}, Debora and {Tacchella}, Sandro and {Trinca}, Alessandro and {Valiante}, Rosa and {Volonteri}, Marta and {Witstok}, Joris and {Zhang}, Saiyang},
        title = "{A direct black hole mass measurement in a Little Red Dot at the Epoch of Reionization}",
      journal = {arXiv e-prints},
         year = 2025,
        month = aug,
          eid = {arXiv:2508.21748},
        pages = {arXiv:2508.21748},
          doi = {10.48550/arXiv.2508.21748},
archivePrefix = {arXiv},
       eprint = {2508.21748},
 primaryClass = {astro-ph.GA},
       adsurl = {https://ui.adsabs.harvard.edu/abs/2025arXiv250821748J}
}

@ARTICLE{Begelman2026,
       author = {{Begelman}, Mitchell C. and {Dexter}, Jason},
        title = "{Little Red Dots as Late-stage Quasi-stars}",
      journal = {\apj},
         year = 2026,
        month = jan,
       volume = {996},
       number = {1},
          eid = {48},
        pages = {48},
          doi = {10.3847/1538-4357/ae274a},
archivePrefix = {arXiv},
       eprint = {2507.09085},
 primaryClass = {astro-ph.GA},
       adsurl = {https://ui.adsabs.harvard.edu/abs/2026ApJ...996...48B}
}

@ARTICLE{Liu2025,
       author = {{Liu}, Hanpu and {Jiang}, Yan-Fei and {Quataert}, Eliot and {Greene}, Jenny E. and {Ma}, Yilun},
        title = "{The Balmer Break and Optical Continuum of Little Red Dots from Super-Eddington Accretion}",
      journal = {\apj},
         year = 2025,
        month = nov,
       volume = {994},
       number = {1},
          eid = {113},
        pages = {113},
          doi = {10.3847/1538-4357/ae0c19},
archivePrefix = {arXiv},
       eprint = {2507.07190},
 primaryClass = {astro-ph.GA},
       adsurl = {https://ui.adsabs.harvard.edu/abs/2025ApJ...994..113L}
}

@ARTICLE{Maiolino2025_QSO1,
       author = {{Maiolino}, Roberto and {Uebler}, Hannah and {D'Eugenio}, Francesco and {Scholtz}, Jan and {Juodzbalis}, Ignas and {Ji}, Xihan and {Perna}, Michele and {Bromm}, Volker and {Dayal}, Pratika and {Koudmani}, Sophie and {Liu}, Boyuan and {Schneider}, Raffaella and {Sijacki}, Debora and {Valiante}, Rosa and {Trinca}, Alessandro and {Zhang}, Saiyang and {Volonteri}, Marta and {Inayoshi}, Kohei and {Carniani}, Stefano and {Nakajima}, Kimihiko and {Isobe}, Yuki and {Witstok}, Joris and {Jones}, Gareth C. and {Tacchella}, Sandro and {Arribas}, Santiago and {Bunker}, Andrew and {Cataldi}, Elisa and {Charlot}, Stephane and {Cresci}, Giovanni and {Curti}, Mirko and {Fabian}, Andrew C. and {Katz}, Harley and {Kumari}, Nimisha and {Laporte}, Nicolas and {Mazzolari}, Giovanni and {Robertson}, Brant and {Sun}, Fengwu and {Rodriguez Del Pino}, Bruno and {Venturi}, Giacomo},
        title = "{A black hole in a near-pristine galaxy 700 million years after the Big Bang}",
      journal = {arXiv e-prints},
         year = 2025,
        month = may,
          eid = {arXiv:2505.22567},
        pages = {arXiv:2505.22567},
          doi = {10.48550/arXiv.2505.22567},
archivePrefix = {arXiv},
       eprint = {2505.22567},
 primaryClass = {astro-ph.GA},
       adsurl = {https://ui.adsabs.harvard.edu/abs/2025arXiv250522567M}
}

@software{Newville2016_LMFIT,
       author = {{Newville}, Matthew and {Stensitzki}, Till and {Allen}, Daniel B. and {Rawlik}, Michal and {Ingargiola}, Antonino and {Nelson}, Andrew},
        title = "{Lmfit: Non-Linear Least-Square Minimization and Curve-Fitting for Python}",
 howpublished = {Astrophysics Source Code Library, record ascl:1606.014},
         year = 2016,
        month = jun,
          eid = {ascl:1606.014},
archivePrefix = {ascl},
       eprint = {1606.014},
       adsurl = {https://ui.adsabs.harvard.edu/abs/2016ascl.soft06014N}
}

@ARTICLE{Luridiana2015_PyNeb,
       author = {{Luridiana}, V. and {Morisset}, C. and {Shaw}, R.~A.},
        title = "{PyNeb: a new tool for analyzing emission lines. I. Code description and validation of results}",
      journal = {\aap},
         year = 2015,
        month = jan,
       volume = {573},
          eid = {A42},
        pages = {A42},
          doi = {10.1051/0004-6361/201323152},
archivePrefix = {arXiv},
       eprint = {1410.6662},
 primaryClass = {astro-ph.IM},
       adsurl = {https://ui.adsabs.harvard.edu/abs/2015A&A...573A..42L}
}

@ARTICLE{emcee2013,
       author = {{Foreman-Mackey}, Daniel and {Hogg}, David W. and {Lang}, Dustin and {Goodman}, Jonathan},
        title = "{emcee: The MCMC Hammer}",
      journal = {\pasp},
         year = 2013,
        month = mar,
       volume = {125},
       number = {925},
        pages = {306},
          doi = {10.1086/670067},
archivePrefix = {arXiv},
       eprint = {1202.3665},
 primaryClass = {astro-ph.IM},
       adsurl = {https://ui.adsabs.harvard.edu/abs/2013PASP..125..306F}
}

@ARTICLE{Jakobsen2022,
       author = {{Jakobsen}, P. and {Ferruit}, P. and {Alves de Oliveira}, C. and {Arribas}, S. and {Bagnasco}, G. and {Barho}, R. and {Beck}, T.~L. and {Birkmann}, S. and {B{\"o}ker}, T. and {Bunker}, A.~J. and {Charlot}, S. and {de Jong}, P. and {de Marchi}, G. and {Ehrenwinkler}, R. and {Falcolini}, M. and {Fels}, R. and {Franx}, M. and {Franz}, D. and {Funke}, M. and {Giardino}, G. and {Gnata}, X. and {Holota}, W. and {Honnen}, K. and {Jensen}, P.~L. and {Jentsch}, M. and {Johnson}, T. and {Jollet}, D. and {Karl}, H. and {Kling}, G. and {K{\"o}hler}, J. and {Kolm}, M.-G. and {Kumari}, N. and {Lander}, M.~E. and {Lemke}, R. and {L{\'o}pez-Caniego}, M. and {L{\"u}tzgendorf}, N. and {Maiolino}, R. and {Manjavacas}, E. and {Marston}, A. and {Maschmann}, M. and {Maurer}, R. and {Messerschmidt}, B. and {Moseley}, S.~H. and {Mosner}, P. and {Mott}, D.~B. and {Muzerolle}, J. and {Pirzkal}, N. and {Pittet}, J.-F. and {Plitzke}, A. and {Posselt}, W. and {Rapp}, B. and {Rauscher}, B.~J. and {Rawle}, T. and {Rix}, H.-W. and {R{\"o}del}, A. and {Rumler}, P. and {Sabbi}, E. and {Salvignol}, J.-C. and {Schmid}, T. and {Sirianni}, M. and {Smith}, C. and {Strada}, P. and {te Plate}, M. and {Valenti}, J. and {Wettemann}, T. and {Wiehe}, T. and {Wiesmayer}, M. and {Willott}, C.~J. and {Wright}, R. and {Zeidler}, P. and {Zincke}, C.},
        title = "{The Near-Infrared Spectrograph (NIRSpec) on the James Webb Space Telescope. I. Overview of the instrument and its capabilities}",
      journal = {\aap},
         year = 2022,
        month = may,
       volume = {661},
          eid = {A80},
        pages = {A80},
          doi = {10.1051/0004-6361/202142663},
archivePrefix = {arXiv},
       eprint = {2202.03305},
 primaryClass = {astro-ph.IM},
       adsurl = {https://ui.adsabs.harvard.edu/abs/2022A&A...661A..80J}
}

@ARTICLE{Dojcinovic2023,
       author = {{Doj{\v{c}}inovi{\'c}}, Ivan and {Kova{\v{c}}evi{\'c}-Doj{\v{c}}inovi{\'c}}, Jelena and {Popovi{\'c}}, Luka {\v{C}}.},
        title = "{The flux ratio of the [N II] {\ensuremath{\lambda}}{\ensuremath{\lambda}} 6548, 6583 {\r{A}} lines in sample of Active Galactic Nuclei Type 2}",
      journal = {Advances in Space Research},
         year = 2023,
        month = jan,
       volume = {71},
       number = {2},
        pages = {1219-1226},
          doi = {10.1016/j.asr.2022.04.041},
archivePrefix = {arXiv},
       eprint = {2204.10036},
 primaryClass = {astro-ph.GA},
       adsurl = {https://ui.adsabs.harvard.edu/abs/2023AdSpR..71.1219D}
}

@ARTICLE{DallaBonta2025,
       author = {{Dalla Bont{\`a}}, E. and {Peterson}, B.~M. and {Grier}, C.~J. and {Berton}, M. and {Brandt}, W.~N. and {Ciroi}, S. and {Corsini}, E.~M. and {Dalla Barba}, B. and {Davies}, R. and {Dehghanian}, M. and {Edelson}, R. and {Foschini}, L. and {Gasparri}, D. and {Ho}, L.~C. and {Horne}, K. and {Iodice}, E. and {Morelli}, L. and {Pizzella}, A. and {Portaluri}, E. and {Shen}, Y. and {Schneider}, D.~P. and {Vestergaard}, M.},
        title = "{Estimating masses of supermassive black holes in active galactic nuclei from the H{\ensuremath{\alpha}} emission line}",
      journal = {\aap},
         year = 2025,
        month = apr,
       volume = {696},
          eid = {A48},
        pages = {A48},
          doi = {10.1051/0004-6361/202452746},
archivePrefix = {arXiv},
       eprint = {2410.21387},
 primaryClass = {astro-ph.GA},
       adsurl = {https://ui.adsabs.harvard.edu/abs/2025A&A...696A..48D}
}

@ARTICLE{Curti2020,
       author = {{Curti}, Mirko and {Mannucci}, Filippo and {Cresci}, Giovanni and {Maiolino}, Roberto},
        title = "{The mass-metallicity and the fundamental metallicity relation revisited on a fully T$_{e}$-based abundance scale for galaxies}",
      journal = {\mnras},
         year = 2020,
        month = jan,
       volume = {491},
       number = {1},
        pages = {944-964},
          doi = {10.1093/mnras/stz2910},
archivePrefix = {arXiv},
       eprint = {1910.00597},
 primaryClass = {astro-ph.GA},
       adsurl = {https://ui.adsabs.harvard.edu/abs/2020MNRAS.491..944C}
}




\appendix

\section{Modelling the NIRSpec/IFS Point Spread Function}\label{sec:IFSPSF}

\begin{figure}
\centering
  {\phantomsubcaption\label{fig:psfstar.a}
   \phantomsubcaption\label{fig:psfstar.b}
   \phantomsubcaption\label{fig:psfstar.c}
   \phantomsubcaption\label{fig:psfstar.d}
   \phantomsubcaption\label{fig:psfstar.e}
   \phantomsubcaption\label{fig:psfstar.f}
   \phantomsubcaption\label{fig:psfstar.g}
   \phantomsubcaption\label{fig:psfstar.h}
   \phantomsubcaption\label{fig:psfstar.i}
   \phantomsubcaption\label{fig:psfstar.j}}
   \includegraphics[width=\linewidth]{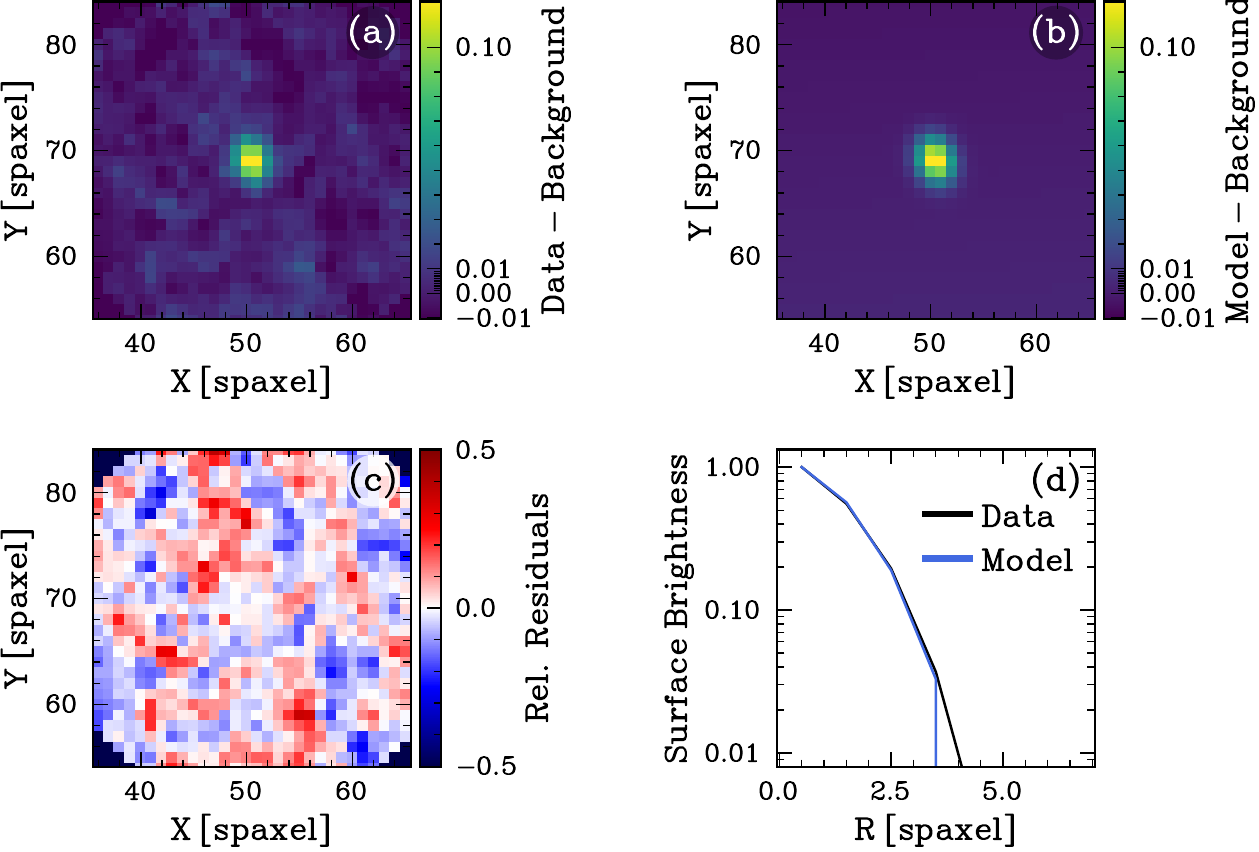}
\caption{Modelling the star near \cliff with \textsc{pysersic}. From left to right, from top to bottom, the panels show the data, best-fit model, relative residuals, and the circularised radial profile, using asinh scaling, to highlight the PSF wings. See Fig.~\ref{fig:surfacebrightnessprofiles} for a comparison of the radial profile of this best-fit model to the radial profile of various emission lines of \cliff.}\label{fig:psfstar}
\end{figure}

Our investigation of \cliff exploits the presence of a star within the instrument FoV (see Fig.~\ref{fig:fov}). To provide independent validation of the radial surface brightness profiles presented in Fig.~\ref{fig:surfacebrightnessprofiles}, we create a synthetic median image of the star in a 0.1-\textmu m spectral window centred on \cliff's \OIIIL emission (Fig.~\ref{fig:psfstar.a}). We then model this image with the Bayesian modelling tool \textsc{pysersic} \citep{PashaMiller2023}, with the setup outlined in \citet{DEugenio2026}, and using a S\'ersic profile with fixed index $n=0.5$ (i.e. representing a Gaussian). We tested a range of $n$ values and convolution methods to assess the stability of this model.

Our results are presented in Fig.~\ref{fig:psfstar}, where we show the PSF model (panel~\subref{fig:psfstar.b}), the relative residuals (panel~\subref{fig:psfstar.c}), and the circularised radial profile (panel~\subref{fig:psfstar.d}).

\section{Modelling the NIRSpec/IFS instrument gradient}\label{sec:instgrad}

\begin{figure}
\centering
   \includegraphics[width=\linewidth]{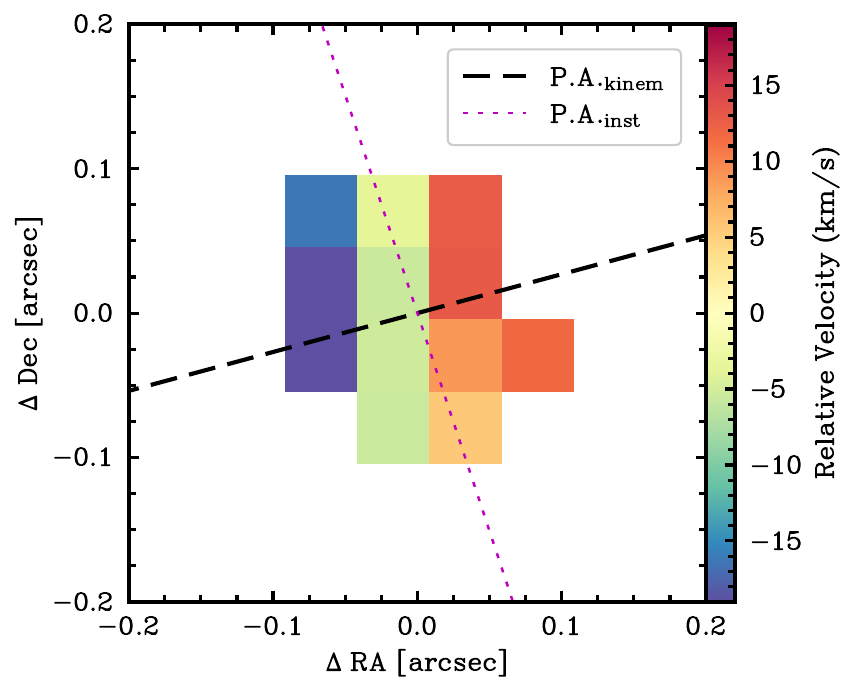}
\caption{Narrow-line \OIIIL relative velocity map of \cliff, presented with a SNR cut of 5. The best-fit kinematic PA is shown by the black dashed line, while the instrumental PA is shown by the dotted magenta line. Even though the observed velocity gradient is perpendicular to the slices (as expected from an instrument artefact), its sign is opposite to the instrumental effect we infer from modelling the interloper \neigh.}
\label{fig:cliffkinem}
\end{figure}

\begin{figure}
\centering
   \includegraphics[width=\linewidth]{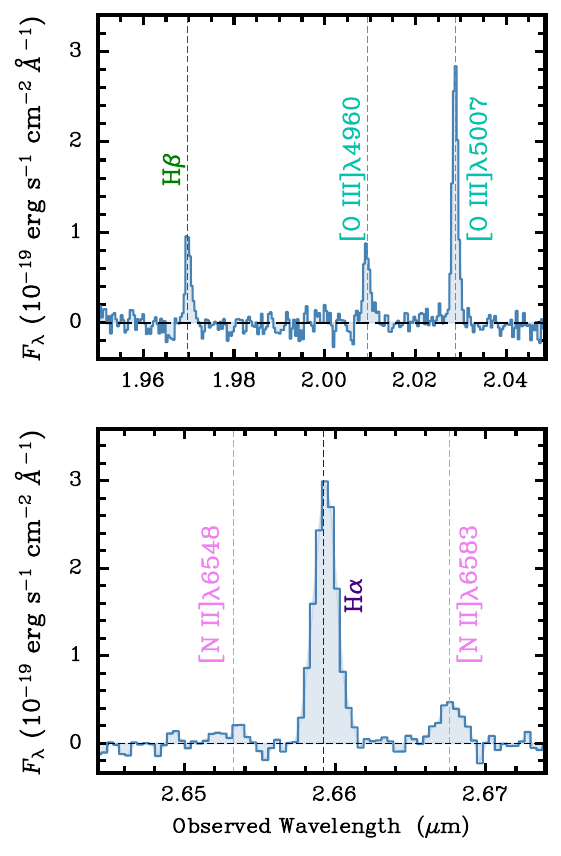}
\caption{Portions of integrated spectra extracted from a $0.15''$ radius circular aperture centred on \neigh (see Fig.~\ref{fig:fov}), showing emission lines detected in this foreground galaxy. \textbf{Top panel:} The \Hbeta-\OIIIL spectral region. \textbf{Bottom panel:} The \Halpha spectral region, also showing a detection of \NIIall.}
\label{fig:foregroundlines}
\end{figure}

At the time of writing this article, reconstructed NIRSpec/IFS datacubes are known to display an artificial velocity gradient\footnote{for more detail, see \hyperlink{https://jwst-docs.stsci.edu/known-issues-with-jwst-data/nirspec-known-issues/nirspec-ifu-known-issues/}{https://jwst-docs.stsci.edu/known-issues-with-jwst-data/nirspec-known-issues/nirspec-ifu-known-issues/} and e.g. \citealt{Pascalau2026_gradient}.}. This gradient is thought to be introduced by slice-to-slice astrometric drift, therefore any kinematic signal aligned with the NIRSpec position angle (P.A.) must be interpreted with caution \citep[as discussed in e.g.,][]{Juodzbalis2025_QSO1}.
\cliff itself was observed with a NIRSpec P.A. of 198.31\textdegree, and we measure a velocity gradient of -15~\kms along the east--west direction, i.e. extremely close to the instrument P.A. (Fig.~\ref{fig:cliffkinem}).
Given this alignment and the compactness of \cliff, there is therefore a non-negligible chance that the observed gradient may be artificial.

To assess this possibility, we exploit the serendipitous presence of \neigh, an inclined galaxy within the NIRSpec FoV (Fig.~\ref{fig:fov}).
This source displays strong \Hbeta, \OIIIall, \Halpha and \NIIall emission (see Fig.~\ref{fig:foregroundlines}), and is therefore suitable for kinematic modelling.
We here present a simple model to assess whether the velocity gradient observed in \cliff is artificial and if galaxy kinematics can be used to constrain the artificial gradient. For this demonstration we combine tilted-disc kinematics, an arctan rotation curve and a constant instrument gradient; we do not take into account the non-uniform light profile of \neigh, nor the instrument PSF. While both effects are important, their careful modelling is beyond the exploratory scope of this appendix.

\begin{figure}
\centering
  {\phantomsubcaption\label{fig:grado3.a}
   \phantomsubcaption\label{fig:grado3.b}
   \phantomsubcaption\label{fig:grado3.c}
   \phantomsubcaption\label{fig:grado3.d}
   \phantomsubcaption\label{fig:grado3.e}
   \phantomsubcaption\label{fig:grado3.f}
   \phantomsubcaption\label{fig:grado3.g}
   \phantomsubcaption\label{fig:grado3.h}
   \phantomsubcaption\label{fig:grado3.i}}
   \includegraphics[width=\linewidth]{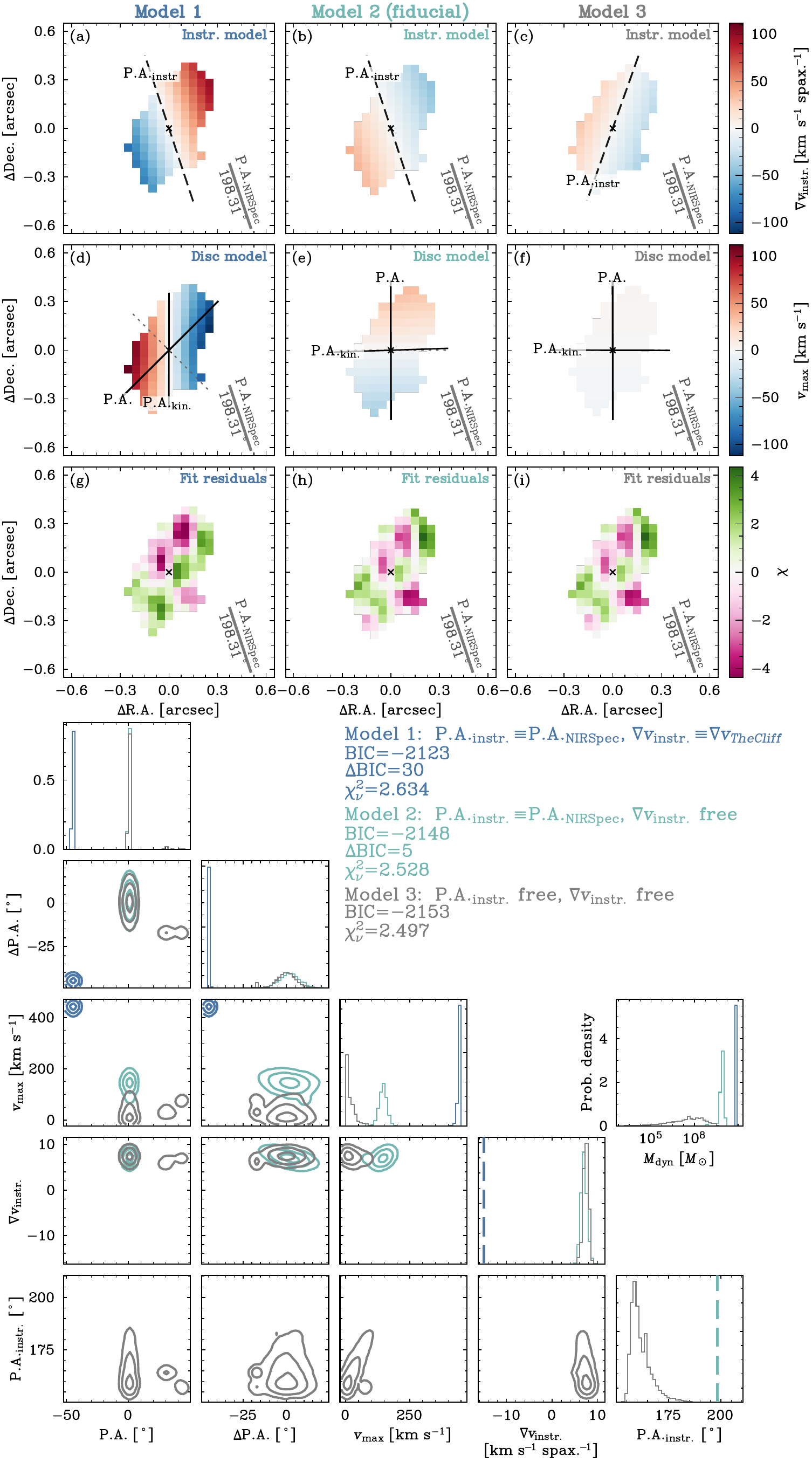}
\caption{Comparison of the fiducial model of the artificial NIRSpec/IFS velocity gradient (Model~2) to the model where the instrument gradient is fixed to what measured in \cliff (Model~1), and to a model where both the gradient and its direction are free (Model~3). Model~2 is the only one that explains the observations, while also yielding a physically plausible intrinsic velocity for the foreground galaxy \neigh (note the rapidly rotating and kinematically twisted map of Model~1, and the non-rotating map of Model~3). Model~2 favours an instrument gradient opposite to what measured in \cliff (panel~\subref{fig:grado3.b}), implying an even stronger velocity gradient than the measured -15~\kms shown in Fig.~\ref{fig:cliffkinem}.}
\label{fig:grado3}
\end{figure}

We consider a fourteen parameter model: seven parameters describe the emission-line map (centre, total flux, half-light semi-major axis, projected axis ratio, S\'ersic index, and position angle), five are for the galaxy kinematics
(kinematic position angle, inclination, intrinsic axis ratio, maximum velocity $v_\mathrm{max}$, and turnover radius), and two for the instrument (i.e. the magnitude and position angle of the artificial velocity gradient, $\nabla v_\mathrm{instr.}$ and P.A.$_\mathrm{instr.}$, respectively). We consider three models, Models~1--3 (Fig.~\ref{fig:grado3}), which differ only in whether these last two parameters are fixed or left free.

We adopt a strong Gaussian probability prior over the kinematic position angle; the mean and standard deviation of $\Delta\,\mathrm{P.A.}\equiv\mathrm{P.A.}_\mathrm{kin.} - \mathrm{P.A.}$ are 0\textdegree\ and 5\textdegree, favouring solutions that are aligned to the morphological minor axis -- as appropriate for a rotating disc. Similarly, we strongly limit the intrinsic axis ratio, constraining in the range 0.05--$q$, where $q$ is the observed axis ratio.
The model is optimised jointly to reproduce the flux and velocity maps, using the \textsc{emcee} MCMC integrator \citep{emcee2013}. Below we discuss only the \OIII-based inference, but we note that \Halpha yields a consistent picture (and statistically consistent results for $\nabla v_\mathrm{instr.}$), while \Hbeta suffers from larger uncertainties, and other lines such as \Hgamma and \NII are too noisy to draw any conclusions.

\begin{figure}
\centering
  {\phantomsubcaption\label{fig:mapo3.a}
   \phantomsubcaption\label{fig:mapo3.b}
   \phantomsubcaption\label{fig:mapo3.c}
   \phantomsubcaption\label{fig:mapo3.d}
   \phantomsubcaption\label{fig:mapo3.e}
   \phantomsubcaption\label{fig:mapo3.f}}
   \includegraphics[width=\linewidth]{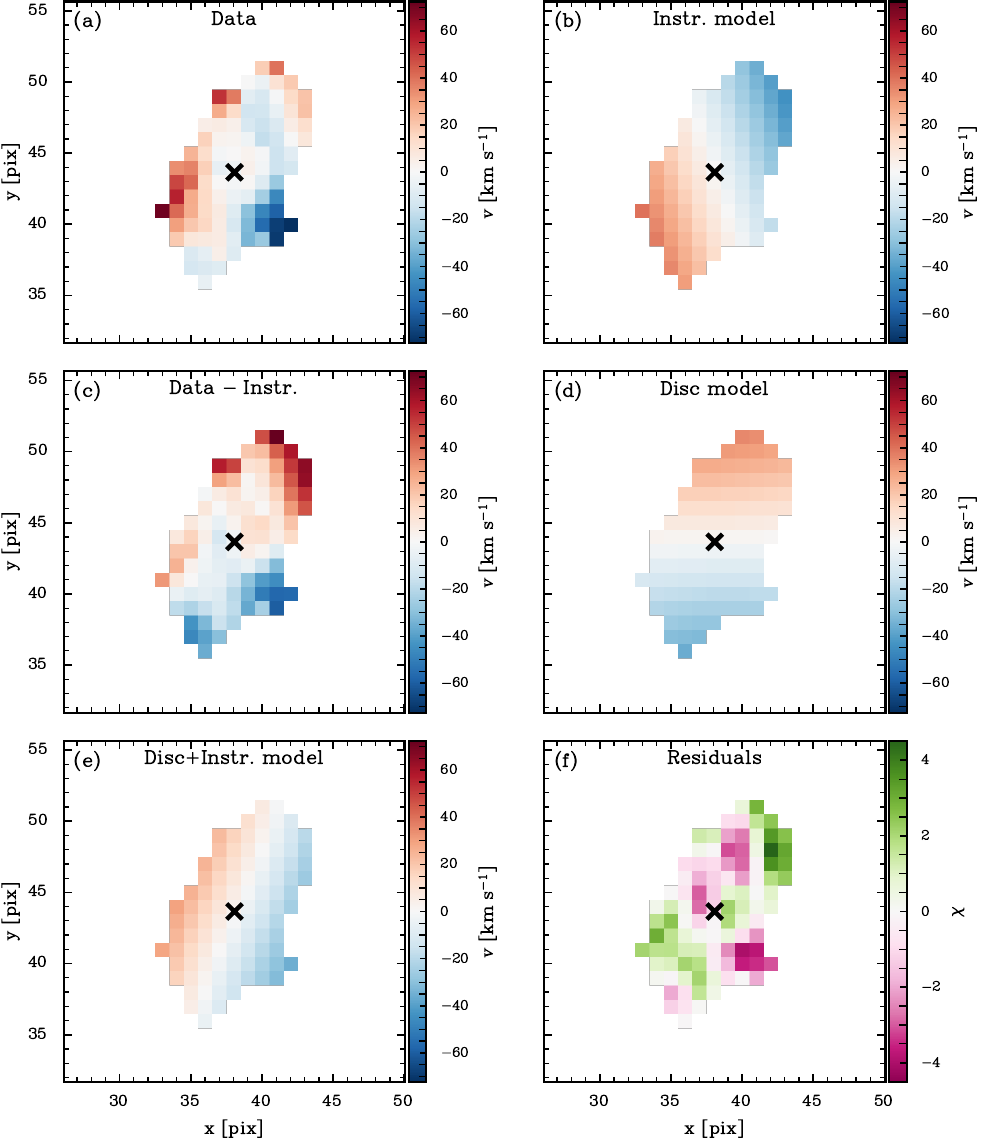}
\caption{The fiducial joint fit of the foreground galaxy \neigh (panel~\subref{fig:mapo3.a}) using a constant instrument gradient 
(panel~\subref{fig:mapo3.b}) plus a standard inclined-disc model.
Middle row shows the data after removing the best-fit instrument model
(panel~\subref{fig:mapo3.c}), which should match the best-fit disc model (panel~\subref{fig:mapo3.d}). Panel~\subref{fig:mapo3.e}) shows the best-fit model of the data (i.e., panels~\subref{fig:mapo3.b}+\subref{fig:mapo3.d}) and the relative residuals (panel~\subref{fig:mapo3.f}).
Note how removing the instrument gradient mitigates the complexity of the observed velocity map. The best-fit instrument gradient has similar orientation but opposite sign to that measured in \cliff (Fig.~\ref{fig:cliffkinem}).}\label{fig:mapo3}
\end{figure}

In the fiducial model, Model~2, we fix the gradient direction to the NIRSpec P.A., 198.31\textdegree.
The resulting kinematic maps are shown in
Fig.~\ref{fig:mapo3}. The galaxy model displays disc-like rotation as appropriate for an inclined disc, which is plausible for a galaxy of its mass and redshift. After adding the best-fit instrument gradient, the resulting model kinematics no longer resemble a disc, but match better the `tumbling' pattern seen in the observations (panels~\subref{fig:mapo3.a} and~\subref{fig:mapo3.e} of Fig.~\ref{fig:mapo3}).
Surprisingly, the instrument gradient of $\nabla v_\mathrm{instr}=+11\pm1~\kms~\mathrm{spaxel}^{-1}$ is opposite to what observed in \cliff (cf. Fig.~\ref{fig:mapo3.b} and~\ref{fig:cliffkinem}).

To validate this result, we compare it to the other two models. In Model~1 we fix both P.A.$_\mathrm{instr.}=198.31$\textdegree, and $\nabla v_\mathrm{instr}=-15$~\kms~spaxel$^{-1}$ (i.e. \cliff value; Fig.~\ref{fig:cliffkinem}).
By comparing this model to the fiducial Model~2 (Fig.~\ref{fig:grado3}), we can reject the hypothesis that \cliff's observed velocity gradient is fully due to the
NIRSpec instrument: Model~1 has considerably worse $\chi^2_\nu$ and $\Delta\text{BIC}$ than Model~2, displays poor convergence, and the best-fit galaxy model is an implausible combination of a rapidly rotating disc ($v_\mathrm{max}=446_{-8}^{+3}~\kms$; reaching the upper bounds of the probability prior) with strong kinematic asymmetry ($\Delta\,\text{P.A.}=-62\pm2$\textdegree, a 6 $\sigma$ deviation from the prior).
In Model~3 we free both $\nabla v_\mathrm{instr.}$ and P.A.$_\mathrm{instr.}$. 
This model finds $\nabla_\mathrm{instr} = +7.7\pm0.5$~\kms~spaxel$^{-1}$ and P.A.$_\mathrm{instr}=159_{-3}^{+5}$\textdegree, different from the NIRSpec P.A. (8~$\sigma$).
While statistically consistent with the data (and achieving $\Delta\,\text{BIC}=-9$ and $-1$ for \OIII and \Halpha, respectively), Model~3 is optimised by collapsing $v_\mathrm{max}$ to only $7_{-6}^{+17}$~\kms, which we deem unlikely for a galaxy with $\log(M_\ast/\Msun)=9.6\pm0.1$ \citep{Muzzin2025}; indeed, the inferred rotation is so small that the implied virial mass is over 30 times smaller than $M_\ast$ (Fig.~\ref{fig:grado3}, bottom panels).
More to the point, the magnitude and orientation of the NIRSpec gradient inferred from Model~3 are still opposite to what seen in \cliff, supporting the hypothesis that the observed gradient may be not only real, but actually even stronger than what is measured. 
Still, substantial uncertainties remain. 
Aside from the limits of our toy galaxy model, the instrument gradient may vary spatially, and in any case a single disc galaxy may be out of stationary equilibrium, as evidenced by residual tumbling motions even after subtracting the instrument model (e.g., Fig.~\ref{fig:mapo3.c}). 
Before drawing scientific conclusions from applying our proposed correction, more observations and exploring more complex models are required.

\section{Aperture Loss Correction Factors}\label{sec:ALC}

\begin{figure}
\centering
   \includegraphics[width=\linewidth]{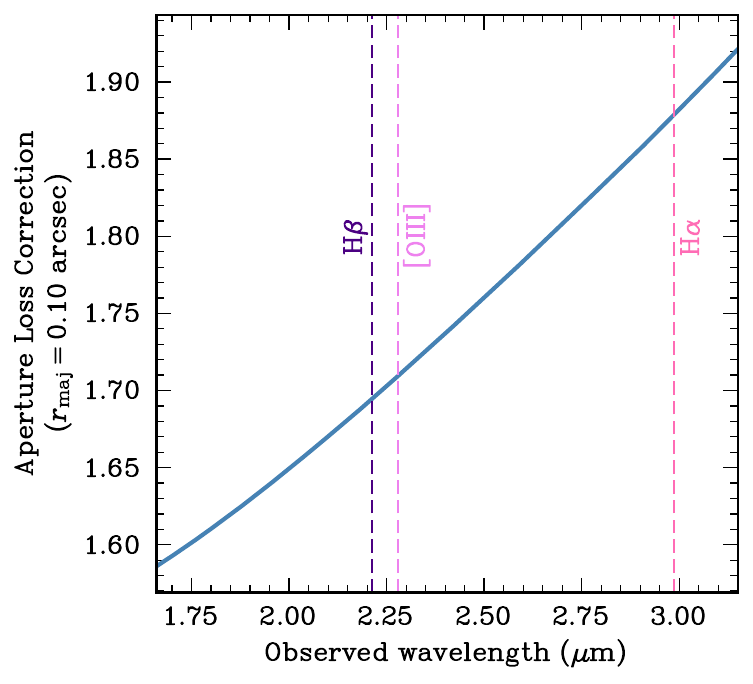}
\caption{Aperture loss correction factors as function of observed wavelength for the aperture shown in Fig.~\ref{fig:aperture}. This was derived using the full PSF model from \textsc{stpsf}, rather than a Gaussian approximation. Values have been rescaled to match our empirically measured ALC factor for \Halpha found through a curve of growth analysis. The wavelengths of some key emission lines have been indicated with vertical dashed lines and labelled.}
\label{fig:ALC}
\end{figure}

In Fig.~\ref{fig:ALC} we present the curve used to calculate ALC factors for each emission line in this study, based on a \Halpha curve of growth analysis as discussed in Section~\ref{sec:Size}. 
We derive this from the empirical ALC curve calculated from the PSF derived from \textsc{stpsf}\footnote{\hyperlink{https://stpsf.readthedocs.io/en/latest/}{https://stpsf.readthedocs.io/en/latest/}; the successor to \textsc{webbpsf} \citep{Perrin2015}. \textsc{stpsf} accounts for many properties of the NIRSpec instrument (e.g. geometric distortion, detector charge transfer effects), and therefore may be used to derive physical models of PSF wavelength dependence.} for our aperture (Fig.~\ref{fig:aperture}), which is rescaled to match the empirically measured \Halpha ALC factor of $1.88\pm0.05$. 
For further detail regarding ALCs derived from the PSF, see \cite{Jones2026}.

\section{Another argument against high density}\label{sec:density2}

\begin{figure}
\centering
   \includegraphics[width=\linewidth]{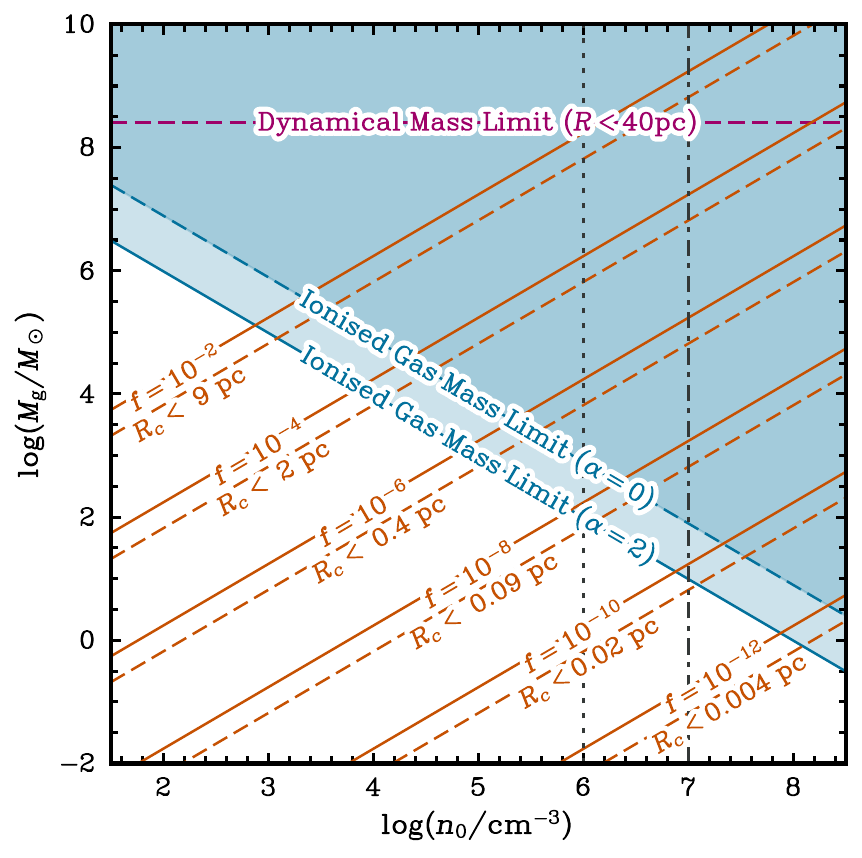}
\caption{Constraints on gas density for narrow-line emitting gas. The orange lines indicate the gas mass $M_\mathrm{g}$ enclosed within the aperture (Fig.~\ref{fig:aperture}) as a function of gas density $n_0$ at $R = 40$ pc, for a power-law radial distribution (Equation~\ref{eq:densprof})  with indices $\alpha = 2$ (solid) and $\alpha = 0$ (dashed). Each line is labelled by its corresponding filling factor and implied upper limit on single-cloud size. The blue lines indicate the mass of ionised gas constrained by the $\Hbeta_\mathrm{N}$ luminosity for the two power indices. The vertical dotted line indicates the density where collisional de-excitation of \OIIIL begins to affect its flux; the vertical dot-dashed line indicates the density required to suppress \OIIIL flux by a factor of 10. Explaining the weakness of \OIIIL via collisional de-excitation would require extremely low filling factors and small cloud sizes. The horizontal pink line indicates the upper limit on dynamical mass derived in Section~\ref{sec:M_dyn}.}
\label{fig:supp_dens_constratint}
\end{figure}

In this section, we provide a supplemental argument against high density as an explanation for \OIIIL weakness, as outlined for QSO1 in \cite{Maiolino2025_QSO1}.

Firstly, we assume the density follows a radial distribution described by a power law,
\begin{equation}
n(r) = n_0\left(\frac{r}{r_0}\right)^{-\alpha},
\label{eq:densprof}
\end{equation}
where $n_0$ is the density at some reference radius $r_0$. Here we take $r_0 = 40$ pc, i.e. the effective radius of \cliff as inferred by \cite{deGraaff2025}; this is smaller than the aperture described in Section~\ref{sec:Size}, but a smaller $r_0$ is a more conservative assumption for obtaining a bound on the gas density. Furthermore, we assume that the clouds within this density profile have a filling factor $f$ defined by
\begin{equation}
    f = \frac{V_\mathrm{clouds}}{V_\mathrm{total}} = \frac{N_\mathrm{clouds}R_\mathrm{c}^3}{R^3_\mathrm{max}},
    \label{eq:fillfactor}
\end{equation}
where $V_\mathrm{total}$ is the total volume enclosed by the aperture, assumed spherical with radius $R_\mathrm{max}$. $V_\mathrm{clouds}$ is the volume occupied by ionised gas clouds of radius $R_\mathrm{c}$, and $N_\mathrm{clouds}$ is the number of clouds. Assuming a medium comprising only hydrogen for simplicity, the mass of gas ($M_\mathrm{g}$) contained within the aperture is therefore given by
\begin{equation}
    M_\mathrm{g} = \int_{R_\mathrm{min}}^{R_\mathrm{max}}4\pi fm_\mathrm{p}n(r)r^2 \ dr,
    \label{eq:gasmass}
\end{equation}
where $m_\mathrm{p}$ is the proton mass. We adopt $R_\mathrm{min} = 5 \ \mathrm{pc}$, as smaller radii would be entirely within the black hole sphere of influence and would therefore result in broader line emission\footnote{As in \cite{Maiolino2025_QSO1}, we note that adopting a smaller inner radius would simply strengthen the results.}. Hence, we plot the inferred gas mass as a function of density for different filling factors (Fig.~\ref{fig:supp_dens_constratint}), and determine the implied maximum cloud size (determined by assuming the extreme, very conservative case of $N_\mathrm{cloud} = 1$) associated with each filling factor. In reality, not all of the gas would be contained within a single cloud, so the clouds would be much smaller.

We can compare the above estimate with the ionised gas mass, obtained from the luminosity of $\Hbeta_\mathrm{N}$
\begin{equation}
    L(\Hbeta_\mathrm{N}) = \int_{R_\mathrm{min}}^{R_\mathrm{max}}4\pi f\varepsilon(T)n(r)^2r^2 \ dr,
\end{equation}
where $\varepsilon(T)$ is the temperature-dependent emissivity of narrow \Hbeta. The total mass of ionised gas can then be inferred from the measured $L(\Hbeta_\mathrm{N})$, and is given by
\begin{equation}
    \frac{M_\mathrm{ion}}{M_\odot} \approx 10^9 \frac{L(\Hbeta_\mathrm{N})}{10^{43} \ \mathrm{erg} \ \mathrm{s}^{-1}} \frac{100 \ \mathrm{cm}^{-3}}{n_0} \frac{3-2\alpha}{3-\alpha} \frac{R^{3-\alpha}_\mathrm{max}-R^{3-\alpha}_\mathrm{min}}{R^{3-2\alpha}_\mathrm{max}-R^{3-2\alpha}_\mathrm{min}} r_0^{-\alpha}, 
    \label{eq:iongasmass}
\end{equation}
where we have assumed a typical temperature $T = 10^4 \ \mathrm{K}$.

Assuming $\alpha=2$, in order for the gas to have a density greater than $10^6~\mathrm{cm}^{-3}$, at which \OIIIL starts to be collisionally suppressed by a factor of 1.5, the ionised clouds would have extremely small filling factors ($f \lesssim 10^{-8} \ N_\mathrm{clouds}^{-1}$), and sizes $<0.09$ pc.
Such extreme filling factors are unrealistic, as no gas phase is known to have such extreme values, especially the ionized phase (not even for the BLR of AGN). Even an extreme scenario with $\alpha=0$, i.e. uniform filling, would still yield $f \lesssim 10^{-7}\ N_\mathrm{clouds}^{-1}$. Adopting the same \textsc{Cloudy} models as \cite{Maiolino2025_QSO1}, we infer that to suppress \OIIIL by a factor of ten, the ISM would need to reach densities $>10^7 \ \mathrm{cm}^{-3}$. This would necessitate even lower filling factors and yet smaller clouds.
We also note that such high density would imply an extremely low mass of ionized gas, less than $10~M_\odot$.

\section{The Aesopica hydrodynamical simulations}\label{sec:sims}

\textsc{Aesopica} is a new suite of large-volume cosmological simulations (Koudmani et al., in prep), built upon the \textsc{Fable} galaxy formation model \citep{Henden2018}; the \textsc{Fable} sub-grid models are themselves largely based on the Illustris galaxy formation model \citep{Vogelsberger2014}. In \textsc{Fable}, some aspects are unchanged from Illustris, namely the models for star formation \citep{SpringelHernquist2003}, radiative cooling \citep{Katz1996, Wiersma2009a} and chemical enrichment \citep{Wiersma2009b}. However, the models of stellar feedback \citep{Volgerssberger2013} and AGN feedback \citep{Sijacki2015} have been updated, incorporating thermal stellar feedback and AGN duty cycles as part of the simulations.

\textsc{Aesopica} introduces targeted updates for modelling the growth of infant SMBHs in the early Universe. In particular, \textsc{Aesopica} explores three key modifications to fiducial galaxy formation models: enabling efficient accretion in the low-mass regime \citep{Koudmani2022}, incorporating super-Eddington accretion, and examining a broad range of seed masses ($10^2\  M_\odot$ to $10^5\ M_\odot$) following seed evolution from early cosmic epochs ($z \sim 20$); while this is earlier than the standard threshold in \textsc{Fable}, we note that seeding could occur even earlier.

\section{Spectral Fitting Posteriors}\label{sec:fit_posteriors}
\begin{figure*}
\centering
   \includegraphics[width=\linewidth]{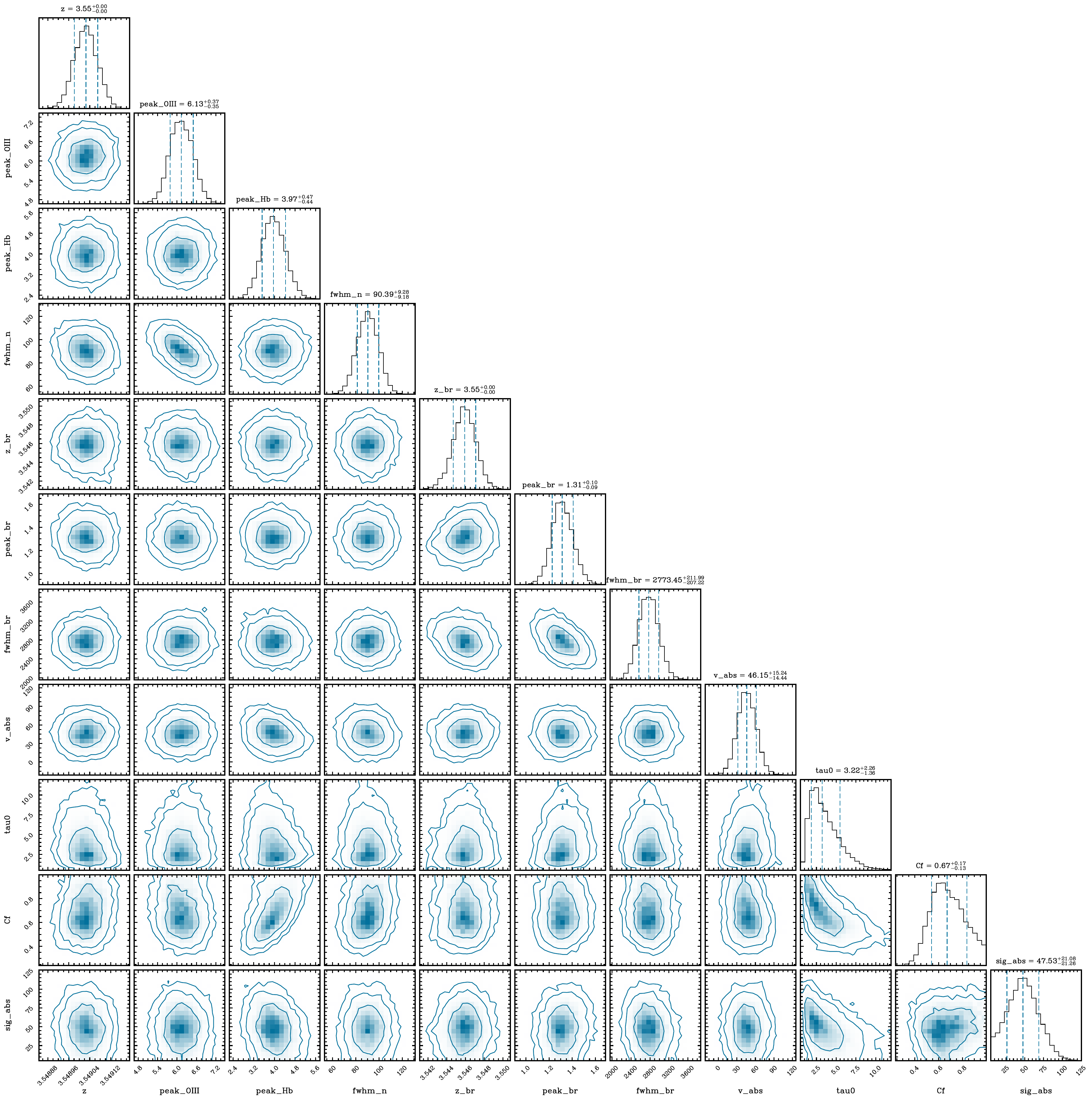}
\caption{Posterior distributions of the fitted parameters in the \Hbeta-\OIIIL model, including the best fit values and their associated errors. Contours enclose 68\%, 95\%, and 99\% of the distributions.}
\label{fig:Hb_OIII_posteriors}
\end{figure*}

\begin{figure*}
\centering
   \includegraphics[width=\linewidth]{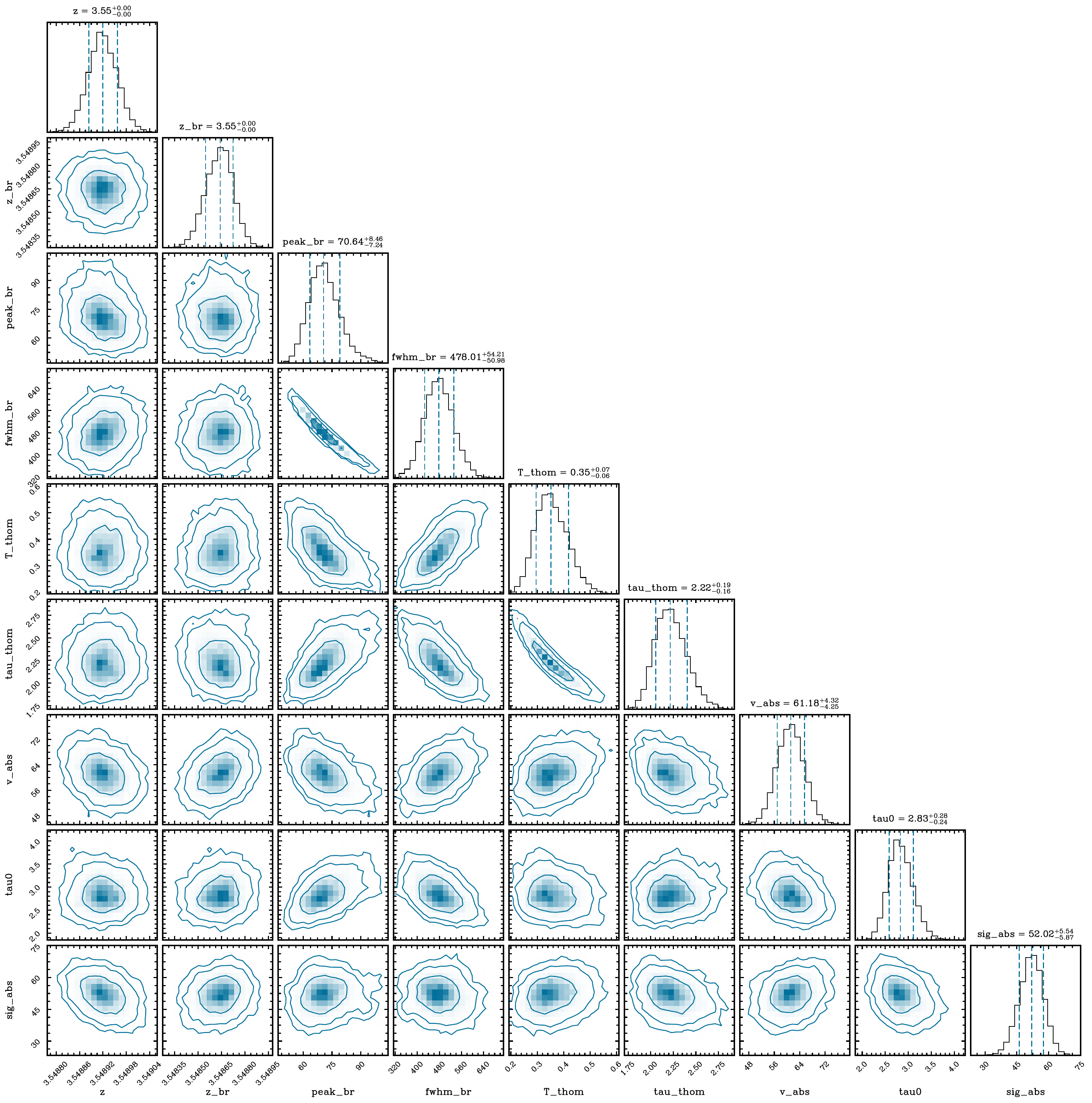}
\caption{Posterior distributions of the fitted parameters in the \textit{Fiducial} \Halpha model, including the best fit values and their associated errors. Contours enclose 68\%, 95\%, and 99\% of the distributions.}
\label{fig:Ha_fiducial_posteriors}
\end{figure*}

\begin{figure*}
\centering
   \includegraphics[width=\linewidth]{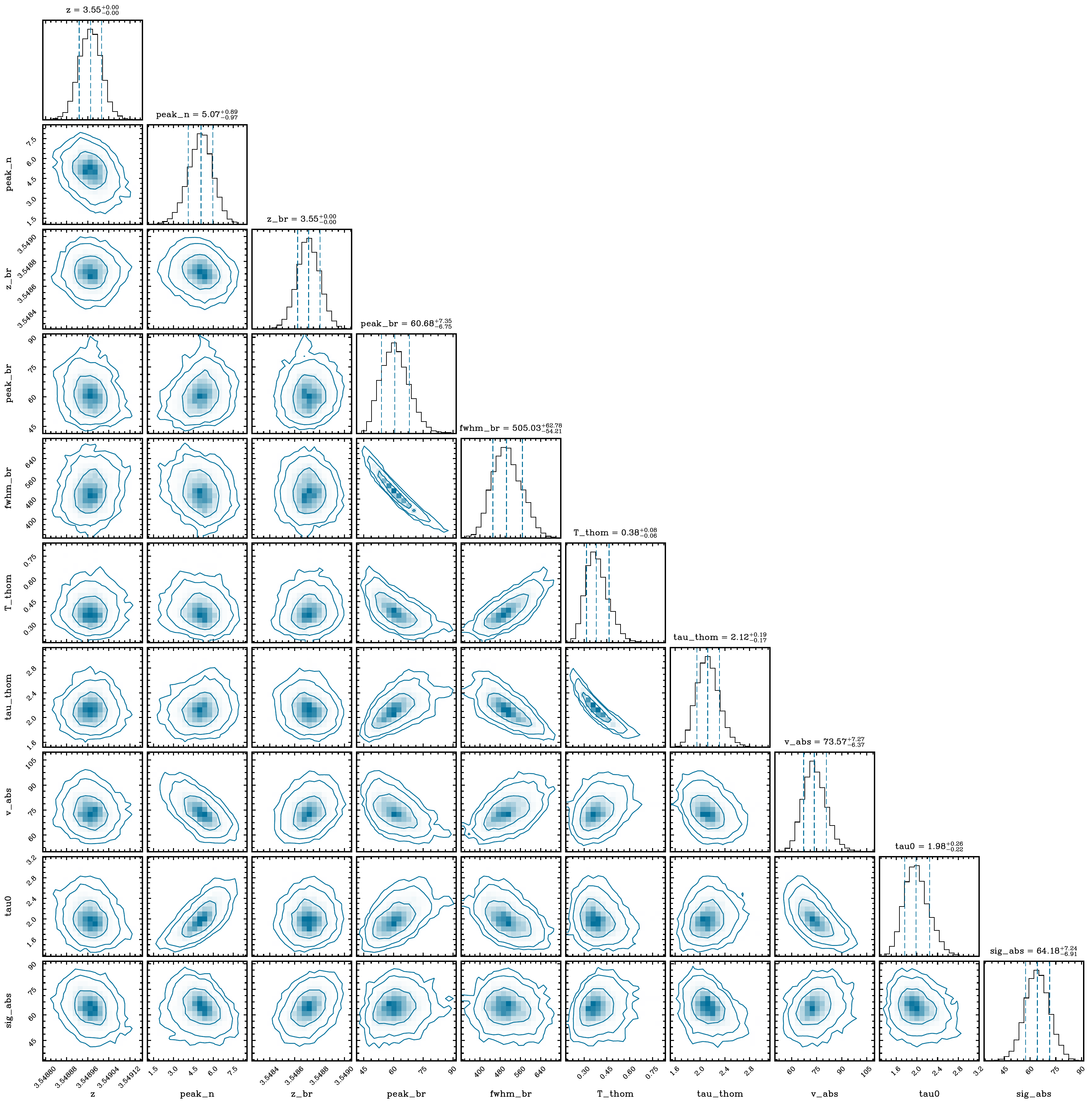}
\caption{Posterior distributions of the fitted parameters in the \textit{Secondary} \Halpha model, including the best fit values and their associated errors. Contours enclose 68\%, 95\%, and 99\% of the distributions.}
\label{fig:Ha_secondary_posteriors}
\end{figure*}

In this section, we present the posteriors from the best-fit spectral models described in Section~\ref{sec:fitting_Hb_OIII} and \ref{sec:Fitting_Ha}. Fig.~\ref{fig:Hb_OIII_posteriors} shows the posteriors from fitting the \Hbeta-\OIIIL complex (Section~\ref{sec:fitting_Hb_OIII}), while Fig.~\ref{fig:Ha_fiducial_posteriors} and \ref{fig:Ha_secondary_posteriors} show the posteriors from the Fiducial and Secondary \Halpha model fits, respectively. We only include posteriors for the key fitted parameters to provide clarity about potential correlations, so for the sake of brevity fixed parameters (e.g. $\mathrm{FWHM}_\mathrm{N}$ for \Halpha) or parameters not key to our analysis (e.g. continuum slopes) have been omitted from these figures.

\bsp	
\label{lastpage}
\end{document}